\documentclass[longauth]{aa}

\usepackage{graphicx}
\usepackage{natbib}
\usepackage{scalerel}
\usepackage[nolist,nohyperlinks]{acronym}
\usepackage[normalem]{ulem}

\usepackage[table]{xcolor}
\usepackage{txfonts}
\usepackage[pdfencoding=auto,psdextra]{hyperref}
\hypersetup{
    colorlinks=true,
    linkcolor=blue,
    filecolor=magenta,      
    urlcolor=blue,
    citecolor=blue
}
\usepackage{orcidlink} 
\newcommand{\orcid}[1]{\orcidlink{#1}}

\usepackage[nameinlink,capitalise]{cleveref}
\crefname{section}{Sect.}{Sects.}
\Crefname{section}{Section}{Sections}
\crefname{figure}{Fig.}{Figs.}
\Crefname{figure}{Figure}{Figures}
\crefname{equation}{Eq.}{Eqs.}
\Crefname{equation}{Equation}{Equations}

\makeatletter
\renewcommand*\aa@pageof{, page \thepage{} of \pageref*{LastPage}}
\makeatother

\usepackage[utf8]{inputenc}

\usepackage[switch, modulo]{lineno}

\usepackage{euclid}

\newcommand{\Phosphoros}{\texttt{Phosphoros}\xspace}
\newcommand{\NNPZ}{\texttt{NNPZ}\xspace}
\newcommand{\PRF}{\texttt{PRF}\xspace}
\newcommand{\RF}{\texttt{RF}\xspace}

\definecolor{darkpink}{rgb}{0.91, 0.33, 0.5}

\begin{document}

\title{Euclid Quick Data Release (Q1)}
\subtitle{Photometric redshifts and physical properties of galaxies through the PHZ processing function}

\author{Euclid Collaboration: M.~Tucci\thanks{\email{marco.tucci@unige.ch}}\inst{\ref{aff1}}
\and S.~Paltani\orcid{0000-0002-8108-9179}\inst{\ref{aff1}}
\and W.~G.~Hartley\inst{\ref{aff1}}
\and F.~Dubath\orcid{0000-0002-6533-2810}\inst{\ref{aff1}}
\and N.~Morisset\inst{\ref{aff1}}
\and M.~Bolzonella\orcid{0000-0003-3278-4607}\inst{\ref{aff2}}
\and S.~Fotopoulou\orcid{0000-0002-9686-254X}\inst{\ref{aff3}}
\and F.~Tarsitano\orcid{0000-0002-5919-0238}\inst{\ref{aff1}}
\and C.~Saulder\orcid{0000-0002-0408-5633}\inst{\ref{aff4},\ref{aff5}}
\and L.~Pozzetti\orcid{0000-0001-7085-0412}\inst{\ref{aff2}}
\and A.~Enia\orcid{0000-0002-0200-2857}\inst{\ref{aff6},\ref{aff2}}
\and Y.~Kang\orcid{0009-0000-8588-7250}\inst{\ref{aff1}}
\and H.~Degaudenzi\orcid{0000-0002-5887-6799}\inst{\ref{aff1}}
\and R.~Saglia\orcid{0000-0003-0378-7032}\inst{\ref{aff5},\ref{aff4}}
\and M.~Salvato\orcid{0000-0001-7116-9303}\inst{\ref{aff4}}
\and O.~Ilbert\orcid{0000-0002-7303-4397}\inst{\ref{aff7}}
\and S.~A.~Stanford\orcid{0000-0003-0122-0841}\inst{\ref{aff8}}
\and W.~Roster\orcid{0000-0002-9149-6528}\inst{\ref{aff4}}
\and F.~J.~Castander\orcid{0000-0001-7316-4573}\inst{\ref{aff9},\ref{aff10}}
\and A.~Humphrey\orcid{0000-0002-0510-2351}\inst{\ref{aff11},\ref{aff12}}
\and H.~Landt\orcid{0000-0001-8391-6900}\inst{\ref{aff13}}
\and M.~Selwood\orcid{0009-0002-3235-0825}\inst{\ref{aff3}}
\and G.~Stevens\orcid{0000-0002-8885-4443}\inst{\ref{aff3}}
\and N.~Aghanim\orcid{0000-0002-6688-8992}\inst{\ref{aff14}}
\and B.~Altieri\orcid{0000-0003-3936-0284}\inst{\ref{aff15}}
\and A.~Amara\inst{\ref{aff16}}
\and S.~Andreon\orcid{0000-0002-2041-8784}\inst{\ref{aff17}}
\and N.~Auricchio\orcid{0000-0003-4444-8651}\inst{\ref{aff2}}
\and H.~Aussel\orcid{0000-0002-1371-5705}\inst{\ref{aff18}}
\and C.~Baccigalupi\orcid{0000-0002-8211-1630}\inst{\ref{aff19},\ref{aff20},\ref{aff21},\ref{aff22}}
\and M.~Baldi\orcid{0000-0003-4145-1943}\inst{\ref{aff6},\ref{aff2},\ref{aff23}}
\and A.~Balestra\orcid{0000-0002-6967-261X}\inst{\ref{aff24}}
\and S.~Bardelli\orcid{0000-0002-8900-0298}\inst{\ref{aff2}}
\and P.~Battaglia\orcid{0000-0002-7337-5909}\inst{\ref{aff2}}
\and A.~N.~Belikov\inst{\ref{aff25},\ref{aff26}}
\and F.~Bernardeau\inst{\ref{aff27},\ref{aff28}}
\and A.~Biviano\orcid{0000-0002-0857-0732}\inst{\ref{aff20},\ref{aff19}}
\and A.~Bonchi\orcid{0000-0002-2667-5482}\inst{\ref{aff29}}
\and E.~Branchini\orcid{0000-0002-0808-6908}\inst{\ref{aff30},\ref{aff31},\ref{aff17}}
\and M.~Brescia\orcid{0000-0001-9506-5680}\inst{\ref{aff32},\ref{aff33}}
\and J.~Brinchmann\orcid{0000-0003-4359-8797}\inst{\ref{aff11},\ref{aff34}}
\and S.~Camera\orcid{0000-0003-3399-3574}\inst{\ref{aff35},\ref{aff36},\ref{aff37}}
\and G.~Ca\~nas-Herrera\orcid{0000-0003-2796-2149}\inst{\ref{aff38},\ref{aff39},\ref{aff40}}
\and V.~Capobianco\orcid{0000-0002-3309-7692}\inst{\ref{aff37}}
\and C.~Carbone\orcid{0000-0003-0125-3563}\inst{\ref{aff41}}
\and J.~Carretero\orcid{0000-0002-3130-0204}\inst{\ref{aff42},\ref{aff43}}
\and S.~Casas\orcid{0000-0002-4751-5138}\inst{\ref{aff44}}
\and M.~Castellano\orcid{0000-0001-9875-8263}\inst{\ref{aff45}}
\and G.~Castignani\orcid{0000-0001-6831-0687}\inst{\ref{aff2}}
\and S.~Cavuoti\orcid{0000-0002-3787-4196}\inst{\ref{aff33},\ref{aff46}}
\and K.~C.~Chambers\orcid{0000-0001-6965-7789}\inst{\ref{aff47}}
\and A.~Cimatti\inst{\ref{aff48}}
\and C.~Colodro-Conde\inst{\ref{aff49}}
\and G.~Congedo\orcid{0000-0003-2508-0046}\inst{\ref{aff50}}
\and C.~J.~Conselice\orcid{0000-0003-1949-7638}\inst{\ref{aff51}}
\and L.~Conversi\orcid{0000-0002-6710-8476}\inst{\ref{aff52},\ref{aff15}}
\and Y.~Copin\orcid{0000-0002-5317-7518}\inst{\ref{aff53}}
\and F.~Courbin\orcid{0000-0003-0758-6510}\inst{\ref{aff54},\ref{aff55}}
\and H.~M.~Courtois\orcid{0000-0003-0509-1776}\inst{\ref{aff56}}
\and M.~Cropper\orcid{0000-0003-4571-9468}\inst{\ref{aff57}}
\and A.~Da~Silva\orcid{0000-0002-6385-1609}\inst{\ref{aff58},\ref{aff59}}
\and G.~De~Lucia\orcid{0000-0002-6220-9104}\inst{\ref{aff20}}
\and A.~M.~Di~Giorgio\orcid{0000-0002-4767-2360}\inst{\ref{aff60}}
\and H.~Dole\orcid{0000-0002-9767-3839}\inst{\ref{aff14}}
\and C.~A.~J.~Duncan\orcid{0009-0003-3573-0791}\inst{\ref{aff50},\ref{aff51}}
\and X.~Dupac\inst{\ref{aff15}}
\and A.~Ealet\orcid{0000-0003-3070-014X}\inst{\ref{aff53}}
\and S.~Escoffier\orcid{0000-0002-2847-7498}\inst{\ref{aff61}}
\and M.~Fabricius\orcid{0000-0002-7025-6058}\inst{\ref{aff4},\ref{aff5}}
\and M.~Farina\orcid{0000-0002-3089-7846}\inst{\ref{aff60}}
\and R.~Farinelli\inst{\ref{aff2}}
\and S.~Ferriol\inst{\ref{aff53}}
\and F.~Finelli\orcid{0000-0002-6694-3269}\inst{\ref{aff2},\ref{aff62}}
\and N.~Fourmanoit\orcid{0009-0005-6816-6925}\inst{\ref{aff61}}
\and M.~Frailis\orcid{0000-0002-7400-2135}\inst{\ref{aff20}}
\and E.~Franceschi\orcid{0000-0002-0585-6591}\inst{\ref{aff2}}
\and M.~Fumana\orcid{0000-0001-6787-5950}\inst{\ref{aff41}}
\and S.~Galeotta\orcid{0000-0002-3748-5115}\inst{\ref{aff20}}
\and K.~George\orcid{0000-0002-1734-8455}\inst{\ref{aff5}}
\and W.~Gillard\orcid{0000-0003-4744-9748}\inst{\ref{aff61}}
\and B.~Gillis\orcid{0000-0002-4478-1270}\inst{\ref{aff50}}
\and C.~Giocoli\orcid{0000-0002-9590-7961}\inst{\ref{aff2},\ref{aff23}}
\and P.~G\'omez-Alvarez\orcid{0000-0002-8594-5358}\inst{\ref{aff63},\ref{aff15}}
\and J.~Gracia-Carpio\inst{\ref{aff4}}
\and B.~R.~Granett\orcid{0000-0003-2694-9284}\inst{\ref{aff17}}
\and A.~Grazian\orcid{0000-0002-5688-0663}\inst{\ref{aff24}}
\and F.~Grupp\inst{\ref{aff4},\ref{aff5}}
\and S.~V.~H.~Haugan\orcid{0000-0001-9648-7260}\inst{\ref{aff64}}
\and J.~Hoar\inst{\ref{aff15}}
\and H.~Hoekstra\orcid{0000-0002-0641-3231}\inst{\ref{aff40}}
\and W.~Holmes\inst{\ref{aff65}}
\and I.~M.~Hook\orcid{0000-0002-2960-978X}\inst{\ref{aff66}}
\and F.~Hormuth\inst{\ref{aff67}}
\and A.~Hornstrup\orcid{0000-0002-3363-0936}\inst{\ref{aff68},\ref{aff69}}
\and P.~Hudelot\inst{\ref{aff28}}
\and K.~Jahnke\orcid{0000-0003-3804-2137}\inst{\ref{aff70}}
\and M.~Jhabvala\inst{\ref{aff71}}
\and B.~Joachimi\orcid{0000-0001-7494-1303}\inst{\ref{aff72}}
\and E.~Keih\"anen\orcid{0000-0003-1804-7715}\inst{\ref{aff73}}
\and S.~Kermiche\orcid{0000-0002-0302-5735}\inst{\ref{aff61}}
\and A.~Kiessling\orcid{0000-0002-2590-1273}\inst{\ref{aff65}}
\and M.~Kilbinger\orcid{0000-0001-9513-7138}\inst{\ref{aff18}}
\and B.~Kubik\orcid{0009-0006-5823-4880}\inst{\ref{aff53}}
\and M.~K\"ummel\orcid{0000-0003-2791-2117}\inst{\ref{aff5}}
\and M.~Kunz\orcid{0000-0002-3052-7394}\inst{\ref{aff74}}
\and H.~Kurki-Suonio\orcid{0000-0002-4618-3063}\inst{\ref{aff75},\ref{aff76}}
\and Q.~Le~Boulc'h\inst{\ref{aff77}}
\and A.~M.~C.~Le~Brun\orcid{0000-0002-0936-4594}\inst{\ref{aff78}}
\and D.~Le~Mignant\orcid{0000-0002-5339-5515}\inst{\ref{aff7}}
\and P.~Liebing\inst{\ref{aff57}}
\and S.~Ligori\orcid{0000-0003-4172-4606}\inst{\ref{aff37}}
\and P.~B.~Lilje\orcid{0000-0003-4324-7794}\inst{\ref{aff64}}
\and V.~Lindholm\orcid{0000-0003-2317-5471}\inst{\ref{aff75},\ref{aff76}}
\and I.~Lloro\orcid{0000-0001-5966-1434}\inst{\ref{aff79}}
\and G.~Mainetti\orcid{0000-0003-2384-2377}\inst{\ref{aff77}}
\and D.~Maino\inst{\ref{aff80},\ref{aff41},\ref{aff81}}
\and E.~Maiorano\orcid{0000-0003-2593-4355}\inst{\ref{aff2}}
\and O.~Mansutti\orcid{0000-0001-5758-4658}\inst{\ref{aff20}}
\and S.~Marcin\inst{\ref{aff82}}
\and O.~Marggraf\orcid{0000-0001-7242-3852}\inst{\ref{aff83}}
\and M.~Martinelli\orcid{0000-0002-6943-7732}\inst{\ref{aff45},\ref{aff84}}
\and N.~Martinet\orcid{0000-0003-2786-7790}\inst{\ref{aff7}}
\and F.~Marulli\orcid{0000-0002-8850-0303}\inst{\ref{aff85},\ref{aff2},\ref{aff23}}
\and R.~Massey\orcid{0000-0002-6085-3780}\inst{\ref{aff86}}
\and D.~C.~Masters\orcid{0000-0001-5382-6138}\inst{\ref{aff87}}
\and S.~Maurogordato\inst{\ref{aff88}}
\and H.~J.~McCracken\orcid{0000-0002-9489-7765}\inst{\ref{aff28}}
\and E.~Medinaceli\orcid{0000-0002-4040-7783}\inst{\ref{aff2}}
\and S.~Mei\orcid{0000-0002-2849-559X}\inst{\ref{aff89},\ref{aff90}}
\and M.~Melchior\inst{\ref{aff91}}
\and Y.~Mellier\inst{\ref{aff92},\ref{aff28}}
\and M.~Meneghetti\orcid{0000-0003-1225-7084}\inst{\ref{aff2},\ref{aff23}}
\and E.~Merlin\orcid{0000-0001-6870-8900}\inst{\ref{aff45}}
\and G.~Meylan\inst{\ref{aff93}}
\and A.~Mora\orcid{0000-0002-1922-8529}\inst{\ref{aff94}}
\and M.~Moresco\orcid{0000-0002-7616-7136}\inst{\ref{aff85},\ref{aff2}}
\and L.~Moscardini\orcid{0000-0002-3473-6716}\inst{\ref{aff85},\ref{aff2},\ref{aff23}}
\and R.~Nakajima\orcid{0009-0009-1213-7040}\inst{\ref{aff83}}
\and C.~Neissner\orcid{0000-0001-8524-4968}\inst{\ref{aff95},\ref{aff43}}
\and S.-M.~Niemi\inst{\ref{aff38}}
\and J.~W.~Nightingale\orcid{0000-0002-8987-7401}\inst{\ref{aff96}}
\and C.~Padilla\orcid{0000-0001-7951-0166}\inst{\ref{aff95}}
\and F.~Pasian\orcid{0000-0002-4869-3227}\inst{\ref{aff20}}
\and K.~Pedersen\inst{\ref{aff97}}
\and W.~J.~Percival\orcid{0000-0002-0644-5727}\inst{\ref{aff98},\ref{aff99},\ref{aff100}}
\and V.~Pettorino\inst{\ref{aff38}}
\and S.~Pires\orcid{0000-0002-0249-2104}\inst{\ref{aff18}}
\and G.~Polenta\orcid{0000-0003-4067-9196}\inst{\ref{aff29}}
\and M.~Poncet\inst{\ref{aff101}}
\and L.~A.~Popa\inst{\ref{aff102}}
\and G.~D.~Racca\inst{\ref{aff38},\ref{aff40}}
\and F.~Raison\orcid{0000-0002-7819-6918}\inst{\ref{aff4}}
\and R.~Rebolo\orcid{0000-0003-3767-7085}\inst{\ref{aff49},\ref{aff103},\ref{aff104}}
\and A.~Renzi\orcid{0000-0001-9856-1970}\inst{\ref{aff105},\ref{aff106}}
\and J.~Rhodes\orcid{0000-0002-4485-8549}\inst{\ref{aff65}}
\and G.~Riccio\inst{\ref{aff33}}
\and E.~Romelli\orcid{0000-0003-3069-9222}\inst{\ref{aff20}}
\and M.~Roncarelli\orcid{0000-0001-9587-7822}\inst{\ref{aff2}}
\and B.~Rusholme\orcid{0000-0001-7648-4142}\inst{\ref{aff107}}
\and Z.~Sakr\orcid{0000-0002-4823-3757}\inst{\ref{aff108},\ref{aff109},\ref{aff110}}
\and A.~G.~S\'anchez\orcid{0000-0003-1198-831X}\inst{\ref{aff4}}
\and D.~Sapone\orcid{0000-0001-7089-4503}\inst{\ref{aff111}}
\and B.~Sartoris\orcid{0000-0003-1337-5269}\inst{\ref{aff5},\ref{aff20}}
\and J.~A.~Schewtschenko\orcid{0000-0002-4913-6393}\inst{\ref{aff50}}
\and M.~Schirmer\orcid{0000-0003-2568-9994}\inst{\ref{aff70}}
\and P.~Schneider\orcid{0000-0001-8561-2679}\inst{\ref{aff83}}
\and A.~Secroun\orcid{0000-0003-0505-3710}\inst{\ref{aff61}}
\and E.~Sefusatti\orcid{0000-0003-0473-1567}\inst{\ref{aff20},\ref{aff19},\ref{aff21}}
\and M.~Seiffert\orcid{0000-0002-7536-9393}\inst{\ref{aff65}}
\and S.~Serrano\orcid{0000-0002-0211-2861}\inst{\ref{aff10},\ref{aff112},\ref{aff9}}
\and P.~Simon\inst{\ref{aff83}}
\and C.~Sirignano\orcid{0000-0002-0995-7146}\inst{\ref{aff105},\ref{aff106}}
\and G.~Sirri\orcid{0000-0003-2626-2853}\inst{\ref{aff23}}
\and A.~Spurio~Mancini\orcid{0000-0001-5698-0990}\inst{\ref{aff113}}
\and L.~Stanco\orcid{0000-0002-9706-5104}\inst{\ref{aff106}}
\and J.~Steinwagner\orcid{0000-0001-7443-1047}\inst{\ref{aff4}}
\and P.~Tallada-Cresp\'{i}\orcid{0000-0002-1336-8328}\inst{\ref{aff42},\ref{aff43}}
\and A.~N.~Taylor\inst{\ref{aff50}}
\and H.~I.~Teplitz\orcid{0000-0002-7064-5424}\inst{\ref{aff87}}
\and I.~Tereno\inst{\ref{aff58},\ref{aff114}}
\and N.~Tessore\orcid{0000-0002-9696-7931}\inst{\ref{aff72}}
\and S.~Toft\orcid{0000-0003-3631-7176}\inst{\ref{aff115},\ref{aff116}}
\and R.~Toledo-Moreo\orcid{0000-0002-2997-4859}\inst{\ref{aff117}}
\and F.~Torradeflot\orcid{0000-0003-1160-1517}\inst{\ref{aff43},\ref{aff42}}
\and I.~Tutusaus\orcid{0000-0002-3199-0399}\inst{\ref{aff109}}
\and E.~A.~Valentijn\inst{\ref{aff25}}
\and L.~Valenziano\orcid{0000-0002-1170-0104}\inst{\ref{aff2},\ref{aff62}}
\and J.~Valiviita\orcid{0000-0001-6225-3693}\inst{\ref{aff75},\ref{aff76}}
\and T.~Vassallo\orcid{0000-0001-6512-6358}\inst{\ref{aff5},\ref{aff20}}
\and G.~Verdoes~Kleijn\orcid{0000-0001-5803-2580}\inst{\ref{aff25}}
\and A.~Veropalumbo\orcid{0000-0003-2387-1194}\inst{\ref{aff17},\ref{aff31},\ref{aff30}}
\and Y.~Wang\orcid{0000-0002-4749-2984}\inst{\ref{aff87}}
\and J.~Weller\orcid{0000-0002-8282-2010}\inst{\ref{aff5},\ref{aff4}}
\and A.~Zacchei\orcid{0000-0003-0396-1192}\inst{\ref{aff20},\ref{aff19}}
\and G.~Zamorani\orcid{0000-0002-2318-301X}\inst{\ref{aff2}}
\and F.~M.~Zerbi\inst{\ref{aff17}}
\and I.~A.~Zinchenko\orcid{0000-0002-2944-2449}\inst{\ref{aff5}}
\and E.~Zucca\orcid{0000-0002-5845-8132}\inst{\ref{aff2}}
\and V.~Allevato\orcid{0000-0001-7232-5152}\inst{\ref{aff33}}
\and M.~Ballardini\orcid{0000-0003-4481-3559}\inst{\ref{aff118},\ref{aff119},\ref{aff2}}
\and C.~Burigana\orcid{0000-0002-3005-5796}\inst{\ref{aff120},\ref{aff62}}
\and R.~Cabanac\orcid{0000-0001-6679-2600}\inst{\ref{aff109}}
\and A.~Cappi\inst{\ref{aff2},\ref{aff88}}
\and P.~Casenove\orcid{0009-0006-6736-1670}\inst{\ref{aff101}}
\and D.~Di~Ferdinando\inst{\ref{aff23}}
\and J.~A.~Escartin~Vigo\inst{\ref{aff4}}
\and G.~Fabbian\orcid{0000-0002-3255-4695}\inst{\ref{aff121}}
\and L.~Gabarra\orcid{0000-0002-8486-8856}\inst{\ref{aff122}}
\and M.~Huertas-Company\orcid{0000-0002-1416-8483}\inst{\ref{aff49},\ref{aff123},\ref{aff124},\ref{aff125}}
\and J.~Mart\'{i}n-Fleitas\orcid{0000-0002-8594-569X}\inst{\ref{aff94}}
\and S.~Matthew\orcid{0000-0001-8448-1697}\inst{\ref{aff50}}
\and M.~Maturi\orcid{0000-0002-3517-2422}\inst{\ref{aff108},\ref{aff126}}
\and N.~Mauri\orcid{0000-0001-8196-1548}\inst{\ref{aff48},\ref{aff23}}
\and R.~B.~Metcalf\orcid{0000-0003-3167-2574}\inst{\ref{aff85},\ref{aff2}}
\and A.~A.~Nucita\inst{\ref{aff127},\ref{aff128},\ref{aff129}}
\and A.~Pezzotta\orcid{0000-0003-0726-2268}\inst{\ref{aff130},\ref{aff4}}
\and M.~P\"ontinen\orcid{0000-0001-5442-2530}\inst{\ref{aff75}}
\and C.~Porciani\orcid{0000-0002-7797-2508}\inst{\ref{aff83}}
\and I.~Risso\orcid{0000-0003-2525-7761}\inst{\ref{aff131}}
\and V.~Scottez\inst{\ref{aff92},\ref{aff132}}
\and M.~Sereno\orcid{0000-0003-0302-0325}\inst{\ref{aff2},\ref{aff23}}
\and M.~Tenti\orcid{0000-0002-4254-5901}\inst{\ref{aff23}}
\and M.~Viel\orcid{0000-0002-2642-5707}\inst{\ref{aff19},\ref{aff20},\ref{aff22},\ref{aff21},\ref{aff133}}
\and M.~Wiesmann\orcid{0009-0000-8199-5860}\inst{\ref{aff64}}
\and Y.~Akrami\orcid{0000-0002-2407-7956}\inst{\ref{aff134},\ref{aff135}}
\and I.~T.~Andika\orcid{0000-0001-6102-9526}\inst{\ref{aff136},\ref{aff137}}
\and S.~Anselmi\orcid{0000-0002-3579-9583}\inst{\ref{aff106},\ref{aff105},\ref{aff138}}
\and M.~Archidiacono\orcid{0000-0003-4952-9012}\inst{\ref{aff80},\ref{aff81}}
\and F.~Atrio-Barandela\orcid{0000-0002-2130-2513}\inst{\ref{aff139}}
\and C.~Benoist\inst{\ref{aff88}}
\and K.~Benson\inst{\ref{aff57}}
\and D.~Bertacca\orcid{0000-0002-2490-7139}\inst{\ref{aff105},\ref{aff24},\ref{aff106}}
\and M.~Bethermin\orcid{0000-0002-3915-2015}\inst{\ref{aff140}}
\and L.~Bisigello\orcid{0000-0003-0492-4924}\inst{\ref{aff24}}
\and A.~Blanchard\orcid{0000-0001-8555-9003}\inst{\ref{aff109}}
\and L.~Blot\orcid{0000-0002-9622-7167}\inst{\ref{aff141},\ref{aff138}}
\and H.~B\"ohringer\orcid{0000-0001-8241-4204}\inst{\ref{aff4},\ref{aff142},\ref{aff143}}
\and S.~Borgani\orcid{0000-0001-6151-6439}\inst{\ref{aff144},\ref{aff19},\ref{aff20},\ref{aff21},\ref{aff133}}
\and M.~L.~Brown\orcid{0000-0002-0370-8077}\inst{\ref{aff51}}
\and S.~Bruton\orcid{0000-0002-6503-5218}\inst{\ref{aff145}}
\and A.~Calabro\orcid{0000-0003-2536-1614}\inst{\ref{aff45}}
\and B.~Camacho~Quevedo\orcid{0000-0002-8789-4232}\inst{\ref{aff10},\ref{aff9}}
\and F.~Caro\inst{\ref{aff45}}
\and C.~S.~Carvalho\inst{\ref{aff114}}
\and T.~Castro\orcid{0000-0002-6292-3228}\inst{\ref{aff20},\ref{aff21},\ref{aff19},\ref{aff133}}
\and R.~Chary\orcid{0000-0001-7583-0621}\inst{\ref{aff87},\ref{aff146}}
\and F.~Cogato\orcid{0000-0003-4632-6113}\inst{\ref{aff85},\ref{aff2}}
\and A.~R.~Cooray\orcid{0000-0002-3892-0190}\inst{\ref{aff147}}
\and O.~Cucciati\orcid{0000-0002-9336-7551}\inst{\ref{aff2}}
\and S.~Davini\orcid{0000-0003-3269-1718}\inst{\ref{aff31}}
\and F.~De~Paolis\orcid{0000-0001-6460-7563}\inst{\ref{aff127},\ref{aff128},\ref{aff129}}
\and G.~Desprez\orcid{0000-0001-8325-1742}\inst{\ref{aff25}}
\and A.~D\'iaz-S\'anchez\orcid{0000-0003-0748-4768}\inst{\ref{aff148}}
\and J.~J.~Diaz\inst{\ref{aff49}}
\and S.~Di~Domizio\orcid{0000-0003-2863-5895}\inst{\ref{aff30},\ref{aff31}}
\and J.~M.~Diego\orcid{0000-0001-9065-3926}\inst{\ref{aff149}}
\and P.-A.~Duc\orcid{0000-0003-3343-6284}\inst{\ref{aff140}}
\and Y.~Fang\inst{\ref{aff5}}
\and A.~M.~N.~Ferguson\inst{\ref{aff50}}
\and A.~G.~Ferrari\orcid{0009-0005-5266-4110}\inst{\ref{aff23}}
\and P.~G.~Ferreira\orcid{0000-0002-3021-2851}\inst{\ref{aff122}}
\and A.~Finoguenov\orcid{0000-0002-4606-5403}\inst{\ref{aff75}}
\and A.~Fontana\orcid{0000-0003-3820-2823}\inst{\ref{aff45}}
\and A.~Franco\orcid{0000-0002-4761-366X}\inst{\ref{aff128},\ref{aff127},\ref{aff129}}
\and K.~Ganga\orcid{0000-0001-8159-8208}\inst{\ref{aff89}}
\and J.~Garc\'ia-Bellido\orcid{0000-0002-9370-8360}\inst{\ref{aff134}}
\and T.~Gasparetto\orcid{0000-0002-7913-4866}\inst{\ref{aff20}}
\and V.~Gautard\inst{\ref{aff150}}
\and E.~Gaztanaga\orcid{0000-0001-9632-0815}\inst{\ref{aff9},\ref{aff10},\ref{aff151}}
\and F.~Giacomini\orcid{0000-0002-3129-2814}\inst{\ref{aff23}}
\and A.~H.~Gonzalez\orcid{0000-0002-0933-8601}\inst{\ref{aff152}}
\and G.~Gozaliasl\orcid{0000-0002-0236-919X}\inst{\ref{aff153},\ref{aff75}}
\and M.~Guidi\orcid{0000-0001-9408-1101}\inst{\ref{aff6},\ref{aff2}}
\and C.~M.~Gutierrez\orcid{0000-0001-7854-783X}\inst{\ref{aff154}}
\and A.~Hall\orcid{0000-0002-3139-8651}\inst{\ref{aff50}}
\and S.~Hemmati\orcid{0000-0003-2226-5395}\inst{\ref{aff107}}
\and C.~Hern\'andez-Monteagudo\orcid{0000-0001-5471-9166}\inst{\ref{aff104},\ref{aff49}}
\and H.~Hildebrandt\orcid{0000-0002-9814-3338}\inst{\ref{aff155}}
\and J.~Hjorth\orcid{0000-0002-4571-2306}\inst{\ref{aff97}}
\and J.~J.~E.~Kajava\orcid{0000-0002-3010-8333}\inst{\ref{aff156},\ref{aff157}}
\and V.~Kansal\orcid{0000-0002-4008-6078}\inst{\ref{aff158},\ref{aff159}}
\and D.~Karagiannis\orcid{0000-0002-4927-0816}\inst{\ref{aff118},\ref{aff160}}
\and K.~Kiiveri\inst{\ref{aff73}}
\and C.~C.~Kirkpatrick\inst{\ref{aff73}}
\and S.~Kruk\orcid{0000-0001-8010-8879}\inst{\ref{aff15}}
\and J.~Le~Graet\orcid{0000-0001-6523-7971}\inst{\ref{aff61}}
\and L.~Legrand\orcid{0000-0003-0610-5252}\inst{\ref{aff161},\ref{aff162}}
\and M.~Lembo\orcid{0000-0002-5271-5070}\inst{\ref{aff118},\ref{aff119}}
\and F.~Lepori\orcid{0009-0000-5061-7138}\inst{\ref{aff163}}
\and G.~Leroy\orcid{0009-0004-2523-4425}\inst{\ref{aff13},\ref{aff86}}
\and G.~F.~Lesci\orcid{0000-0002-4607-2830}\inst{\ref{aff85},\ref{aff2}}
\and J.~Lesgourgues\orcid{0000-0001-7627-353X}\inst{\ref{aff44}}
\and L.~Leuzzi\orcid{0009-0006-4479-7017}\inst{\ref{aff85},\ref{aff2}}
\and T.~I.~Liaudat\orcid{0000-0002-9104-314X}\inst{\ref{aff164}}
\and A.~Loureiro\orcid{0000-0002-4371-0876}\inst{\ref{aff165},\ref{aff166}}
\and J.~Macias-Perez\orcid{0000-0002-5385-2763}\inst{\ref{aff167}}
\and G.~Maggio\orcid{0000-0003-4020-4836}\inst{\ref{aff20}}
\and M.~Magliocchetti\orcid{0000-0001-9158-4838}\inst{\ref{aff60}}
\and E.~A.~Magnier\orcid{0000-0002-7965-2815}\inst{\ref{aff47}}
\and F.~Mannucci\orcid{0000-0002-4803-2381}\inst{\ref{aff168}}
\and R.~Maoli\orcid{0000-0002-6065-3025}\inst{\ref{aff169},\ref{aff45}}
\and C.~J.~A.~P.~Martins\orcid{0000-0002-4886-9261}\inst{\ref{aff170},\ref{aff11}}
\and L.~Maurin\orcid{0000-0002-8406-0857}\inst{\ref{aff14}}
\and M.~Miluzio\inst{\ref{aff15},\ref{aff171}}
\and P.~Monaco\orcid{0000-0003-2083-7564}\inst{\ref{aff144},\ref{aff20},\ref{aff21},\ref{aff19}}
\and C.~Moretti\orcid{0000-0003-3314-8936}\inst{\ref{aff22},\ref{aff133},\ref{aff20},\ref{aff19},\ref{aff21}}
\and G.~Morgante\inst{\ref{aff2}}
\and S.~Nadathur\orcid{0000-0001-9070-3102}\inst{\ref{aff151}}
\and K.~Naidoo\orcid{0000-0002-9182-1802}\inst{\ref{aff151}}
\and A.~Navarro-Alsina\orcid{0000-0002-3173-2592}\inst{\ref{aff83}}
\and S.~Nesseris\orcid{0000-0002-0567-0324}\inst{\ref{aff134}}
\and M.~Oguri\orcid{0000-0003-3484-399X}\inst{\ref{aff172},\ref{aff173}}
\and F.~Passalacqua\orcid{0000-0002-8606-4093}\inst{\ref{aff105},\ref{aff106}}
\and K.~Paterson\orcid{0000-0001-8340-3486}\inst{\ref{aff70}}
\and L.~Patrizii\inst{\ref{aff23}}
\and A.~Pisani\orcid{0000-0002-6146-4437}\inst{\ref{aff61},\ref{aff174}}
\and D.~Potter\orcid{0000-0002-0757-5195}\inst{\ref{aff163}}
\and S.~Quai\orcid{0000-0002-0449-8163}\inst{\ref{aff85},\ref{aff2}}
\and M.~Radovich\orcid{0000-0002-3585-866X}\inst{\ref{aff24}}
\and P.-F.~Rocci\inst{\ref{aff14}}
\and G.~Rodighiero\orcid{0000-0002-9415-2296}\inst{\ref{aff105},\ref{aff24}}
\and S.~Sacquegna\orcid{0000-0002-8433-6630}\inst{\ref{aff127},\ref{aff128},\ref{aff129}}
\and M.~Sahl\'en\orcid{0000-0003-0973-4804}\inst{\ref{aff175}}
\and D.~B.~Sanders\orcid{0000-0002-1233-9998}\inst{\ref{aff47}}
\and E.~Sarpa\orcid{0000-0002-1256-655X}\inst{\ref{aff22},\ref{aff133},\ref{aff21}}
\and C.~Scarlata\orcid{0000-0002-9136-8876}\inst{\ref{aff176}}
\and J.~Schaye\orcid{0000-0002-0668-5560}\inst{\ref{aff40}}
\and A.~Schneider\orcid{0000-0001-7055-8104}\inst{\ref{aff163}}
\and D.~Sciotti\orcid{0009-0008-4519-2620}\inst{\ref{aff45},\ref{aff84}}
\and E.~Sellentin\inst{\ref{aff177},\ref{aff40}}
\and L.~C.~Smith\orcid{0000-0002-3259-2771}\inst{\ref{aff178}}
\and K.~Tanidis\orcid{0000-0001-9843-5130}\inst{\ref{aff122}}
\and G.~Testera\inst{\ref{aff31}}
\and R.~Teyssier\orcid{0000-0001-7689-0933}\inst{\ref{aff174}}
\and S.~Tosi\orcid{0000-0002-7275-9193}\inst{\ref{aff30},\ref{aff31},\ref{aff17}}
\and A.~Troja\orcid{0000-0003-0239-4595}\inst{\ref{aff105},\ref{aff106}}
\and C.~Valieri\inst{\ref{aff23}}
\and A.~Venhola\orcid{0000-0001-6071-4564}\inst{\ref{aff179}}
\and D.~Vergani\orcid{0000-0003-0898-2216}\inst{\ref{aff2}}
\and G.~Verza\orcid{0000-0002-1886-8348}\inst{\ref{aff180}}
\and P.~Vielzeuf\orcid{0000-0003-2035-9339}\inst{\ref{aff61}}
\and N.~A.~Walton\orcid{0000-0003-3983-8778}\inst{\ref{aff178}}
\and J.~R.~Weaver\orcid{0000-0003-1614-196X}\inst{\ref{aff181}}
\and J.~G.~Sorce\orcid{0000-0002-2307-2432}\inst{\ref{aff182},\ref{aff14}}
\and E.~Soubrie\orcid{0000-0001-9295-1863}\inst{\ref{aff14}}
\and D.~Scott\orcid{0000-0002-6878-9840}\inst{\ref{aff183}}}
										   
\institute{Department of Astronomy, University of Geneva, ch. d'Ecogia 16, 1290 Versoix, Switzerland\label{aff1}
\and
INAF-Osservatorio di Astrofisica e Scienza dello Spazio di Bologna, Via Piero Gobetti 93/3, 40129 Bologna, Italy\label{aff2}
\and
School of Physics, HH Wills Physics Laboratory, University of Bristol, Tyndall Avenue, Bristol, BS8 1TL, UK\label{aff3}
\and
Max Planck Institute for Extraterrestrial Physics, Giessenbachstr. 1, 85748 Garching, Germany\label{aff4}
\and
Universit\"ats-Sternwarte M\"unchen, Fakult\"at f\"ur Physik, Ludwig-Maximilians-Universit\"at M\"unchen, Scheinerstrasse 1, 81679 M\"unchen, Germany\label{aff5}
\and
Dipartimento di Fisica e Astronomia, Universit\`a di Bologna, Via Gobetti 93/2, 40129 Bologna, Italy\label{aff6}
\and
Aix-Marseille Universit\'e, CNRS, CNES, LAM, Marseille, France\label{aff7}
\and
Department of Physics and Astronomy, University of California, Davis, CA 95616, USA\label{aff8}
\and
Institute of Space Sciences (ICE, CSIC), Campus UAB, Carrer de Can Magrans, s/n, 08193 Barcelona, Spain\label{aff9}
\and
Institut d'Estudis Espacials de Catalunya (IEEC),  Edifici RDIT, Campus UPC, 08860 Castelldefels, Barcelona, Spain\label{aff10}
\and
Instituto de Astrof\'isica e Ci\^encias do Espa\c{c}o, Universidade do Porto, CAUP, Rua das Estrelas, PT4150-762 Porto, Portugal\label{aff11}
\and
DTx -- Digital Transformation CoLAB, Building 1, Azur\'em Campus, University of Minho, 4800-058 Guimar\~aes, Portugal\label{aff12}
\and
Department of Physics, Centre for Extragalactic Astronomy, Durham University, South Road, Durham, DH1 3LE, UK\label{aff13}
\and
Universit\'e Paris-Saclay, CNRS, Institut d'astrophysique spatiale, 91405, Orsay, France\label{aff14}
\and
ESAC/ESA, Camino Bajo del Castillo, s/n., Urb. Villafranca del Castillo, 28692 Villanueva de la Ca\~nada, Madrid, Spain\label{aff15}
\and
School of Mathematics and Physics, University of Surrey, Guildford, Surrey, GU2 7XH, UK\label{aff16}
\and
INAF-Osservatorio Astronomico di Brera, Via Brera 28, 20122 Milano, Italy\label{aff17}
\and
Universit\'e Paris-Saclay, Universit\'e Paris Cit\'e, CEA, CNRS, AIM, 91191, Gif-sur-Yvette, France\label{aff18}
\and
IFPU, Institute for Fundamental Physics of the Universe, via Beirut 2, 34151 Trieste, Italy\label{aff19}
\and
INAF-Osservatorio Astronomico di Trieste, Via G. B. Tiepolo 11, 34143 Trieste, Italy\label{aff20}
\and
INFN, Sezione di Trieste, Via Valerio 2, 34127 Trieste TS, Italy\label{aff21}
\and
SISSA, International School for Advanced Studies, Via Bonomea 265, 34136 Trieste TS, Italy\label{aff22}
\and
INFN-Sezione di Bologna, Viale Berti Pichat 6/2, 40127 Bologna, Italy\label{aff23}
\and
INAF-Osservatorio Astronomico di Padova, Via dell'Osservatorio 5, 35122 Padova, Italy\label{aff24}
\and
Kapteyn Astronomical Institute, University of Groningen, PO Box 800, 9700 AV Groningen, The Netherlands\label{aff25}
\and
ATG Europe BV, Huygensstraat 34, 2201 DK Noordwijk, The Netherlands\label{aff26}
\and
Institut de Physique Th\'eorique, CEA, CNRS, Universit\'e Paris-Saclay 91191 Gif-sur-Yvette Cedex, France\label{aff27}
\and
Institut d'Astrophysique de Paris, UMR 7095, CNRS, and Sorbonne Universit\'e, 98 bis boulevard Arago, 75014 Paris, France\label{aff28}
\and
Space Science Data Center, Italian Space Agency, via del Politecnico snc, 00133 Roma, Italy\label{aff29}
\and
Dipartimento di Fisica, Universit\`a di Genova, Via Dodecaneso 33, 16146, Genova, Italy\label{aff30}
\and
INFN-Sezione di Genova, Via Dodecaneso 33, 16146, Genova, Italy\label{aff31}
\and
Department of Physics "E. Pancini", University Federico II, Via Cinthia 6, 80126, Napoli, Italy\label{aff32}
\and
INAF-Osservatorio Astronomico di Capodimonte, Via Moiariello 16, 80131 Napoli, Italy\label{aff33}
\and
Faculdade de Ci\^encias da Universidade do Porto, Rua do Campo de Alegre, 4150-007 Porto, Portugal\label{aff34}
\and
Dipartimento di Fisica, Universit\`a degli Studi di Torino, Via P. Giuria 1, 10125 Torino, Italy\label{aff35}
\and
INFN-Sezione di Torino, Via P. Giuria 1, 10125 Torino, Italy\label{aff36}
\and
INAF-Osservatorio Astrofisico di Torino, Via Osservatorio 20, 10025 Pino Torinese (TO), Italy\label{aff37}
\and
European Space Agency/ESTEC, Keplerlaan 1, 2201 AZ Noordwijk, The Netherlands\label{aff38}
\and
Institute Lorentz, Leiden University, Niels Bohrweg 2, 2333 CA Leiden, The Netherlands\label{aff39}
\and
Leiden Observatory, Leiden University, Einsteinweg 55, 2333 CC Leiden, The Netherlands\label{aff40}
\and
INAF-IASF Milano, Via Alfonso Corti 12, 20133 Milano, Italy\label{aff41}
\and
Centro de Investigaciones Energ\'eticas, Medioambientales y Tecnol\'ogicas (CIEMAT), Avenida Complutense 40, 28040 Madrid, Spain\label{aff42}
\and
Port d'Informaci\'{o} Cient\'{i}fica, Campus UAB, C. Albareda s/n, 08193 Bellaterra (Barcelona), Spain\label{aff43}
\and
Institute for Theoretical Particle Physics and Cosmology (TTK), RWTH Aachen University, 52056 Aachen, Germany\label{aff44}
\and
INAF-Osservatorio Astronomico di Roma, Via Frascati 33, 00078 Monteporzio Catone, Italy\label{aff45}
\and
INFN section of Naples, Via Cinthia 6, 80126, Napoli, Italy\label{aff46}
\and
Institute for Astronomy, University of Hawaii, 2680 Woodlawn Drive, Honolulu, HI 96822, USA\label{aff47}
\and
Dipartimento di Fisica e Astronomia "Augusto Righi" - Alma Mater Studiorum Universit\`a di Bologna, Viale Berti Pichat 6/2, 40127 Bologna, Italy\label{aff48}
\and
Instituto de Astrof\'{\i}sica de Canarias, V\'{\i}a L\'actea, 38205 La Laguna, Tenerife, Spain\label{aff49}
\and
Institute for Astronomy, University of Edinburgh, Royal Observatory, Blackford Hill, Edinburgh EH9 3HJ, UK\label{aff50}
\and
Jodrell Bank Centre for Astrophysics, Department of Physics and Astronomy, University of Manchester, Oxford Road, Manchester M13 9PL, UK\label{aff51}
\and
European Space Agency/ESRIN, Largo Galileo Galilei 1, 00044 Frascati, Roma, Italy\label{aff52}
\and
Universit\'e Claude Bernard Lyon 1, CNRS/IN2P3, IP2I Lyon, UMR 5822, Villeurbanne, F-69100, France\label{aff53}
\and
Institut de Ci\`{e}ncies del Cosmos (ICCUB), Universitat de Barcelona (IEEC-UB), Mart\'{i} i Franqu\`{e}s 1, 08028 Barcelona, Spain\label{aff54}
\and
Instituci\'o Catalana de Recerca i Estudis Avan\c{c}ats (ICREA), Passeig de Llu\'{\i}s Companys 23, 08010 Barcelona, Spain\label{aff55}
\and
UCB Lyon 1, CNRS/IN2P3, IUF, IP2I Lyon, 4 rue Enrico Fermi, 69622 Villeurbanne, France\label{aff56}
\and
Mullard Space Science Laboratory, University College London, Holmbury St Mary, Dorking, Surrey RH5 6NT, UK\label{aff57}
\and
Departamento de F\'isica, Faculdade de Ci\^encias, Universidade de Lisboa, Edif\'icio C8, Campo Grande, PT1749-016 Lisboa, Portugal\label{aff58}
\and
Instituto de Astrof\'isica e Ci\^encias do Espa\c{c}o, Faculdade de Ci\^encias, Universidade de Lisboa, Campo Grande, 1749-016 Lisboa, Portugal\label{aff59}
\and
INAF-Istituto di Astrofisica e Planetologia Spaziali, via del Fosso del Cavaliere, 100, 00100 Roma, Italy\label{aff60}
\and
Aix-Marseille Universit\'e, CNRS/IN2P3, CPPM, Marseille, France\label{aff61}
\and
INFN-Bologna, Via Irnerio 46, 40126 Bologna, Italy\label{aff62}
\and
FRACTAL S.L.N.E., calle Tulip\'an 2, Portal 13 1A, 28231, Las Rozas de Madrid, Spain\label{aff63}
\and
Institute of Theoretical Astrophysics, University of Oslo, P.O. Box 1029 Blindern, 0315 Oslo, Norway\label{aff64}
\and
Jet Propulsion Laboratory, California Institute of Technology, 4800 Oak Grove Drive, Pasadena, CA, 91109, USA\label{aff65}
\and
Department of Physics, Lancaster University, Lancaster, LA1 4YB, UK\label{aff66}
\and
Felix Hormuth Engineering, Goethestr. 17, 69181 Leimen, Germany\label{aff67}
\and
Technical University of Denmark, Elektrovej 327, 2800 Kgs. Lyngby, Denmark\label{aff68}
\and
Cosmic Dawn Center (DAWN), Denmark\label{aff69}
\and
Max-Planck-Institut f\"ur Astronomie, K\"onigstuhl 17, 69117 Heidelberg, Germany\label{aff70}
\and
NASA Goddard Space Flight Center, Greenbelt, MD 20771, USA\label{aff71}
\and
Department of Physics and Astronomy, University College London, Gower Street, London WC1E 6BT, UK\label{aff72}
\and
Department of Physics and Helsinki Institute of Physics, Gustaf H\"allstr\"omin katu 2, 00014 University of Helsinki, Finland\label{aff73}
\and
Universit\'e de Gen\`eve, D\'epartement de Physique Th\'eorique and Centre for Astroparticle Physics, 24 quai Ernest-Ansermet, CH-1211 Gen\`eve 4, Switzerland\label{aff74}
\and
Department of Physics, P.O. Box 64, 00014 University of Helsinki, Finland\label{aff75}
\and
Helsinki Institute of Physics, Gustaf H{\"a}llstr{\"o}min katu 2, University of Helsinki, Helsinki, Finland\label{aff76}
\and
Centre de Calcul de l'IN2P3/CNRS, 21 avenue Pierre de Coubertin 69627 Villeurbanne Cedex, France\label{aff77}
\and
Laboratoire d'etude de l'Univers et des phenomenes eXtremes, Observatoire de Paris, Universit\'e PSL, Sorbonne Universit\'e, CNRS, 92190 Meudon, France\label{aff78}
\and
SKA Observatory, Jodrell Bank, Lower Withington, Macclesfield, Cheshire SK11 9FT, UK\label{aff79}
\and
Dipartimento di Fisica "Aldo Pontremoli", Universit\`a degli Studi di Milano, Via Celoria 16, 20133 Milano, Italy\label{aff80}
\and
INFN-Sezione di Milano, Via Celoria 16, 20133 Milano, Italy\label{aff81}
\and
University of Applied Sciences and Arts of Northwestern Switzerland, School of Computer Science, 5210 Windisch, Switzerland\label{aff82}
\and
Universit\"at Bonn, Argelander-Institut f\"ur Astronomie, Auf dem H\"ugel 71, 53121 Bonn, Germany\label{aff83}
\and
INFN-Sezione di Roma, Piazzale Aldo Moro, 2 - c/o Dipartimento di Fisica, Edificio G. Marconi, 00185 Roma, Italy\label{aff84}
\and
Dipartimento di Fisica e Astronomia "Augusto Righi" - Alma Mater Studiorum Universit\`a di Bologna, via Piero Gobetti 93/2, 40129 Bologna, Italy\label{aff85}
\and
Department of Physics, Institute for Computational Cosmology, Durham University, South Road, Durham, DH1 3LE, UK\label{aff86}
\and
Infrared Processing and Analysis Center, California Institute of Technology, Pasadena, CA 91125, USA\label{aff87}
\and
Universit\'e C\^{o}te d'Azur, Observatoire de la C\^{o}te d'Azur, CNRS, Laboratoire Lagrange, Bd de l'Observatoire, CS 34229, 06304 Nice cedex 4, France\label{aff88}
\and
Universit\'e Paris Cit\'e, CNRS, Astroparticule et Cosmologie, 75013 Paris, France\label{aff89}
\and
CNRS-UCB International Research Laboratory, Centre Pierre Binetruy, IRL2007, CPB-IN2P3, Berkeley, USA\label{aff90}
\and
University of Applied Sciences and Arts of Northwestern Switzerland, School of Engineering, 5210 Windisch, Switzerland\label{aff91}
\and
Institut d'Astrophysique de Paris, 98bis Boulevard Arago, 75014, Paris, France\label{aff92}
\and
Institute of Physics, Laboratory of Astrophysics, Ecole Polytechnique F\'ed\'erale de Lausanne (EPFL), Observatoire de Sauverny, 1290 Versoix, Switzerland\label{aff93}
\and
Aurora Technology for European Space Agency (ESA), Camino bajo del Castillo, s/n, Urbanizacion Villafranca del Castillo, Villanueva de la Ca\~nada, 28692 Madrid, Spain\label{aff94}
\and
Institut de F\'{i}sica d'Altes Energies (IFAE), The Barcelona Institute of Science and Technology, Campus UAB, 08193 Bellaterra (Barcelona), Spain\label{aff95}
\and
School of Mathematics, Statistics and Physics, Newcastle University, Herschel Building, Newcastle-upon-Tyne, NE1 7RU, UK\label{aff96}
\and
DARK, Niels Bohr Institute, University of Copenhagen, Jagtvej 155, 2200 Copenhagen, Denmark\label{aff97}
\and
Waterloo Centre for Astrophysics, University of Waterloo, Waterloo, Ontario N2L 3G1, Canada\label{aff98}
\and
Department of Physics and Astronomy, University of Waterloo, Waterloo, Ontario N2L 3G1, Canada\label{aff99}
\and
Perimeter Institute for Theoretical Physics, Waterloo, Ontario N2L 2Y5, Canada\label{aff100}
\and
Centre National d'Etudes Spatiales -- Centre spatial de Toulouse, 18 avenue Edouard Belin, 31401 Toulouse Cedex 9, France\label{aff101}
\and
Institute of Space Science, Str. Atomistilor, nr. 409 M\u{a}gurele, Ilfov, 077125, Romania\label{aff102}
\and
Consejo Superior de Investigaciones Cientificas, Calle Serrano 117, 28006 Madrid, Spain\label{aff103}
\and
Universidad de La Laguna, Departamento de Astrof\'{\i}sica, 38206 La Laguna, Tenerife, Spain\label{aff104}
\and
Dipartimento di Fisica e Astronomia "G. Galilei", Universit\`a di Padova, Via Marzolo 8, 35131 Padova, Italy\label{aff105}
\and
INFN-Padova, Via Marzolo 8, 35131 Padova, Italy\label{aff106}
\and
Caltech/IPAC, 1200 E. California Blvd., Pasadena, CA 91125, USA\label{aff107}
\and
Institut f\"ur Theoretische Physik, University of Heidelberg, Philosophenweg 16, 69120 Heidelberg, Germany\label{aff108}
\and
Institut de Recherche en Astrophysique et Plan\'etologie (IRAP), Universit\'e de Toulouse, CNRS, UPS, CNES, 14 Av. Edouard Belin, 31400 Toulouse, France\label{aff109}
\and
Universit\'e St Joseph; Faculty of Sciences, Beirut, Lebanon\label{aff110}
\and
Departamento de F\'isica, FCFM, Universidad de Chile, Blanco Encalada 2008, Santiago, Chile\label{aff111}
\and
Satlantis, University Science Park, Sede Bld 48940, Leioa-Bilbao, Spain\label{aff112}
\and
Department of Physics, Royal Holloway, University of London, TW20 0EX, UK\label{aff113}
\and
Instituto de Astrof\'isica e Ci\^encias do Espa\c{c}o, Faculdade de Ci\^encias, Universidade de Lisboa, Tapada da Ajuda, 1349-018 Lisboa, Portugal\label{aff114}
\and
Cosmic Dawn Center (DAWN)\label{aff115}
\and
Niels Bohr Institute, University of Copenhagen, Jagtvej 128, 2200 Copenhagen, Denmark\label{aff116}
\and
Universidad Polit\'ecnica de Cartagena, Departamento de Electr\'onica y Tecnolog\'ia de Computadoras,  Plaza del Hospital 1, 30202 Cartagena, Spain\label{aff117}
\and
Dipartimento di Fisica e Scienze della Terra, Universit\`a degli Studi di Ferrara, Via Giuseppe Saragat 1, 44122 Ferrara, Italy\label{aff118}
\and
Istituto Nazionale di Fisica Nucleare, Sezione di Ferrara, Via Giuseppe Saragat 1, 44122 Ferrara, Italy\label{aff119}
\and
INAF, Istituto di Radioastronomia, Via Piero Gobetti 101, 40129 Bologna, Italy\label{aff120}
\and
School of Physics and Astronomy, Cardiff University, The Parade, Cardiff, CF24 3AA, UK\label{aff121}
\and
Department of Physics, Oxford University, Keble Road, Oxford OX1 3RH, UK\label{aff122}
\and
Instituto de Astrof\'isica de Canarias (IAC); Departamento de Astrof\'isica, Universidad de La Laguna (ULL), 38200, La Laguna, Tenerife, Spain\label{aff123}
\and
Universit\'e PSL, Observatoire de Paris, Sorbonne Universit\'e, CNRS, LERMA, 75014, Paris, France\label{aff124}
\and
Universit\'e Paris-Cit\'e, 5 Rue Thomas Mann, 75013, Paris, France\label{aff125}
\and
Zentrum f\"ur Astronomie, Universit\"at Heidelberg, Philosophenweg 12, 69120 Heidelberg, Germany\label{aff126}
\and
Department of Mathematics and Physics E. De Giorgi, University of Salento, Via per Arnesano, CP-I93, 73100, Lecce, Italy\label{aff127}
\and
INFN, Sezione di Lecce, Via per Arnesano, CP-193, 73100, Lecce, Italy\label{aff128}
\and
INAF-Sezione di Lecce, c/o Dipartimento Matematica e Fisica, Via per Arnesano, 73100, Lecce, Italy\label{aff129}
\and
INAF - Osservatorio Astronomico di Brera, via Emilio Bianchi 46, 23807 Merate, Italy\label{aff130}
\and
INAF-Osservatorio Astronomico di Brera, Via Brera 28, 20122 Milano, Italy, and INFN-Sezione di Genova, Via Dodecaneso 33, 16146, Genova, Italy\label{aff131}
\and
ICL, Junia, Universit\'e Catholique de Lille, LITL, 59000 Lille, France\label{aff132}
\and
ICSC - Centro Nazionale di Ricerca in High Performance Computing, Big Data e Quantum Computing, Via Magnanelli 2, Bologna, Italy\label{aff133}
\and
Instituto de F\'isica Te\'orica UAM-CSIC, Campus de Cantoblanco, 28049 Madrid, Spain\label{aff134}
\and
CERCA/ISO, Department of Physics, Case Western Reserve University, 10900 Euclid Avenue, Cleveland, OH 44106, USA\label{aff135}
\and
Technical University of Munich, TUM School of Natural Sciences, Physics Department, James-Franck-Str.~1, 85748 Garching, Germany\label{aff136}
\and
Max-Planck-Institut f\"ur Astrophysik, Karl-Schwarzschild-Str.~1, 85748 Garching, Germany\label{aff137}
\and
Laboratoire Univers et Th\'eorie, Observatoire de Paris, Universit\'e PSL, Universit\'e Paris Cit\'e, CNRS, 92190 Meudon, France\label{aff138}
\and
Departamento de F{\'\i}sica Fundamental. Universidad de Salamanca. Plaza de la Merced s/n. 37008 Salamanca, Spain\label{aff139}
\and
Universit\'e de Strasbourg, CNRS, Observatoire astronomique de Strasbourg, UMR 7550, 67000 Strasbourg, France\label{aff140}
\and
Center for Data-Driven Discovery, Kavli IPMU (WPI), UTIAS, The University of Tokyo, Kashiwa, Chiba 277-8583, Japan\label{aff141}
\and
Ludwig-Maximilians-University, Schellingstrasse 4, 80799 Munich, Germany\label{aff142}
\and
Max-Planck-Institut f\"ur Physik, Boltzmannstr. 8, 85748 Garching, Germany\label{aff143}
\and
Dipartimento di Fisica - Sezione di Astronomia, Universit\`a di Trieste, Via Tiepolo 11, 34131 Trieste, Italy\label{aff144}
\and
California Institute of Technology, 1200 E California Blvd, Pasadena, CA 91125, USA\label{aff145}
\and
University of California, Los Angeles, CA 90095-1562, USA\label{aff146}
\and
Department of Physics \& Astronomy, University of California Irvine, Irvine CA 92697, USA\label{aff147}
\and
Departamento F\'isica Aplicada, Universidad Polit\'ecnica de Cartagena, Campus Muralla del Mar, 30202 Cartagena, Murcia, Spain\label{aff148}
\and
Instituto de F\'isica de Cantabria, Edificio Juan Jord\'a, Avenida de los Castros, 39005 Santander, Spain\label{aff149}
\and
CEA Saclay, DFR/IRFU, Service d'Astrophysique, Bat. 709, 91191 Gif-sur-Yvette, France\label{aff150}
\and
Institute of Cosmology and Gravitation, University of Portsmouth, Portsmouth PO1 3FX, UK\label{aff151}
\and
Department of Astronomy, University of Florida, Bryant Space Science Center, Gainesville, FL 32611, USA\label{aff152}
\and
Department of Computer Science, Aalto University, PO Box 15400, Espoo, FI-00 076, Finland\label{aff153}
\and
Instituto de Astrof\'\i sica de Canarias, c/ Via Lactea s/n, La Laguna 38200, Spain. Departamento de Astrof\'\i sica de la Universidad de La Laguna, Avda. Francisco Sanchez, La Laguna, 38200, Spain\label{aff154}
\and
Ruhr University Bochum, Faculty of Physics and Astronomy, Astronomical Institute (AIRUB), German Centre for Cosmological Lensing (GCCL), 44780 Bochum, Germany\label{aff155}
\and
Department of Physics and Astronomy, Vesilinnantie 5, 20014 University of Turku, Finland\label{aff156}
\and
Serco for European Space Agency (ESA), Camino bajo del Castillo, s/n, Urbanizacion Villafranca del Castillo, Villanueva de la Ca\~nada, 28692 Madrid, Spain\label{aff157}
\and
ARC Centre of Excellence for Dark Matter Particle Physics, Melbourne, Australia\label{aff158}
\and
Centre for Astrophysics \& Supercomputing, Swinburne University of Technology,  Hawthorn, Victoria 3122, Australia\label{aff159}
\and
Department of Physics and Astronomy, University of the Western Cape, Bellville, Cape Town, 7535, South Africa\label{aff160}
\and
DAMTP, Centre for Mathematical Sciences, Wilberforce Road, Cambridge CB3 0WA, UK\label{aff161}
\and
Kavli Institute for Cosmology Cambridge, Madingley Road, Cambridge, CB3 0HA, UK\label{aff162}
\and
Department of Astrophysics, University of Zurich, Winterthurerstrasse 190, 8057 Zurich, Switzerland\label{aff163}
\and
IRFU, CEA, Universit\'e Paris-Saclay 91191 Gif-sur-Yvette Cedex, France\label{aff164}
\and
Oskar Klein Centre for Cosmoparticle Physics, Department of Physics, Stockholm University, Stockholm, SE-106 91, Sweden\label{aff165}
\and
Astrophysics Group, Blackett Laboratory, Imperial College London, London SW7 2AZ, UK\label{aff166}
\and
Univ. Grenoble Alpes, CNRS, Grenoble INP, LPSC-IN2P3, 53, Avenue des Martyrs, 38000, Grenoble, France\label{aff167}
\and
INAF-Osservatorio Astrofisico di Arcetri, Largo E. Fermi 5, 50125, Firenze, Italy\label{aff168}
\and
Dipartimento di Fisica, Sapienza Universit\`a di Roma, Piazzale Aldo Moro 2, 00185 Roma, Italy\label{aff169}
\and
Centro de Astrof\'{\i}sica da Universidade do Porto, Rua das Estrelas, 4150-762 Porto, Portugal\label{aff170}
\and
HE Space for European Space Agency (ESA), Camino bajo del Castillo, s/n, Urbanizacion Villafranca del Castillo, Villanueva de la Ca\~nada, 28692 Madrid, Spain\label{aff171}
\and
Center for Frontier Science, Chiba University, 1-33 Yayoi-cho, Inage-ku, Chiba 263-8522, Japan\label{aff172}
\and
Department of Physics, Graduate School of Science, Chiba University, 1-33 Yayoi-Cho, Inage-Ku, Chiba 263-8522, Japan\label{aff173}
\and
Department of Astrophysical Sciences, Peyton Hall, Princeton University, Princeton, NJ 08544, USA\label{aff174}
\and
Theoretical astrophysics, Department of Physics and Astronomy, Uppsala University, Box 515, 751 20 Uppsala, Sweden\label{aff175}
\and
Minnesota Institute for Astrophysics, University of Minnesota, 116 Church St SE, Minneapolis, MN 55455, USA\label{aff176}
\and
Mathematical Institute, University of Leiden, Einsteinweg 55, 2333 CA Leiden, The Netherlands\label{aff177}
\and
Institute of Astronomy, University of Cambridge, Madingley Road, Cambridge CB3 0HA, UK\label{aff178}
\and
Space physics and astronomy research unit, University of Oulu, Pentti Kaiteran katu 1, FI-90014 Oulu, Finland\label{aff179}
\and
Center for Computational Astrophysics, Flatiron Institute, 162 5th Avenue, 10010, New York, NY, USA\label{aff180}
\and
Department of Astronomy, University of Massachusetts, Amherst, MA 01003, USA\label{aff181}
\and
Univ. Lille, CNRS, Centrale Lille, UMR 9189 CRIStAL, 59000 Lille, France\label{aff182}
\and
Department of Physics and Astronomy, University of British Columbia, Vancouver, BC V6T 1Z1, Canada\label{aff183}}    

%
%
\abstract{
The ESA \Euclid mission will measure the photometric redshifts of billions of galaxies in order to provide an accurate 3D view of the Universe at optical and near-infrared wavelengths. Photometric redshifts are determined by the PHZ processing function on the basis of the multi-wavelength photometry of \Euclid and ground-based observations. In this paper, we describe in detail the so-called PHZ processing used for the first `quick' (Q1) \Euclid data release, the output products, and their validation with respect to the \Euclid requirements. The PHZ pipeline is responsible for the following main tasks: i) source classification into star, galaxy, and quasar (or QSO) classes based on photometric colours; ii) determination of photometric redshifts for the core science; iii) determination of physical properties of galaxies for non-cosmological science. The classification is able to provide a star sample with a high level of purity, a highly complete galaxy sample, and reliable probabilities of belonging to those classes. The identification of QSOs is instead more problematic: photometric information available in the Euclid Wide Survey alone seems to be insufficient to accurately separate QSOs from galaxies. The performance of the pipeline in the determination of photometric redshifts has been tested using the COSMOS2020 catalogue and a large sample of spectroscopic redshifts. The results in both cases are in line with expectations: the precision of the estimates are compatible with \Euclid requirements, while, as expected, a bias correction is needed to achieve the accuracy level required for the cosmological probes. Finally, the pipeline provides reliable estimates of the physical properties of galaxies, in good agreement with findings from the COSMOS2020 catalogue, except for an unrealistically large fraction of very young galaxies with very high specific star-formation rates. The application of appropriate priors is, however, sufficient to obtain reliable physical properties for those problematic objects. We present several areas for improvement for future \Euclid data releases.}
%
%
%
%
   \titlerunning{Q1 PHZ processing and products}
   \authorrunning{Euclid Collaboration: Tucci et al.}
   
   \maketitle
%
%
%
%
   
\section{\label{sc:Intro}Introduction}

\Euclid is a space mission that is conducting a survey of the extragalactic sky over 14\,000\,deg$^2$ with optical and near-infrared imaging \citep{Scaramella-EP1,EuclidSkyOverview}. It is equipped with two instruments: the VIS optical camera \citep{EuclidSkyVIS}, which provides high-resolution images of the sky; and the Near Infrared Spectrometer and Photometer near-infrared instrument \citep[NISP][]{Maciaszek22,EuclidSkyNISP}, designed for photometry in three near-infrared (NIR) bands, $\YE,\,\JE,$ and $\HE$, as well as for NIR spectroscopy. The resulting Euclid Wide Survey (EWS) reaches a minimum depth of $24.5$\,AB\,magnitudes in the visible band ($\IE$) with a signal-to-noise ratio (S/N) of 10 for extended sources, and in the NIR bands a depth better than $24.0$ for S/N=5 for point sources  \citep{Scaramella-EP1}. The main objective of the mission is to study dark matter and dark energy by measuring the evolution of large-scale structures. In particular, as discussed in \citet{EuclidSkyOverview}, the mission is optimised for two primary cosmological probes, namely weak gravitational lensing (WL) and galaxy clustering (GC). 

The \Euclid cosmological analysis requires the determination of redshifts for an extremely large number of galaxies to enable the tomographic study of the mass distribution. \Euclid spectroscopy can be performed only for a limited number of galaxies,\footnote{\Euclid will provide about 35 million spectroscopic redshifts for galaxies in the range $0.9\le z\le1.8$ \citep{Q1-TP007}.} and high-quality photometric redshifts are therefore crucial for WL and photometric GC studies. The \Euclid requirements on photometric redshifts are quite stringent: tomographic analyses require that galaxies are placed in the redshift bins with a precision substantially better than the bin width. Assuming a baseline configuration of 13 bins in the range $0.2<z<2.5$, the dispersion of photometric redshifts must be $\sigma_z<0.05(1+z)$, with an outlier fraction $<10$\% \citep{ama07,EuclidSkyOverview}. In addition, the mean redshift in tomographic bins must be known better than $0.002(1+z)$ \citep{ma06,hut06,ama07,kit08}. Achieving this level of accuracy is not, however, a goal for this \Euclid data release.

The \Euclid broad-band photometry alone is not sufficient to fulfill the above requirements. Simulations have shown the need for additional ground-based data with at least four filters in the wavelength range 420--930\,nm (corresponding to the common $g,\,r,\,i,\,z$ filter set), with minimal 5\,$\sigma$ depths of 25.7, 25.1, 24.8, and 24.6 AB\,magnitudes, respectively, for point-like sources \citep{Scaramella-EP1}. To this aim, a large coordinated campaign of ground-based observations is being conducted to provide multi-band photometry across the \Euclid sky areas. In the southern sky, the Dark Energy Survey \citep[DES;][]{abb21} is currently used until deeper data from the Legacy Survey of Space and Time (LSST) by the Rubin Observatory \citep{ive19} will be available. In the northern hemisphere, a new collaboration has been set up, the Ultraviolet Near Infrared Optical Northern Survey (UNIONS; Gwyn et al., in prep.),\footnote{See also the web site \url{https://www.skysurvey.cc}.} with the aim to survey the sky in the $u,\,g,\,r,\,i,\,z$ bands. This is a joint effort between: the Canada-France Imaging Survey \citep[CFIS;][]{iba17} for the {\it u} and {\it r} bands; the Panoramic Survey Telescope and Rapid Response System \citep[Pan-STARRS;][]{cha16} for the {\it i} band; and the Subaru Hyper Suprime Camera \citep[HSC;][]{miy18} for the {\it g} band, through the Waterloo-Hawaii-IfA $g$-band Survey (WHIGS), and for the {\it z} band through the Wide Imaging with Subaru-Hyper Suprime-Cam Euclid Sky (WISHES) survey.

In addition to the main survey, a significant fraction of observation time will be spent on specific fields that, thanks to repeated visits, will accumulate greater depth than EWS, up to a gain of about 2 magnitudes \citep{EuclidSkyOverview}. These deeper fields form the Euclid Deep Survey (EDS) and the Euclid Auxiliary fields (EAFs), with the aim to calibrate the instruments and to characterise the source population \citep{Q1-TP001}.

The \citet{Q1cite} is the first public release of EWS data \citep{Q1-TP001}. It covers 63.1\,deg$^2$, and consists of single-visit observations of the Euclid Deep Field (EDF) North (EDF-N; 22.9\,deg$^2$), Fornax (EDF-F; 12.1\,deg$^2$) and South (EDF-S; 28.1\,deg$^2$). The main purpose of the Q1 release is to provide to the astronomy community with a first set of \Euclid data ready for scientific exploitation. The Q1 products are suitable for non-cosmological science, but not for the core cosmology objectives of \Euclid, due to the small covered area and to the lack of deep observations in the EDS and in the EAFs. Three main Data Releases (DRs) of the EWS are planned in the next years to address the cosmological science objectives, with increasingly large sky areas.

Data products of public releases will contain the processing of \Euclid optical and NIR imaging and of ground-based observations, performed by the \Euclid Science Ground Segment (SGS). The SGS is responsible for carrying out and archiving the entire data processing, from the satellite telemetry to the production of science-ready data. Data processing is constituted and connected by processing functions (PFs) that are self-contained processing modules with specific tasks to be performed, and represent the highest-level breakdown of the SGS data-processing pipeline \citep[for more details, see][]{EuclidSkyOverview,Q1-TP001}. In a nutshell, visible and NIR imaging data are processed to produce fully calibrated images by VIS and NIR PFs \citep{Q1-TP002,Q1-TP003}. External data, derived from ground-based surveys and external missions, are reformatted and recalibrated in order to be consistently handled with \Euclid data (EXT PF). \Euclid and external data are then recollected and merged by the MER PF into stacked images and source catalogues \citep{Q1-TP004}. MER products are the main input to compute photometric redshifts (PHZ PF). The development of PFs is coordinated by the Organisation Units (OUs), which are teams of \Euclid scientists and engineers with the appropriate scientific and software-development competences for each individual PF.

The topic of this paper is the PHZ PF, which is in charge of computing photometric redshifts of galaxies from the multi-wavelength photometry of \Euclid and ground-based observations. Other relevant tasks are the classification of the detected sources based on their photometric colours, and the determination of physical properties of \Euclid sources.

The main products of the classification are the object class -- star, galaxy, and QSO -- and the class probabilities, useful to build samples with levels of purity and completeness different from those defined in the pipeline. For the \Euclid core science, the classification is needed to identify a reliable sample of bright stars, independently of the morphology, as required for the modelling of the VIS point spread function \citep[PSF;][]{EuclidSkyOverview}. For the non-cosmological science, it is used to separate objects in different classes and to determine the corresponding physical parameters. As shown in the paper, identifying the QSO population is quite challenging if only the colours available in the EWS are taken into account. The selection of QSOs and active galactic nuclei (AGN) in the Q1 release is, however, the main topic of several works, using multi-wavelength data \citep{Q1-SP003,Q1-SP013,Q1-SP027}, or focusing on a particular class of objecs \citep[e.g., red AGN or quasars;][]{Q1-SP011,Q1-SP023}, or directly on imaging data \citep{Q1-SP015,Q1-SP009}.

The Q1 release allows us to evaluate the PHZ PF performance in determining photometric redshifts in view of the first \Euclid cosmological studies with the DR1 release. We first test the performance of our pipeline by applying it on the COSMOS2020 catalogue \citep[][hereafter W22]{wea22} degraded to the EWS covering and depth. We then compare directly the Q1 results for objects with spectroscopic redshifts. In addition to redshifts, the PHZ PF determines physical properties of galaxies such as luminosity, star-formation rate, stellar mass, etc. They are relevant for the study of galaxy evolution, which is an important scientific objective of the Q1 and future releases \citep[see, e.g., the Q1 results on this topic;][]{Q1-SP016,Q1-SP017,Q1-SP031,Q1-SP049}. This paper discusses their validation by comparing the Q1 products to the results from the COSMOS2020 catalogue. We also discuss the issues found in the processing, and the reliability for scientific analyses.

The structure of the paper is as follows. \Cref{sc:overview} introduces the PHZ pipeline for Q1 and the products that will be generated and stored in the Euclid Archive System \citep[EAS;][]{Q1-TP001}. In \cref{sc:mer} we describe the information of the MER source catalogues used in the PHZ PF. The following sections provide a more detailed description of the main steps of the PHZ pipeline, including their validation and Q1 results: the source classification (\cref{sc:class}); the determination of photometric redshifts for the \Euclid core science (\cref{sc:phz}), and the products for the non-cosmological science (\cref{sc:legacy}). 

In this paper, as in all \Euclid papers, we adopt the {\it Planck} 2016 flat $\Lambda$CDM cosmology with $H_0=67.74$\,\kmsMpc, $\Omm=0.3089$, and $\OmLa=0.6911$ \citep{planck15}.

\section{\label{sc:overview} The PHZ processing function for Q1}

In this section we give a general overview of the PHZ pipeline developed for the Q1 release, keeping to the next sections a detailed description of the single steps. The pipeline ingests and analyses the EWS source catalogues produced by the MER PF. Source photometry and colours are used to produce the main outputs of the PHZ PF: source classification; spectral energy distribution (SED) of stars and galaxies; redshift probability density functions (PDFs); and physical properties. In particular, for these purposes, two tools have been developed by OU-PHZ: a template-fitting package called \Phosphoros{} (see \cref{ssc:phz:wl}), used for the redshift determination; and a machine-learning method called Nearest-Neighbour Photometric Redshifts (\NNPZ; see \cref{ssc:pp:gal}), used to estimate physical properties of galaxies. Output products are provided in the EAS through separate catalogues (see also \cref{app:sc:cat}). 

\begin{figure*}
  \includegraphics[width=17cm]{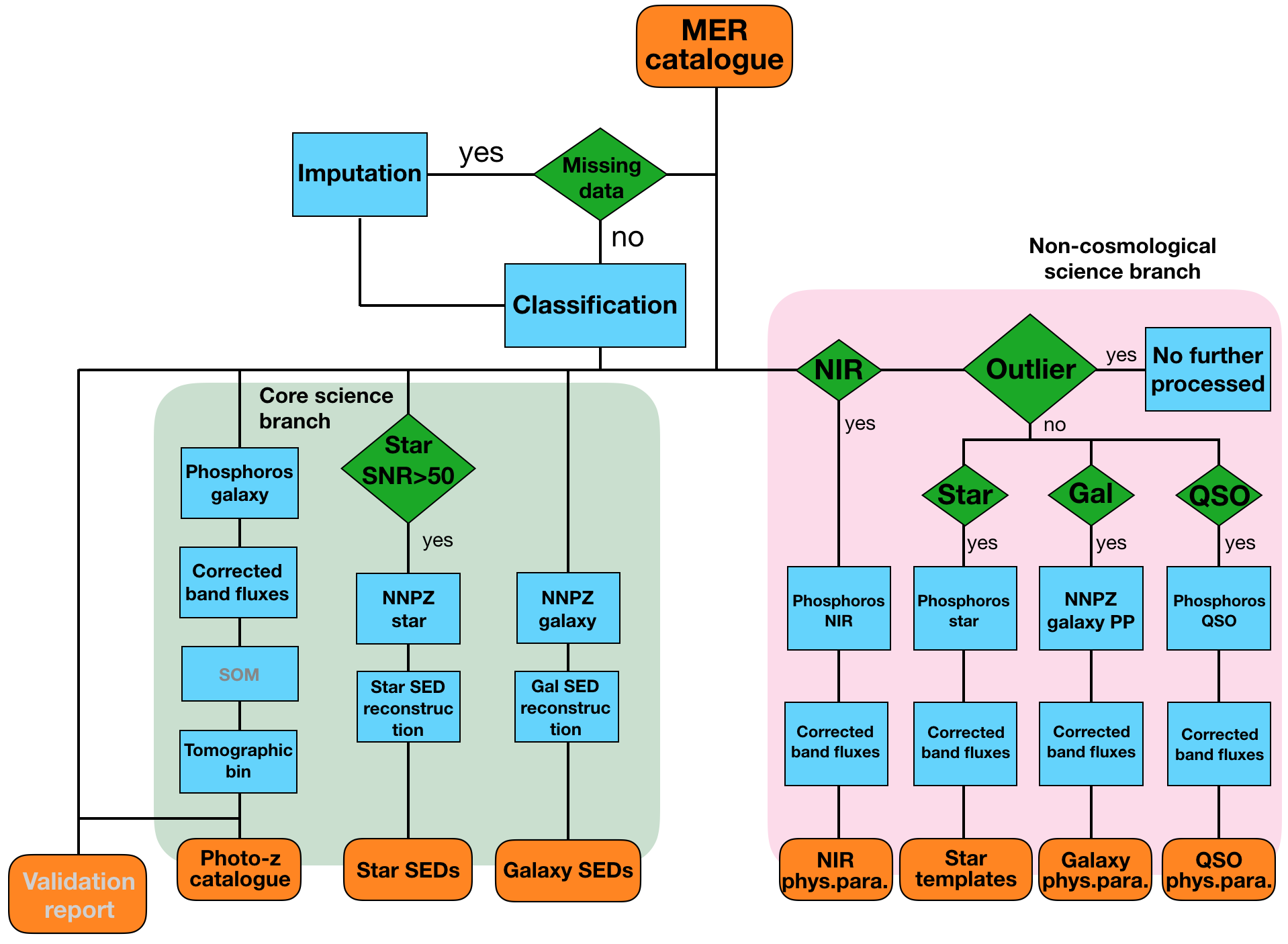}
  \caption{Diagram of the Production pipeline for the Q1 release. Orange boxes are for input or output products; green diamonds for conditional statements; and blue boxes for specific tasks. It shows in grey all the steps of the pipeline that are performed but whose products are not included in the public release.}
  \label{f:prod}
\end{figure*}

The diagram of the pipeline is shown in \cref{f:prod}. Given a source catalogue, the first task of the pipeline is to classify all the detected objects based on their colours (see \cref{sc:class}). Imputation is applied to missing fluxes before classification, but imputed fluxes are used only for classification purposes. After the classification, the pipeline is split into two branches according to the output target: the primary cosmology or core science; and the non-cosmological science. All the detected objects are considered in both branches. 

The core-science branch (green shaded in \cref{f:prod}) consists of three sub-branches that produces three different catalogues. The main product is the photometric redshift catalogue, which includes redshift PDFs computed by \Phosphoros and the related statistics, plus the reconstructed photometric fluxes in the \Euclid, DES, UNIONS, and LSST bands corrected for Galactic extinction. Here, all the detected objects are treated with the \Phosphoros{} configuration tuned for galaxies, regardless of their classification, to avoid selection effects in the cosmological analyses. The pipeline also performs two tasks related to the calibration of the photometric redshifts: it assigns each object to a tomographic bin based on the photometric redshift (see \cref{app:sc:tom}), and place the object in a cell in colour space reorganized by a `self-organizing map' (see \citealt{mas15} for more details; this information is not used in the Q1 release and is not discussed further in this paper). The other two sub-branches provide two additional products that are also used only for the core science: (1) a catalogue of star SEDs for the star-classified objects with ${\rm S/N}>50$ in the \IE detection band, used to model the VIS PSF; (2) a catalogue of galaxy SEDs for all the objects. The SED reconstruction is performed using the \NNPZ algorithm (see \cref{app:sc:sed} and Euclid Collaboration: Tarsitano et al., in prep.).

The second branch of the pipeline (pink shaded in \cref{f:prod}) computes the physical properties of sources for non-cosmological science (see \cref{sc:legacy}). First of all, the pipeline checks for objects detected only in the NIR stack images (hereafter, simply NIR-only objects). These sources, which are expected to be bright galaxies or quasars at very high redshifts or cool stars, are treated in a dedicated sub-branch. All the objects (including NIR-only ones) are then separated according to the class assigned by the classification. If an object is accepted by more than one classifier (see \cref{sc:class}), it will be found in more than one sub-branch. On the other hand, outliers (i.e., objects without any assigned class) are not further processed in the branch (they are typically a few per cent, depending on the ground-based configuration, see \cref{ssc:class:valid}). The star, QSO, and NIR-only sub-branches are processed by the \Phosphoros tool, which provides best-fit redshift, SED template, and intrinsic reddening. Physical properties of galaxies are instead determined by the \NNPZ algorithm. Products of this branch are gathered in four different catalogues, one for each type (namely, star, galaxy, QSO, and NIR-only), which also include the photometry corrected for Galactic extinction, based on the best-fitting SED.

Finally, the Production pipeline produces a report that is used internally to validate the PHZ PF results for each single MER source catalogue, covering a tile of the sky (see \cref{sc:mer}). Validation aims to identify problematic tiles in terms of photometry (too many missing fluxes, too low object surface density, peculiar noise features in the images, etc.), or PHZ data processing (e.g., too many outliers or extended stars, unreliable redshift distributions, etc.). Tiles that do not pass validation criteria are automatically classified as dubious by the pipeline, and are visually checked by OU-PHZ. Tiles with critical issues are set as invalid and not included in the PHZ data products. In total, 28 tiles were removed from the Q1 release (8\% of the tiles, but less than 1\% of the detected sources). They are typically tiles on the border of the EWS, which contain a very low number of sources (from a few hundreds to a few ten of thousands) and a large fraction of missing fluxes.

\section{\label{sc:mer} EWS source catalogues}

\begin{table}
  \caption{Average 5\,$\sigma$ depths of the MER catalogues, the values of the zero-point multiplicative correction (ZPC) applied to fluxes, and the average magnitude correction for Galactic extinction.}
\label{t:mer}
\centering
\begin{tabular}{cccccc}
\hline
\hline 
\noalign{\hskip 2pt}
Survey & Band & Depth & ZPC & \multicolumn{2}{c}{Mag correction\tablefootmark{a}}\\
 & & & & EDF-N & -S,\,-F \\
  \hline 
  \Euclid & $\IE$ & 26.0 & 1.05 & 0.12 & 0.04 \\ 
  \Euclid & $\YE$ & 23.8 & 0.95 & 0.06 & 0.02 \\
  \Euclid & $\JE$ & 24.0 & 0.97 & 0.04 & 0.01 \\
  \Euclid & $\HE$ & 24.0 & 0.95 & 0.03 & 0.01 \\
  \hline 
  DES & $g$ & 25.4 & 1.00 & & 0.06 \\
  DES & $r$ & 25.1 & 1.00 & & 0.04 \\
  DES & $i$ & 24.4 & 1.00 & & 0.03 \\
  DES & $z$ & 23.7 & 1.015 & & 0.02 \\
  \hline
  CFIS & $u$ & 24.1 & 1.00 & 0.26 & \\
  HSC & $g$ & 25.8 & 1.00 & 0.20 & \\
  CFIS & $r$ & 24.8 & 0.99 & 0.14 & \\
  Pan-STARRS & $i$ & 23.9 & 1.09 & 0.11 & \\
  HSC & $z$ & 24.1 & 1.00 & 0.08 & \\
  \hline
\end{tabular}
\tablefoot{
\tablefoottext{a}{The magnitude correction is computed using the average value of $E_{\rm B-V}$ in the fields (namely 0.056 in the EDF-N, 0.017 in the EDF-S, and 0.016 in the EDF-F), the reddening law from \citet{gor23}, and a $5700\,\mathrm{K}$ blackbody SED.}
}
\end{table}

The source catalogues produced by the MER PF are the input data for the PHZ PF. They contain all the objects detected in the \IE images or in the NIR-stacked images obtained by a combination of $\YE$, $\JE$, and $\HE$ data. 
Each of these catalogues covers an area of the sky (or `tile') of about 0.25\,deg$^2$. The Q1 release consists of 344 tiles with almost 30 million sources, and an object density of around 80\,(40)\,arcmin$^{-2}$ for VIS-detected objects with a detection ${\rm S/N}>5$ (and $\IE<24.5$). The details of the photometric catalogue can be found in \citet{Q1-TP004}.

For each detected source, the MER PF provides different kinds of photometric measurements, including aperture, model- and template-fitting photometry \citep[see,][]{Q1-TP004}. For the PHZ PF (and more specifically for the redshift determination), we are interested in flux estimates that measure the total emission of the source, even for extended galaxies. For the VIS or NIR-stacked detection fluxes, we consider the total flux computed within a Kron aperture using the \texttt{T-PHOT} algorithm \citep{mer15, mer16}. In the other bands, the total flux is obtained by scaling the detection flux with a `colour' term that is the ratio between the aperture flux in a given band and that in the detection image:
\be
F_{\rm band}({\rm TOT})=F_{\rm det}({\rm TOT})\,\frac{F_{\rm band}({\rm APER})}{F_{\rm det}({\rm APER})}\,.
\ee
Here, aperture fluxes are measured on images that are PSF-matched to the image with the worst resolution and are computed within a circular aperture that is twice the worst FWHM.

The performance of the photometric-redshift measurements strongly depends on the quality of the photometric observations. We report in \cref{t:mer} the average 5\,$\sigma$ depths of the EWS catalogue in the different bands. Those are estimated from the S/N distribution of the VIS-detected sources using total fluxes for the \IE band and (\texttt{2FWHM}) aperture fluxes for the other bands \citep[they are in good agreement with the depths measured in the imaging data from][]{Q1-TP001,Q1-TP004}. The depths are very close to or better than the minimum depths required for cosmology, both in the \IE and NIR bands.\footnote{The depth requirements for cosmology are for point-like sources, while here most of the sources are extended.} On the other hand, the ground-based data are a bit shallower, especially for the $i$ and $z$ bands of the northern sky and for the $z$ band of the southern sky. The HSC $g$ band data in the EDF-N is instead a bit deeper than it will be in the full EWS North area. 

Other relevant information for the PHZ pipeline included in the source catalogues are: the provenance of sources (i.e., whether they are detected in the VIS or NIR map); and the Milky Way colour excess $E(B-V)$ at the source position estimated from the {\it Planck} dust map \citep{planckdust}, needed to correct fluxes for Galactic extinction (\cref{t:mer} reports the `average' correction in the different bands). Morphological properties of the \Euclid sources are not used in the pipeline, except for validation purposes.

Redshift determination with a template-fitting method such as \Phosphoros (see \cref{sc:phz}) requires an accurate and consistent photometric calibration. Mismatches can arise for a number of reasons, for example, residual photometric calibration errors, imperfect knowledge of the transmission curves, or aperture corrections for the PSF model used. To address any remaining inconsistencies in the photometry, for each band we determine the zero-point (ZP) offset (listed in \cref{t:mer}), and we apply it as a multiplicative correction on the measured fluxes. The ZP offset is negligible or of the order of 1\%, except for the Pan-STARRS $i$-band (9\%) and for \Euclid photometry (3--5\%).

The ZP offsets are determined by OU-PHZ using \Phosphoros{} on the Q1 source catalogues following the usual approach \citep{ilb06}. Offsets measured in this way are naturally model dependent, and we proceed in three steps in order to make our determinations more robust. Firstly, we identify bright compact sources in the EDF-S using the criterion \texttt{MUMAX\_MINUS\_MAG}<$-$2.6\footnote{\texttt{MUMAX\_MINUS\_MAG} is a MER morphological parameter that gives the difference between the peak surface brightness and the magnitude in the detection band.}, and that have been matched to a {\it Gaia} counterpart. We fit the fluxes of these objects with a large set of stellar templates (see \cref{ssc:pp:star}). The medians of the ratios between measured and predicted star fluxes give us the fiducial corrections for the \Euclid and DES bands in this first step. This is done first on the southern sky because the results seem to be more stable than in the north. In the second step, we apply the offsets previously computed for the Euclid bands, and then follow the same procedure for the EDF-N, which allows us to recover the corrections required for the UNIONS bands. 

The effective zero-point adjustments computed so far are appropriate for point sources, and are dependent on the set of stellar SED templates used. While they should, in principle, be applicable also to small galaxies, we nevertheless perform a third step to refine the values and to make them more appropriate for the set of galaxy SED templates that will be used to determine photometric redshifts. In this final step, we apply the offsets found thus far and use a cross-match of known spectroscopic galaxies with the Q1 catalogues from all three EDFs, mostly drawn from PRIMUS \citep{coi2013} and DESI \citep{desi24}. For this refinement step, we fix the galaxy redshift to the spectroscopic value, and compute the flux ratios using the same set of template SEDs described in \cref{sc:phz}. The ZP offsets depend on the extension and light profile of the objects. The values listed in \cref{t:mer} are therefore appropriate for a typical object in the spectroscopic galaxy sample. 

\section{\label{sc:class} Classification}

Source classification is an important task of the \Euclid pipeline for the full exploitation of its processed data. A classification based on the source morphology is performed by the MER PF, giving the probability of objects to be point like or extended \citep{Q1-TP004}. To avoid any spurious correlations with the VIS PSF, in the PHZ PF we aim to classify objects according to their photometry into three classes: star; galaxy; and QSO.\footnote{For Q1, the target of the `QSO' class is extended to luminous AGN. No morphological information is taken into account.} One of the main goals is to identify a very pure sample of bright stars spanning the full range of stellar colours. The VIS PSF must be modelled using stars detected in each exposure, and the stringent requirements on its accuracy and stability impose a $99$\% purity level for star-classified objects with detection ${\rm S/N}>50$. Classification is also relevant for non-cosmological purposes. For example, the study of the evolution of the galaxy physical properties demands high levels of completeness in the galaxy sample, while the identification of quasars is needed for analyses of the AGN population.

The classification is performed using a supervised machine learning method called \texttt{Probabilistic Random Forest} \citep[\PRF;][]{rei19}, a modification of the standard \texttt{Random Forest} (\RF) algorithm \citep{bre01}. PRF has been applied to several astrophysical problems \citep[e.g.,][]{kin21,gua22,muz22,rod23}. The main feature of the method is the capability to take into account uncertainties in the measurements (or features) and in the assigned classes (or labels). In \PRF, features and labels are treated as probability distribution functions, typically with a Gaussian shape, rather than fixed quantities. As an example, in the classical \RF{} an object propagates through the nodes of a specific tree in a deterministic way, and ends up in a single terminal node. In a \PRF{} tree, instead, objects can propagate into both branches of each node with some probability, and reach all the terminal nodes (although in the standard implementations, a probability threshold is applied, and branches with probability below that threshold are pruned from the tree).
As \citet{rei19} have shown, the \PRF{} outperforms the \RF{}. In addition, the use of \PRF{} allows us to take flux uncertainties into account in a seamless manner.

In our implementation, the classification is performed using three binary classifiers that compute the probability of the objects to belong to a specific class. We assign to an object all the classes whose probabilities exceed pre-determined thresholds. A single object can therefore have multiple assigned classes, or, in the case that all the probabilities are below the thresholds, no assigned class (i.e., it is an outlier). The probability thresholds are defined during the training process based on the classification performance of half of the training sample (the other half being used for training), and on the required purity level (see \cref{t:cl:thres}). 

Before running the classification, few pre-processing steps are applied to object fluxes: firstly, we correct observed fluxes for Galactic extinction, based on an empirical SED-dependent approach (\cref{app:ssc:gred}). Then, we impute missing fluxes using the $k$-nearest neighbours algorithm (\cref{app:ssc:impute}). Objects with missing fluxes in more than 3 bands are, however, rejected and no longer processed in the pipeline. Finally, we convert fluxes into `asinh' magnitudes \citep{lup99} in order to deal with negative fluxes, and we compute colours and colour uncertainties (\cref{app:asinhmag}). 

The features for the PRF classifiers are the source colours. We consider all the possible colour combinations, giving 28 (36) features in the DES (UNIONS) configuration. We assume colours to have a Gaussian distribution, with standard deviation given by the colour uncertainty. Labels (i.e., the class of sources) instead do not have any uncertainty. Most of the \PRF{} hyperparameters are equivalent to the standard \RF{} ones. We optimise the most important ones using a grid search method, and the simulations developed by the \Euclid SGS for the Scientific Challenge 8 \citep{EP-Serrano}. The same hyperparameters are used for all the classifiers.

\begin{table}
\caption{Probability thresholds for the PRF classifiers, and the purity achieved in the half training set.}
\label{t:cl:thres}
\centering
\begin{tabular}{lccc}
\hline \hline
Configuration & Star & Galaxy & QSO \\
\hline
DES & 0.73 & 0.4 & 0.85 \\
UNIONS & 0.58 & 0.22 & 0.66 \\
\hline
Purity & 99\% & 90\% & 95\% \\
\hline
\end{tabular}
\end{table}

The results of classification are collected in the EAS in the
\texttt{phz\_classification} catalogue. The main products are the
probabilities of each object to belong to each class, and the
classification flag indicating which classifier accepts the object.

\subsection{\label{ssc:class:train} Training sample for classification}

\begin{table}
    \caption{Spectroscopic surveys used for the classification training set, the number of high quality ($Q$=3,4 in a VVDS-like system) spectra, and the number of selected objects.}
    \centering
    \begin{tabular}{llrr} \hline \hline
    Field & Survey  & $Q$=3-4 & Selected \\ \hline
    ELAIS-S1 & OzDES-DR2\tablefootmark{a} & 1234 & 1192 \\ \hline
    CDFS &     VVDS\tablefootmark{b} & 416 & 416 \\
     & VANDELS\tablefootmark{c} & 197 & 188 \\
     & OzDES-DR2\tablefootmark{a} & 2574 & 2068 \\
    & VUDS-DR1\tablefootmark{d} & 57 & 56 \\\hline
    XMM-LSS &     SDSS-DR16\tablefootmark{e} & 1775 & 1438 \\
    & OzDES-DR2\tablefootmark{a} & 852 & 759 \\
    & VIPERS-PDR2\tablefootmark{f} & 2694 & 2280 \\
    & VVDS\tablefootmark{b} & 1699 & 1694 \\
    & VANDELS\tablefootmark{c} & 207 & 205 \\\hline
    COSMOS &  SDSS-DR16\tablefootmark{e} & 455 & 197 \\
    & VUDS-DR1\tablefootmark{d} & 63 & 53 \\
    & VVDS\tablefootmark{b} & 125 & 56 \\
    & zCOSMOS DR3\tablefootmark{g} & 3437 & 2663 \\
    \hline
    \end{tabular}
    \tablefoot{
    \tablefoottext{a}{\citet[][]{lid20}.}
    \tablefoottext{b}{\citet[][]{lef13}.}
    \tablefoottext{c}{\citet[][]{gar21}.}
    \tablefoottext{d}{\citet[][]{tas17}.}
    \tablefoottext{e}{\citet[][]{ahu20}.}
    \tablefoottext{f}{\citet[][]{sco18}.}
    \tablefoottext{g}{\citet[][]{lil09}.}
    }
    \label{t:specz_surveys}
\end{table}

\PRF{} classifiers need to be trained through a source sample that includes all the features used for the classification plus the classes of the objects. The training sample must be representative of the EWS and labels must be reliable. It thus requires photometric data in the same wavelength range covered by \Euclid with comparable depth, complemented by additional photometric and spectroscopic information to provide reliable estimates of their classes. To this aim, we have gathered data from two of the deepest near-infrared surveys that are available to date: (1) the Visible and Infrared Survey Telescope for Astronomy (VISTA) Deep Extragalactic Observations (VIDEO) DR5 catalogue \citep{jar13}, covering three fields (XMM-LSS, CDFS, and ELAIS-S1) of 12\,deg$^2$ total area in the $z,\,Y,\,J,\,H,$ and $K_{\rm s}$ bands; (2) the UltraVISTA DR4 catalogue \citep{mcc12}, covering the 2\,deg$^2$ area of the COSMOS field in the $Y,\,J,\,H,$ and $K_{\rm s}$ bands. Both surveys have comparable or better depths with respect to the EWS. NIR data are complemented with the optical ($g,\,r,\,i,\,z$) photometry from DES DR2 \citep{abb21} for the VIDEO fields, from DES Deep Fields \citep{har22} for the COSMOS field, and with X-ray measurements from XMM-{\it Newton} and {\it Chandra} \citep{mar16,che18,ni21}.

We build the training sample by selecting the objects with spectroscopic measurements. We consider all the spectroscopic surveys that overlap the VIDEO and UltraVISTA areas and that have accessible 1\,D spectra (see \cref{t:specz_surveys}). The match between photometric and spectroscopic data results in about 21\,000 shared sources with high-quality spectra. All these spectra were visually inspected by OU-PHZ to verify the quality of spectroscopic redshifts. We end up with 13\,265 sources after rejecting objects with more than one missing flux across the optical and NIR ($Y,\,J,\,H$) bands.

Objects are labelled as stars if $z_{\rm spec}<0.0025$ \citep[this corresponds to a velocity of 750\,km\,s$^{-1}$, consistent with radial velocities of stars observed by {\it Gaia};][]{gaiadr3,kat23}. Sources detected in X-rays with $L_{2-10{\rm keV}}>10^{43.75}\,{\rm erg\,s^{-1}}$ are labelled as QSOs. All other objects are labelled as galaxies. The high $L_{2-10{\rm keV}}$ threshold for QSOs is chosen in order to select sources that are expected to be dominated by AGN in the optical. We obtain 2651 stars, 10\,263 galaxies, and 351 QSOs. Because of the small number of QSOs, we decided to label as QSOs all the X-ray detected sources with $L_{2-10{\rm keV}}>10^{43.75}\,{\rm erg\,s^{-1}}$ in the four fields even if they do not have spectroscopic redshifts, in which case photometric redshifts were used to determine luminosity. In this way, we can add another 1978 QSOs to the sample, for a total of 2329. The final training sample consists therefore of 15\,243 objects with known labels.

\begin{figure}
  \resizebox{\hsize}{!}{\includegraphics{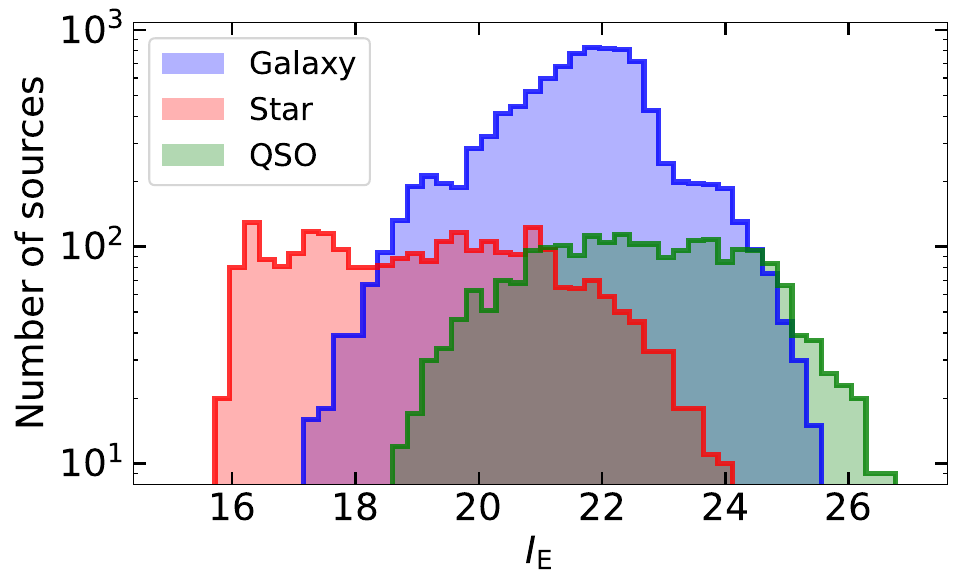}}
  \caption{\IE magnitude distribution per class of objects in the training sample for classification.}
  \label{f:cl:train}
\end{figure}

Before using the sample to train the classifiers, we apply the same pre-processing steps as done for \Euclid data (see \cref{app:preprocess}). In addition, we reconstruct the photometry corresponding to the UNIONS and \Euclid bands (DES photometry are already present in the sample) using the best-fitting SEDs obtained from \Phosphoros{} (see \cref{app:ssc:bandphot}). In \cref{f:cl:train} we show the distribution of objects in the training sample with respect to the $\IE$ magnitude: the distribution of galaxies peaks at $\IE\sim22$; stars and QSOs have a flatter distribution in the range 16--22 and 21--25, respectively. Although not fully representative of the \Euclid data in terms of magnitudes, the training sample is able to provide a good coverage of the colour distributions for the three object classes (see \cref{f:cl:color}).

\subsection{\label{ssc:class:valid} Classification results}

\begin{figure*}
  \resizebox{\hsize}{!}{\includegraphics{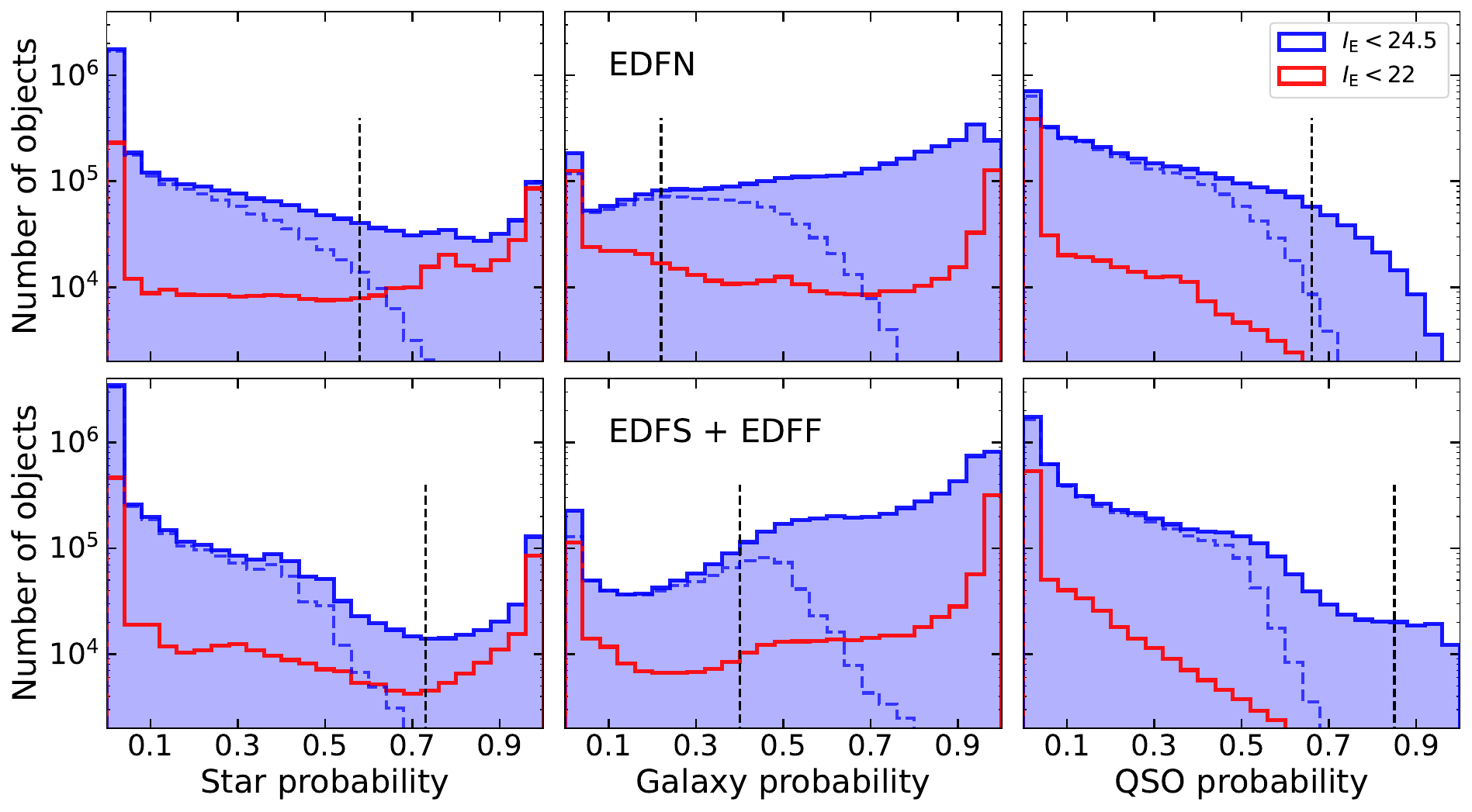}}
  \caption{Distributions of the class probabilities for sources in the north (top panels) and in the south (bottom panels) at $\IE<24.5$ (blue histograms) and at $\IE<22$ (red lines). Blue dashed lines are for sources that have a probability larger than 0.5 in a different class. Vertical lines correspond to the probability thresholds used for the classification (see \cref{t:cl:thres}).}
  \label{f:cl:prob}
\end{figure*}

\begin{figure}
  \resizebox{\hsize}{!}{\includegraphics{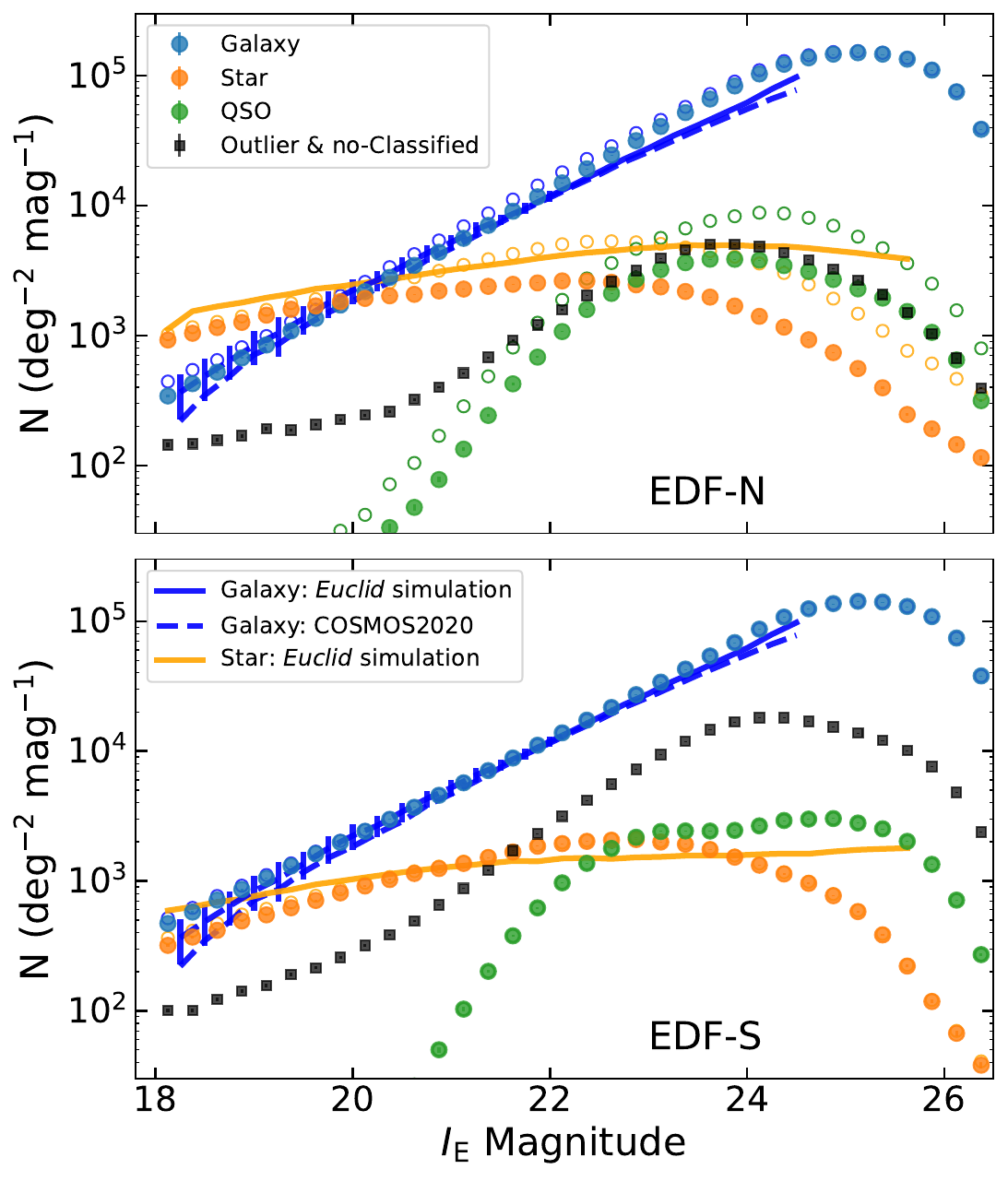}}
  \caption{Number counts per type as a function of the $\IE$ magnitude for the EDF-N (top panel) and EDF-S (bottom panel), compared with counts from the \Euclid SGS simulations (solid blue lines for galaxies and orange lines for stars), and from the COSMOS field (dashed blue lines for galaxies; \IE magnitudes have been reconstructed using the \Phosphoros{} best-fitting SEDs). Solid points are for single-class objects, while open points include multiple classified objects. Results from the EDF-F are similar to those of the EDF-S, except for a slightly lower number counts for stars, and are not reported.}
  \label{f:cl:ns}
\end{figure}

\begin{table}
    \caption{Percentage of VIS-detected objects per type.}
    \begin{tabular}{ccccccc}
    \hline \hline
 & Star & Galaxy & QSO & Multi & Outlier & Reject \\
 North & & & & Class & &  \\
    \hline
S/N>5 & 3.7 & 86.6 & 1.9 & 5.1 & 1.6 & 1.2 \\
\IE<24.5 & 7.9 & 77.7 & 2.2 & 7.8 & 2.4 & 2.0 \\
\IE<22 & 31.0 & 51.8 & 0.4 & 9.6 & 2.0 & 5.1 \\
    \hline
South & & & & & & \\
    \hline
S/N>5 & 2.3 & 87.9 & 1.6 & 0.1 & 6.9 & 1.2 \\
\IE<24.5 & 4.7 & 86.1 & 1.2 & 0.2 & 5.9 & 1.9 \\
\IE<22 & 17.2 & 74.0 & 0.0 & 1.1 & 2.9 & 4.7 \\
    \hline
    \end{tabular}
    \label{t:cl:fract}
\end{table}

In this section we present the classification results for VIS-detected objects with detection ${\rm S/N}>5$. \Cref{f:cl:prob} shows the distributions of the class probabilities. The peaks of the distributions are always found close to 0 and 1. If we consider only bright objects (e.g., $\IE<22$), the distributions tend to be still more concentrated at the edges of the probabilities. Moreover, very few objects have large probability ($>0.5$) in more than one class (see the dashed lines in the figure). These features prove the ability of the pipeline to separate the sources in the different classes and to assign reliable labels for most of them.

As expected, galaxies are the dominant class, almost 90\% of the total (see \cref{t:cl:fract}). Stars represent only a few per cent, but their fraction significantly increases with the detection flux: for example, they are 31\% of the sources at $\IE<22$ in the northern tiles and 17\% in the southern tiles. This difference between north and south can be attributed to the distribution of stars on the sky. For example, the Besan\c{c}on model\footnote{\url{https://model.obs-besancon.fr/modele\_home.php}} \citep{rob03} predicts about twice as many stars in the EDF-N with respect to the EDF-S and EDF-F. QSOs are approximately 1\% of the sources, but the percentage drops well below $1$\% among bright sources. Outliers and rejected objects are less than 10\%, while the number of objects with multiple classes is negligible in the southern sky and 5--10\% in the north, depending on the detection magnitude. 

\begin{figure*}
  \resizebox{\hsize}{!}{\includegraphics{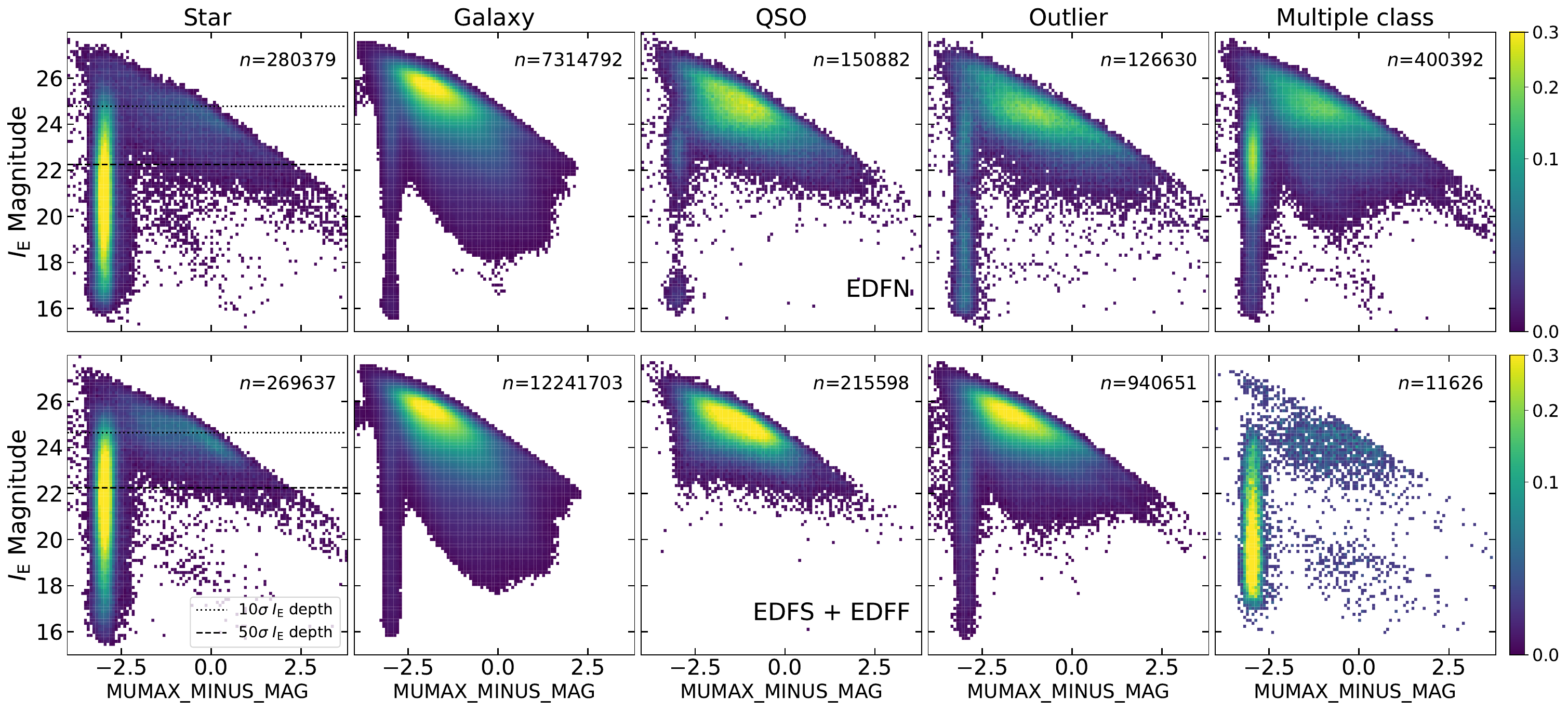}}
  \caption{Density plots of $\IE$ magnitude as a function of \texttt{MUMAX\_MINUS\_MAG} for the different classes of objects, in the EDF-N (top panels) and in the EDF-S and EDF-F (bottom panels). The number of sources per type is reported in each panel.}
  \label{f:cl:morpho}
\end{figure*}

The number counts of the different object classes are shown in \cref{f:cl:ns}. They can be compared with the estimates from the \Euclid SGS simulations. In particular, for stars we use the simulated catalogues for the EDFs, which combine real bright stars and the Besan\c{c}on model simulations. For galaxies, we also consider the number counts obtained from the COSMOS catalogue (W22). As we can see from the figure, galaxy counts follow our expectations quite well. Star counts are in general quite compatible with the Besan\c{c}on model at magnitudes between 20 and 24, while at fainter fluxes the star completeness quickly decreases. According to the model, we also lose about 20\% of the brightest stars ($\IE<20$). Part of these stars are misclassied as galaxies and can be easily identified in the galaxy sample through the morphology (see \cref{f:cl:morpho}). In the EDF-N, star counts are 20--30\% below the model at $\IE<23$, but they become consistent with it if stars with multiple classifications are added, suggesting that most of those sources are actually stars. This may be due to the very low galaxy probability threshold in the EDF-N that causes stars to be classified as galaxies too. Finally, both in the north and in the south, QSOs seem to be identified only among faint or noise-dominated objects, while essentially no bright QSOs are found. We note that rejected objects and outliers can give additional extra contributions to number counts, especially among bright sources (around 5--6\% at $\IE<22$).

\begin{figure}
  \resizebox{\hsize}{!}{\includegraphics{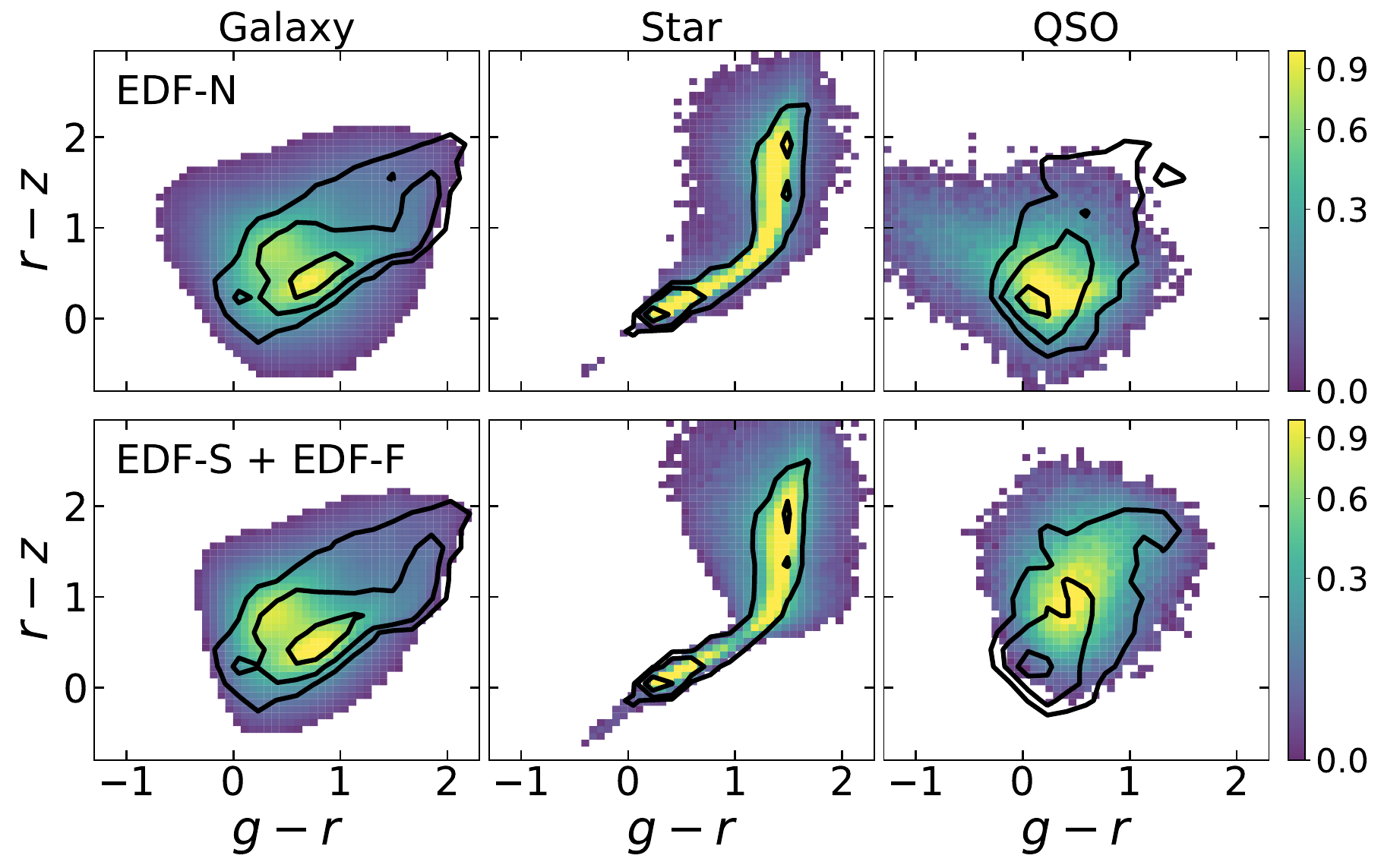}}
  \resizebox{\hsize}{!}{\includegraphics{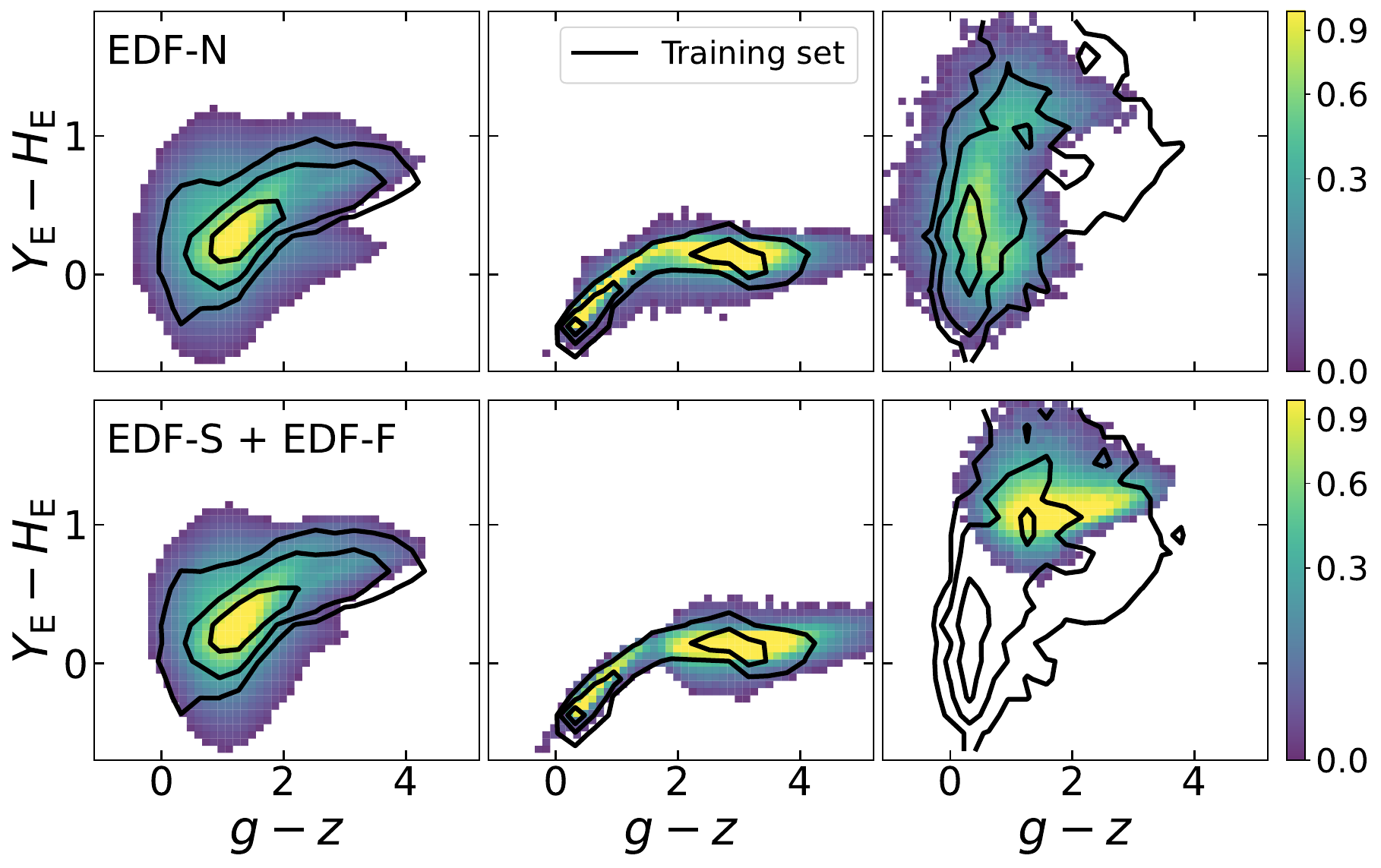}}
  \caption{Two-colour density plots for the different types of object in the northern and southern tiles, compared with the training set (black contour curves). Colours are corrected for Galactic extinction as described in \cref{app:ssc:gred}.}
  \label{f:cl:color}
\end{figure}

In \cref{f:cl:color}, we show two-colour plots for the different types of objects, in comparison with the training set. As expected, the peak of the colour distributions for galaxies and stars match well those from the training set. Their 2D distributions are, however, more widespread in the colour space due to noise effects. In contrast, colours of QSO-classified objects are partially compatible with the training set. For example, in the south, we only find QSOs with very red NIR colours ($\YE-\HE>1$), in the tail of the distribution of the training set, while no QSOs are found in the expected location for this class of objects (i.e., $0<g-z<1$ and $\YE-\HE\simeq0$). This is due to the high probability threshold for the QSO definition in the southern tiles (i.e., 0.85), and to the fact that such high probability can be obtained only in colour regions not covered by galaxies. In the northern tiles, the presence of the $u$ band allows the PRF to achieve the same purity level with a lower threshold, and consequently to find QSOs in a wider colour region.

For validation purposes, it is interesting to check the morphology of the classified objects, given by the parameter \texttt{MUMAX\_MINUS\_MAG}, as a function of the $\IE$ magnitude (\cref{f:cl:morpho}). Star-classified objects are mostly compact, especially at bright or intermediate magnitudes. The fraction of `extended stars' with ${\rm S/N}>50$ is less than 1\%, as required for the PSF modelling. Looking at the galaxies, we observe that, particularly in the southern tiles, there are several compact, bright galaxy-classified objects that might be stars or QSOs with wrong classification. We check the photometric redshift of point-like (\texttt{MUMAX\_MINUS\_MAG}<$-$2.5) galaxy-classified objects at $\IE<23$: more than half have $z<0.5$ (7\% and 28\% with $z<0.2$ in the southern and northern tiles, respectively), and about one third are at $0.5<z<1$. QSO-classified objects are typically faint and extended sources, making their classification quite doubtful. In the north, we find a small set of compact bright ($\IE<24$) QSOs that are probably correctly classified. Finally, if we consider outliers and multiply classified objects, there are significant differences between tiles in the north and in the south, as a result of the different probability thresholds used for classification.

To conclude, the classification seems to succeed in providing reliable class probabilities for stars and galaxies, and in extracting a pure sample of bright stars. Improvements are instead needed in the identification of QSOs, which is currently not reliable. Photometric information alone is likely not sufficient to distinguish QSOs from galaxies, which are much more numerous and overlap the colour space covered by QSOs \citep{EP-Bisigello}. The presence of the $u$ band in the northern configuration helps the QSO classification. More investigation is also needed on the outliers, a relevant fraction of which could be real objects with peculiar locations in the colour space.

\section{\label{sc:phz} Photometric redshifts for the \Euclid core science}

The main product of the processing function is the photo-$z$ catalogue (\texttt{catalogue.phz\_photo\_z} in the EAS) that includes the point estimates and the PDFs of photometric redshifts for all the objects present in the MER catalogues. In this section, we present the computation of the photometric redshifts, their validation, and the redshift distributions in the EDFs.

\subsection{\label{ssc:phz:wl} Building the PHZ catalogue through \Phosphoros}

Photometric redshifts are computed through the template-fitting \Phosphoros tool (Paltani et al., in prep.). With respect to previous template-fitting codes, the main innovation of \Phosphoros is the implementation of a fully Bayesian framework, with flexible priors on all parameters, and sampling from multi-dimensional and marginalised posteriors. It includes most of the advanced features found in similar codes, such as the use of upper limits, zero-point corrections, emission lines, etc. \citep[see, e.g.,][]{ben00,bol00,ilb06,bra08}, with the addition of new features such as: complex user-defined priors (e.g., from luminosity functions); an SED-dependent treatment of Galactic reddening \citep{gal17}; different intergalactic medium prescriptions; correction of the effects of the photometric passband variations \citep{EP-Paltani}; and the sampling of the posterior distributions. \Phosphoros has been developed by the OU-PHZ to achieve a maximal computational efficiency and to run in a computer-intensive processing environment. It was used and fully validated in the DC2 photometric-redshift challenge of the \Euclid OU-PHZ \citep{Desprez-EP10} and on the HSC-CLAUDS survey \citep{des23}.

Compared to the machine-learning methods typically employed for the photometric redshift estimates \citep[see, e.g.,][]{sal19,Desprez-EP10}, the template-fitting approach is considered more appropriate for the \Euclid core science. Machine-learning algorithms, in fact, critically depend on training sets that must be representative of the population of objects under study. They typically perform well in regions of the colour space that are well covered by spectroscopic redshifts, but their performance significantly drops for regions with scarce spectroscopic information, for instance for $z>1$ \citep{Desprez-EP10}. On the other hand, template-fitting algorithms are significantly slower than machine-learning ones, making their use for the full EWS quite challenging in terms of computational time. In the Q1 release, the use of \Phosphoros{} is still feasible in terms of computational time due to the limited number of objects involved. 

Template-fitting algorithms compare observed fluxes with a grid of modelled photometry, spanning over a set of parameters. In the case of \Phosphoros, the parameters are: redshift $z$; galaxy SED; intrinsic reddening curve; and intrinsic attenuation $E(B-V)$. The grid of models is built before applying \Phosphoros to the source catalogues.

Model photometry values are derived using a set of galaxy templates based on the COSMOS library \citep{ilb09}. It consists of 33 templates, eight templates for elliptical and lenticular galaxies and 11 for spiral galaxies from the \citet{pol07} library, and two for exponentially declining star-formation history (SFH) and 12 for young blue star-forming galaxies generated with the \citet[][hereafter BC03]{bru03} stellar population synthesis models. To obtain a finer grid, we added additional templates by interpolating between `adjacent' templates in colour space. Emission lines (\ha, the [\ion{O}{ii}] doublet at 3726+3728\,\AA, the \hb, \ion{H}{$\gamma$} and \ion{H}{$\delta$} Balmer lines, and the [\ion{O}{iii}] lines at 4958\,\AA{} and 5006\,\AA) are added to these templates using the \citet{ken98} relation between \ha{} and the ultraviolet luminosity, and an empirical relation between \ha{} and the emission line fluxes (see Paltani et al., in prep., for more details). A velocity dispersion of 200\,\kms{} is applied to the emission lines. 

The \Phosphoros model grid is computed between redshifts $z=0$ and 6 with 0.01 step. No internal attenuation is applied to early type (elliptical and S0) and (exponentially declining) passive galaxies, while for younger galaxies we consider an internal attenuation with $E(B-V)$ values in the range [0,\,0.5] with 0.05 step. The internal dust attenuation curve is modelled with several attenuation laws: \citet{pre84}; \citet{cal00}; and modified Calzetti law including a bump at 2175\,\AA, as in \citet{ilb09}. Intergalactic medium attenuation is also taken into account following the \citet{ino14} prescription. Finally, Galactic reddening is applied to modelled fluxes in an SED-dependent way, as described in \citet{gal17}. We adopt the Milky Way absorption law of \citet{gor23}.

Before running \Phosphoros, we add a systematic error to the statistical errors in the form of an additional uncertainty proportional to the flux, in order to fix an upper limit to the S/N,
\be
\sigma^2_k({\rm new})=\sigma^2_k({\rm obs})+\beta^2_k\,F^2_k\,,
\ee
where $F_k$ and $\sigma_k$ are the source flux and error at the $k$ band, and $\beta_k$ is equal to 0.02 (0.05) for the {\it g,\,r,\,i,\,z},\,\IE ({\it u},\,\YE,\,\JE,\,\HE) bands. This corresponds to setting the maximum S/N to 50 (20) in those bands. This systematic contribution should take into account for residual errors in the flux scale not absorbed by the ZP correction.

Finally, we apply the following priors to the \Phosphoros models: a `volume' prior proportional to the redshift-dependent differential comoving volume, with an effectiveness ${\rm eff}=0.3$,\footnote{The prior effectiveness is a value between 0 and 1 that modifies the prior as $p_{\rm eff}\propto p^{\,\rm eff}$.} to disfavour low-redshift solutions where the volume
sampled is very small; and a top-hat prior on the source absolute AB\,magnitudes estimated in the DES $r$ band, that is flat in the [$-$24,0] range and 0 otherwise. This prior may be a bit severe and a few bright galaxies at high $z$ could be lost \citep[see, e.g.,][]{ilb05}; however, it helps to avoid many outliers at $z>2$. 

The main products in the \texttt{photo\_z} catalogue are the redshift posterior probability distributions ($z$PDFs) of the sources, derived by marginalising the full posterior distribution over the template and reddening dimensions. These are provided as a vector containing the $z$PDF values for redshifts in the range [0,\,6] with 0.01 step. In addition, different point estimates of the source redshift are computed:
\begin{itemize}
\item the median of the $z$PDF; 

\item the centroid of the main peak of the $z$PDF;

\item the main mode of the full posterior distribution.
\end{itemize}
In general, the point estimates of photometric redshifts are quite consistent with each other. A slightly larger fraction of outliers is observed using the centroid of the main peak or the main mode of the posterior. So, hereafter, unless otherwise stated, we use the $z$PDF median as our redshift estimate.

\subsection{\label{ssc:phz:valid} Validation of photometric redshifts}

The performance of \Phosphoros photometric redshifts ($z_{\rm phos}$) is evaluated on the basis of the residuals $\Delta z=(z_{\rm phos}-z_{\rm ref})/(1+z_{\rm ref})$, where $z_{\rm ref}$ are the reference `true' redshifts (typically from spectroscopic data). We use the following quality metrics: the median of the residuals, ${\rm median}(\Delta z)$, as a measure of the accuracy; the normalized median absolute deviation, $\sigma_{\rm NMAD}=1.4826\,{\rm median}(|\Delta z-{\rm median}(\Delta z)|)$, as a measure of the precision \citep{hoa83}; and the fraction of outliers ($\eta$), defined as the fraction of objects with $|\Delta z|\ge0.15$.

\begin{figure*}
  \resizebox{\hsize}{!}{\includegraphics{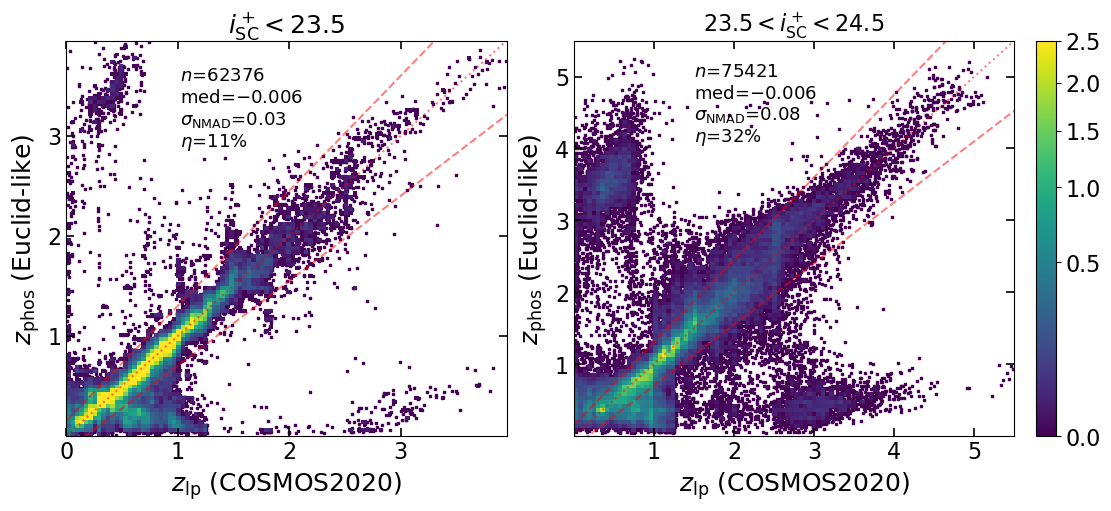}}
  \caption{Density plots comparing the photometric redshifts of the COSMOS2020 sources as computed by \Phosphoros in a \Euclid-like photometry configuration with the photometric redshifts provided in W22 using the \texttt{LePhare} code and the CLASSIC catalogue. Two ranges of magnitude are considered: Suprime-Cam $i^+<23.5$ (left panel); and $23.5\le i^+<24.5$ (right panel). \Phosphoros{} redshifts are the median point estimates. In each plot we report the number of sources and the values of the quality metrics. The dotted red lines show the 1:1 line and the dashed lines show $z_{\rm phot}=z_{\rm spec}\pm0.15(1+z_{\rm spec})$, corresponding to the limits for outliers.}
  \label{f:v_cosmos}
\end{figure*}

Photometric redshifts obtained from the \Phosphoros tool are tested and validated using both photometric and spectroscopic redshifts. While spectroscopic redshifts allow us to compare our predictions with high-accuracy and high-precision redshifts, the spectroscopic sample is not representative in terms of depth and galaxy types. Therefore, we also compare the \Phosphoros{} predictions with high-quality photometric redshifts from COSMOS2020 obtained from deep, wide-band photometric observations (W22). The catalogue contains 1.7 million sources measured in 36 photometric bands from the UV to the infrared. Based on the available spectroscopic data, W22 found that photometric redshifts for galaxies at $i<24$ have a precision of the order of $0.01(1+z)$ and an absolute bias $\le0.004(1+z)$, with 3--4\% of outliers. In order to mimic the \Euclid conditions, we extract the Subaru $g,\,r,\,i,\,z$ HSC optical bands, the Suprime-Cam $i^+$ band and the VIRCAM UltraVISTA $Y,\,J,\,H$ bands to build a mock EWS sample. The $u$ band is not used. In addition, because the COSMOS2020 catalogue is substantially deeper than the \Euclid Q1 data, we add random Gaussian noise to the source photometry to match the nominal Q1 depth of each band. The redshift determination is performed on sources with $i^+<24.5$ and ${\rm S/N}>10$. \Phosphoros is applied with the configuration used in the pipeline and described in the previous section.

\Cref{f:v_cosmos} shows a comparison between the \Phosphoros photometric redshifts and those obtained in W22 using the \texttt{LePhare} code \citep{arn02,ilb06} and the CLASSIC COSMOS2020 catalogue. We consider sources in two ranges of $i^+$ magnitude, brighter and fainter than 23.5. In general, \Phosphoros{} redshifts match quite well with the COSMOS2020 ones, in spite of the significant smaller number of bands involved and noisier photometry. For sources at $i^+<23.5$, we find quality metrics in agreement with the \Euclid requirements for core science. Outliers are mainly found at \texttt{LePhare} redshifts $z_{\rm lp}\simeq0.5$--1 and $z_{\rm phos}<0.5$. This is likely due to the degeneracy between blue galaxies at $z\simeq0.6$, which have a weak 4000\,\AA{} break in the $r$ band, and red galaxies with (Balmer) break blueward of the $g$ band and not captured because of the lack of the $u$ band at $z\simeq0.2$. For faint sources ($i^+>23.5$), we have a clear increase in the dispersion ($\sigma_{\rm NMAD}$ changes from 0.03 to 0.08), while the median of the residuals remains quite small. The fraction of outliers grows up to 32\%. Two (almost) symmetrical clouds are present at $z_{\rm lp}\la0.5$ and $z_{\rm phos}=3$--4, and vice versa, due to a confusion between the Balmer and Lyman breaks. 

\begin{table*}
    \caption{External spectroscopic catalogues with matches in Q1}
    \begin{tabular}{rlllr}
    \hline \hline
    Matches & Survey & References & Quality cuts & EDF- \\
    \hline
    5 &SDSS DR16 & \cite{ahu20} & \texttt{zwarning=0 \& class$\ne$'STAR'} & N \\
    7759 &OzDES & \cite{lid20} & \texttt{2$\leq$qop$<$6 \& z$>$$-$1} & F \\
    16\,851 &PRIMUS & \cite{coi2011} & \texttt{zprimus\_zconf $\geq$ 3 \& duplicate removal} & F \\
    & & \cite{coi2013} & & \\
    2937 &3D-HST & \cite{mom2016} & \texttt{z\_best\_s $\in$ $\lbrace$1,2$\rbrace$} & F \\
    263 &JADES & \cite{eis2023} & \texttt{z\_Spec\_flag $\in$ $\lbrace$A,B$\rbrace$} & F \\
    & & \cite{deu2024} & & \\
    54 &2MRS & \cite{huc2012} & & F,S \\
    1212 &2dFGRS  & \cite{col03} & \texttt{q\_z $\geq$ 3} & F,S \\
    145 &6dFGS  & \cite{jon2009} & \texttt{q\_cz $\geq$ 3} & F,S \\
    716 &VVDS  & \cite{lef13} & \texttt{zflags $\in$ $\lbrace$3,4$\rbrace$} & F \\
    52 &MOSDEF & \cite{kri2015} & \texttt{target=1 \& z\_mosfire $>$0 \& z\_mosfire\_zqual=7} & F \\
    227 &2dflens  & \cite{bla2016} &  \texttt{qual $\in$ $\lbrace$3,4$\rbrace$} & F \\
    699 &VANDELS  & \cite{gar21} & \texttt{zflg $\in$ $\lbrace$3,4,9,13,14,19$\rbrace$} & F \\
    42\,706 &DESI\_EDR  & \cite{desi24} & \texttt{z$>$0.001 \& objtype=TGT \& spectype$\ne$STAR} & N \\
    & & & \texttt{\& deltachi2$>$10 \& zcat\_primary=1} & \\
    & & & \& \texttt{coadd\_fiberstatus=0 \& zwarn$<$4} & \\
    \hline
    \end{tabular}
    \label{tab:specztable}
\end{table*}

\begin{figure*}
  \resizebox{\hsize}{!}{\includegraphics{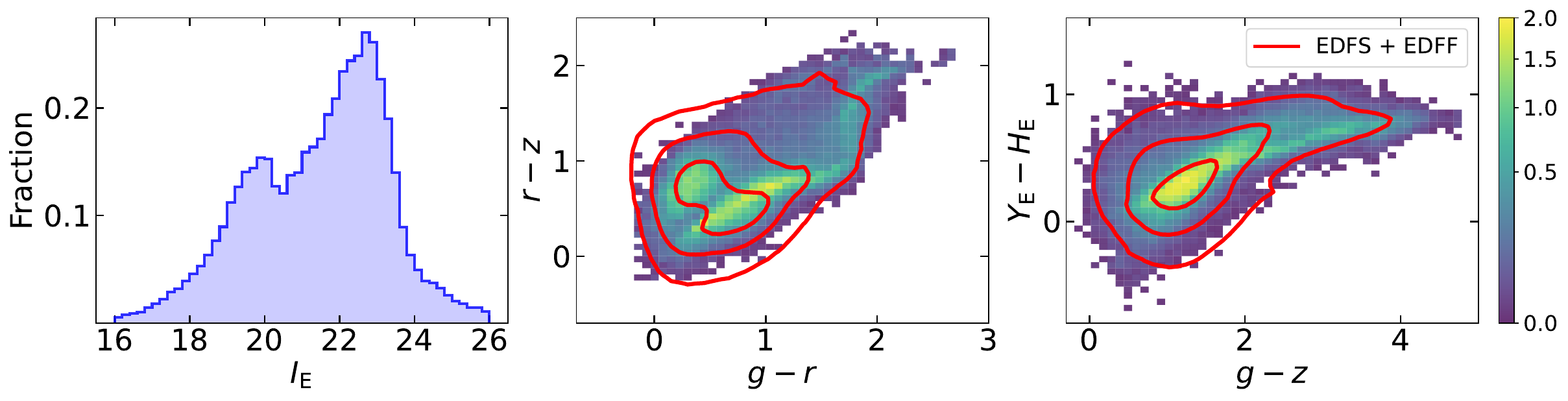}}
  \caption{\IE magnitude distribution (left panel) and two-colour density maps (central and right panels) of Q1 sources with spectroscopic redshifts from external surveys. Red contours are the 2D colour distributions for all the Q1 sources with $\IE<24.5$ in the southern tiles.}
  \label{f:v_specz_col}
\end{figure*}

\begin{figure*}
  \resizebox{\hsize}{!}{\includegraphics{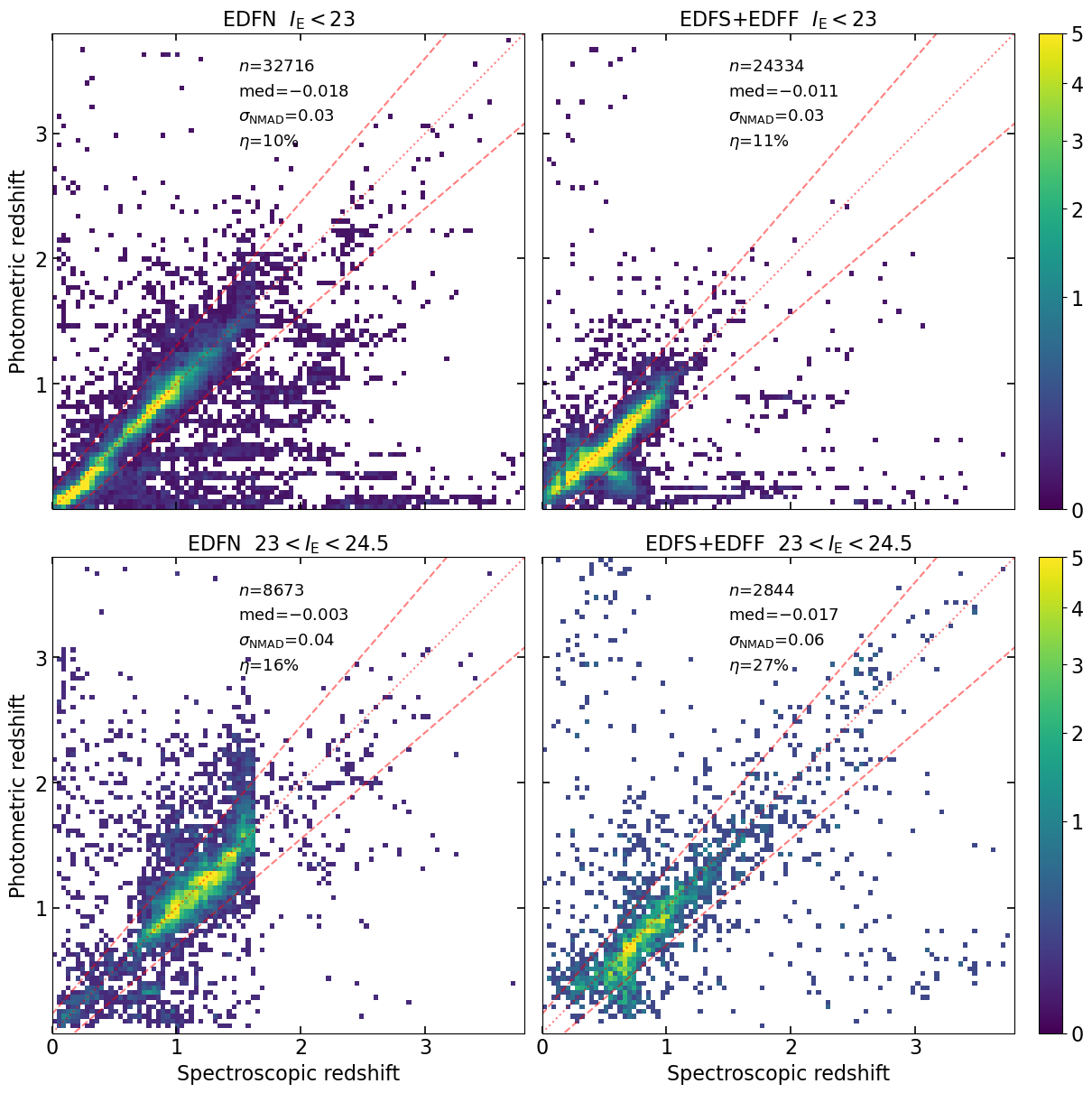}}
  \caption{Density plots comparing \Phosphoros photometric redshifts for Q1 sources with spectroscopic redshifts available in public surveys. Sources are separated according to the ground-based observations (in the northern and southern sky), and for the range of magnitude, $\IE<23$ and $23\le\IE<24.5$.}
  \label{f:v_specz}
\end{figure*}

The quality of photometric redshifts computed by the PHZ pipeline can be directly checked for those Q1 objects whose spectroscopic redshifts are known. To this aim, we have compiled a large set of external surveys with publicly available spectroscopic data and an overlapping footprint with Q1 (see \cref{tab:specztable}). For the match with Q1 sources, we take a tolerance of 1$\arcsecond$ for ground-based data sets and \ang{;;0.4} for space-based data sets such as 3D-HST \citep{mom2016} and JWST Advanced Deep Extragalactic Survey \citep[JADES;][]{eis2023,deu2024}. We use only redshifts with high-quality flags. In total, we find 72\,263 matched \Euclid sources, 42\,731 in the EDF-N (mainly from DESI), 29\,361 in the EDF-F, and only 171 in the EDF-S. 

In \cref{f:v_specz_col} we can see the \IE magnitude distribution of sources with spectroscopic redshifts. As expected, they are mostly bright objects ($\IE<23$). Nevertheless, in terms of colours, the spectroscopic sample is well representative of the full Q1 sample, without selecting or excluding particular regions of the colour space. \Cref{f:v_specz} shows a comparison between the photometric redshifts computed by the PHZ pipeline and the spectroscopic redshifts available for the EDFs. This is done separately for the EDF-N and for the EDF-F and EDF-S, and for sources with $\IE<23$ and $23\le\IE<24.5$. The match is quite good independently of the ground-based configuration and of the source magnitudes. The accuracy of photometric redshifts is typically of the order of 1--2\%, while the precision is in line with the requirements, with $\sigma_{\rm NMAD}=0.03$ at $\IE<23$ and 0.04--0.06 at $23<\IE<24.5$. Quality metrics significantly degrade with magnitude only for the fraction of outliers, which changes from 10\% (12\%) to 16\% (27\%) in the north (south).

The location of the outliers is quite different between north and
south. In the south a large fraction of the outliers are found at
$z_{\rm spec}=0.5$--0.8 and $z_{\rm phot}=0.2$--0.3, very similarly to
what is observed in the COSMOS2020 comparison (\cref{f:v_cosmos}). The
degeneracy between blue and red galaxies seems to break in the UNIONS
configuration, probably thanks to the $u$ band, especially for bright
objects. Outliers in the north appear instead more scattered, as a
possible effect of the Balmer and Lyman break confusion, AGN or QSO
contamination, or possible photometric issues.

\begin{figure}
  \resizebox{\hsize}{!}{\includegraphics{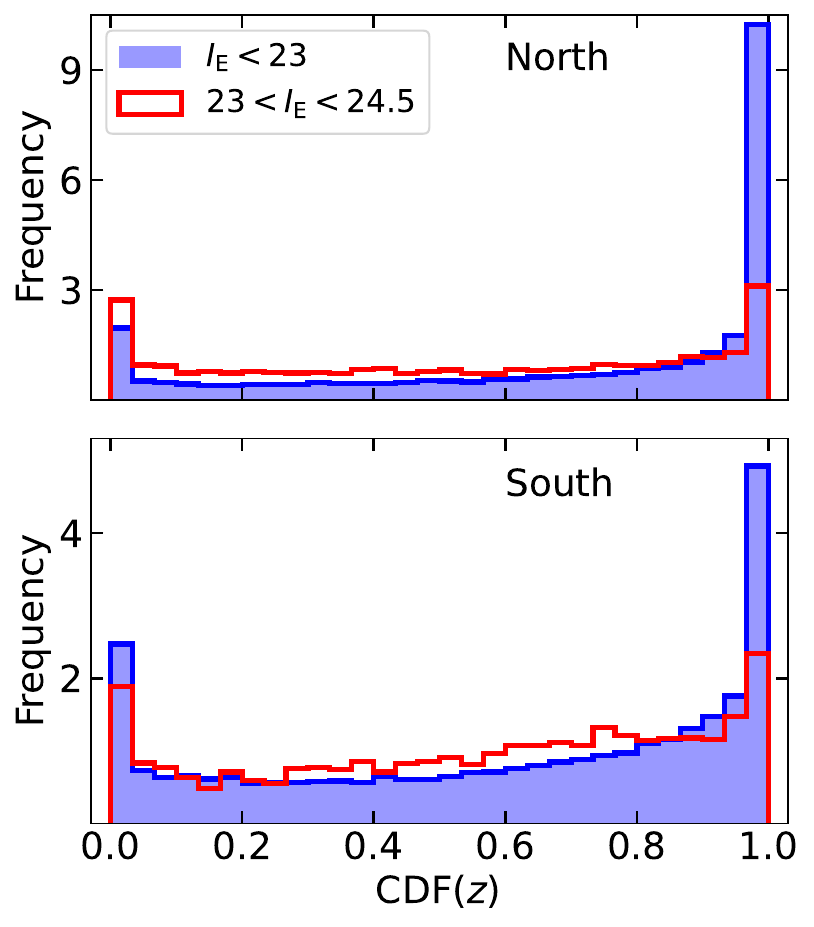}}
  \caption{Probability integral transform (PIT) plots for sources in the northern and southern sky, and with $\IE<23$ and $23\le\IE<24.5$.}
  \label{f:v_pit}
\end{figure}

Information on the quality of the \Phosphoros{} $z$PDFs can be obtained using the probability integral transform (PIT) \citep{daw84,dis18}, which shows the histogram of the cumulative distribution functions (CDFs) at the spectroscopic redshifts (see \cref{f:v_pit}). The histogram should be flat if the $z$PDFs correctly represent the probability distribution of the sources. We observe instead a `U' shape in the plots, especially for bright sources, which indicates that our $z$PDFs are too narrow, leading to an underestimation of the errors (this issue has already been reported in \citealt{Desprez-EP10} for \Phosphoros{} $z$PDFs). The strong peaks at the edges of the histograms are mainly produced by the outliers, whose PIT values are close to 0 or to 1. The PIT is asymmetric and increases towards 1, meaning that the $z$PDFs tend to underestimate the redshifts, in accordance with what is observed in \cref{f:v_specz}. 

\subsection{\label{ssc:phz:res} Redshift distribution of Q1 sources}

\begin{figure}
  \resizebox{\hsize}{!}{\includegraphics{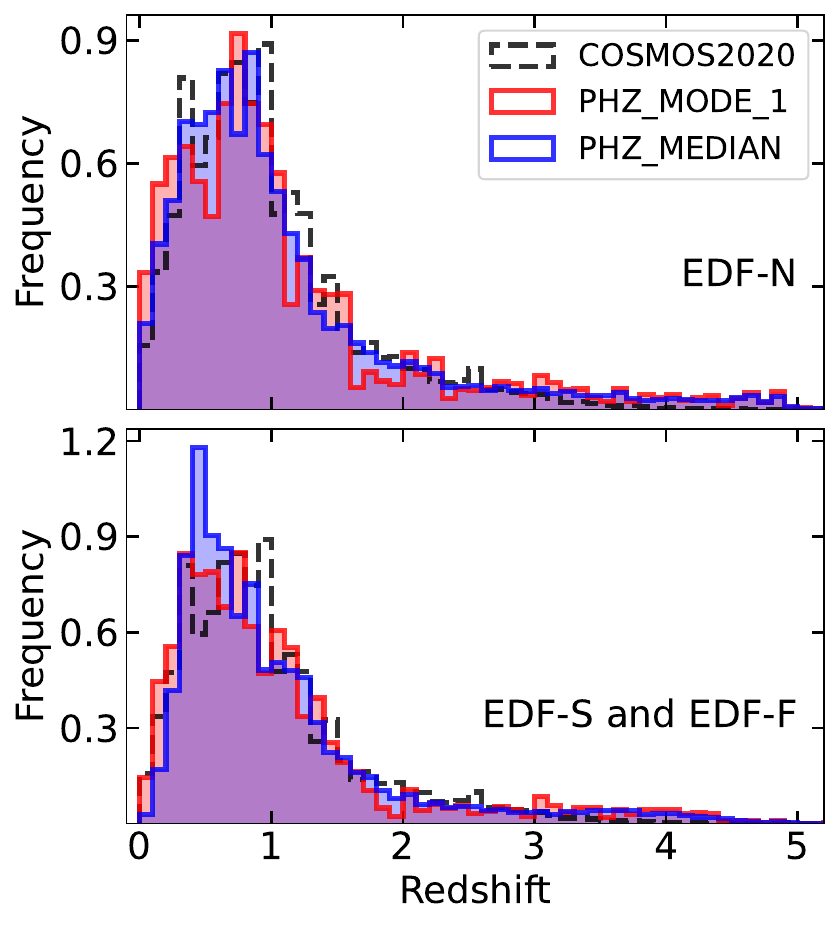}}
  \caption{Redshift distribution for VIS-detected $\IE<24.5$ objects from the EDF-N (top panel), and from the EDF-S and EDF-F (bottom panel). We use as redshift estimate the median of the $z$PDFs (blue histograms) and the centroid of the main peak (red histograms). Dashed histograms are obtained from the COSMOS2020 catalogue.}
  \label{f:nz}
\end{figure}

In \cref{f:nz} we show the redshift distributions for sources in the northern and southern sky. The distributions are computed for all VIS-detected objects that verify the following conditions: $\IE<24.5$; $\IE\,{\rm S/N}>5$; and not star-classified. Such conditions reduce by more than half the number of objects, but keep the object density larger than $30\,$per\,arcmin$^2$. The redshift distributions are obtained using two different point estimates: the median of the $z$PDFs; and the centroid of the main peak. In the figure, we also include the redshift distribution obtained from  the COSMOS2020 catalogue, after applying a magnitude cut at $\IE<24.5$.

The redshift distributions we find show reasonable agreement with COSMOS2020, taking into account the possible effects of sampling variance, independently of the point estimate and the sky area. As expected, they peak at redshift 0.5--0.8, with a sharp decrease at $z>1$. The $N(z)$ distributions tend to overestimate the number of high-$z$ galaxies in the range of 3--5, which is due to low-redshift galaxy outliers with photometric redshift larger than 2--3 (as seen for example in \cref{f:v_cosmos}).

\section{\label{sc:legacy} Products for non-cosmological science}

\begin{table}
\caption{Physical parameters computed by the PHZ PF}
\label{t:pp:list}
\centering
\begin{tabular}{lc}
\hline \hline 
Parameters & Comments\\ \hline
\multicolumn{2}{l}{Galaxies (Method: \NNPZ)} \\
  \hline
  Redshift & \\
  AGN bolometric luminosity \\
  Intrinsic attenuation $A_V$ & not in the catalogue\\
  Absolute AB magnitude\tablefootmark{a} & \\
  Stellar metallicity & [Fe/H] \\
  (SFH) Stellar age of galaxies & \\
  (SFH) Characteristic SF timescale & \\
  Star formation rate (SFR) & $\log_{10}({\rm SFR}/{\rm M_{\odot}\,yr}^{-1})$ \\
  Stellar mass & $\log_{10}({\rm M}_*/{\rm M}_{\odot})$ \\
  Stellar mass formed during SFH & $\log_{10}({\rm M}^{\,\rm form}_*/{\rm M}_{\odot})$ \\
\hline
\multicolumn{2}{l}{Stars, QSOs, NIR-only objects (Method: \Phosphoros)} \\
\hline
  Redshift & not for stars\\
  SED normalization in $L_{\odot}$ & \\
  Luminosity in $L_{\odot}$ & not for stars\\
  SED template & \\
  Reddening curve & not for stars\\
  Internal color excess $E_{B-V}$ & not for stars\\
\hline 
\end{tabular}
\tablefoot{
\tablefoottext{a}{Absolute magnitudes are reconstructed for the
  following passbands: GALEX FUV and NUV; Johnson $U,\,B,\,V,\,R$, and $I$; \Euclid $\IE,\,\YE,\,\JE$, and $\HE$; SLOAN $z$; 2MASS $K_{\rm s}$.}
}
\end{table}

The non-cosmological branch of the production pipeline aims to extract from source photometry relevant information for the scientific cases other than cosmological probes, such as galaxy properties and evolution, the identification of galaxies and quasars at high redshifts ($z\ga6$), the evolution of active galaxies across redshift, and the study of the stellar population in the Milky Way. 

Sources from MER catalogues are treated differently depending on the assigned class (as shown in \cref{f:prod}). Physical properties of galaxy-classified objects are computed through the \NNPZ algorithm, while objects classified as star, QSO, or NIR-only are handled by the \Phosphoros tool. The products of this branch are four different catalogues, one for each class of objects,\footnote{In the EAS, they are the \texttt{phz\_physical\_parameters, phz\_star\_template, phz\_qso\_physical\_parameters}, and \texttt{phz\_nir\_physical\_parameters} catalogues.} which include the computed physical properties and the source fluxes corrected for Galactic reddening.

\subsection{\label{ssc:pp:gal} Physical properties of galaxies}

Physical properties of galaxies are determined using the \NNPZ algorithm (Paltani et al., in prep.). This is a supervised learning algorithm based on the nearest neighbours method (see also \citealt{tan18}, for a first application on the HSC-SSP survey). In the version used by the pipeline, \NNPZ searches in flux space for the $k$ closest neighbours from a reference sample, and computes the required attributes (typically in the form of PDFs) as a weighted average from the $k$ neighbours. The weights are equal to $\exp(-\chi^2/2)$, where the $\chi^2$ (or squared Mahalanobis) distance between the observed photometry and the photometry of the neighbours is
\be
\chi^2=\sum_i(F_{i,{\rm obs}}-F_{i,{\rm neig}})^2/\sigma_{F_{i,{\rm obs}}}^2
\ee
(the sum is over the fluxes in the different passbands). This method does not require any training phase, and it can be efficiently modified to compensate effects that vary on an object-by-object basis \citep[e.g., Galactic reddening and bandpass variations;][]{gal17,EP-Paltani}. The algorithm was tested and validated for the determination of photometric redshifts and physical properties using simulations \citep{EP-Enia} and \Euclid observations \citep{Q1-SP031}.

The $\chi^2$ distance we use in \NNPZ quantifies the likelihood that the galaxy in question is a copy of a given neighbour. In the cases where there are no close neighbours, the object is given a \texttt{PHYS\_PARAM\_FLAGS} value of 1. The general concept for \NNPZ within the \Euclid pipeline is for the reference sample of possible neighbours to be drawn from the PHZ EAFs \citep[see][]{EuclidSkyOverview}, where the much deeper photometry and the many additional bands make the physical property predictions much better than what is possible in the EWS. In addition, after applying a selection similar to that of the EWS, the population of galaxies acts as a natural prior on the physical properties of the ESA galaxies. Because of the unavailability of deep Euclid observations of the EAFs, for Q1 the reference sample is built from a set of more than 1 million stellar population synthesis models. In this case, \NNPZ acts as an emulator of a traditional template-fitting algorithm. 

We generate the model SEDs using the \texttt{bagpipes} package \citep{car18,car19}, with the 2016 version of BC03 models and delayed exponential SFHs.\footnote{The delayed exponential SFH is described according to $(t/\tau^2)\exp(-t/\tau)$, where $t$ is the galaxy age and $\tau$ is the characteristic timescale of star formation (BC03).} The parameters that we vary, their ranges and sampling (log space or linear) are given in \cref{t:pp:para}. In order to sample the six-dimensional space of model parameters, we generate a Halton sequence \citep{hal60}. The Halton sequence is a quasi-random sampling of a given space that provides a semi-regular sampling of the space, and hence it minimizes the maximum distance that a point in the continuous space can be from a sample compared with a regular grid or true random distribution. We generate models for two dust laws, namely the \citet{cal00} and the SMC \citet{pre84} curves, and re-used the same Halton sequence in the other six parameters for each. All SEDs are generated with a fixed total mass in stars formed $M^{\,\rm form}_* = 10^{10} {\rm M_{\odot}}$, which is then scaled according to the object photometry. In addition to the parameters governing the generation of the SEDs, \texttt{bagpipes} returns observed and rest-frame photometry through a desired set of passband filters (see \cref{t:pp:list} for the rest-frame band outputs). 

For objects classified as galaxies, we derive a large set of physical properties, including photometric redshift, star formation rate (SFR), stellar mass, and metallicity (the full list is reported in \cref{t:pp:list}). We stress that a new estimate of photometric redshifts is provided by \NNPZ; the comparison with \Phosphoros{} values is discussed later (see \cref{f:pp:gal:phos_nnpz}). Properties such as stellar mass and SFR are typically determined by fitting synthetic stellar population synthesis SED models \citep[e.g., BC03;][]{mar05,con10}. Posterior distributions of these and other properties show correlations that arise naturally through their dependence on luminosity distance (if redshift is allowed to vary) and mass-to-light ratio, for example. Capturing and storing these correlated posteriors in full for the billions of galaxies that \Euclid will observe is impractical, and so we provide instead an approximation of a sampling of the posterior given by the 30 closest neighbours in the reference data set.

\begin{table}
\caption{Parameters of the model galaxy SEDs, their range, and type of sampling.}
\label{t:pp:para}
\centering
\begin{tabular}{cccc}
\hline \hline
Property & Min value & Max value & Sampling \\
\hline
Redshift & 0.01 & 7 & linear in $(1+z)$\\
Age & $0.01~$Gyr & $13.74~$Gyr & log \\
$\tau$ & $0.01~$Gyr& $31.6~$Gyr & log \\
Stellar metallicity & $0.1~$Z$_{\odot}$ & $2~$Z$_{\odot}$ & linear \\
$A_V$ & $0~$mag & $2.98$~mag& log\\
$\log_{10}U$ & $-4$ & $-2$ & linear\\
\hline
Dust law & \multicolumn{3}{c}{Calzetti / SMC}\\
IMF & \multicolumn{3}{c}{Kroupa}\\
IGM & \multicolumn{3}{c}{Inoue}\\
Stellar spectra & \multicolumn{3}{c}{MILES}\\
\hline
\end{tabular}
\end{table}

\begin{figure*}
  \resizebox{\hsize}{!}{\includegraphics{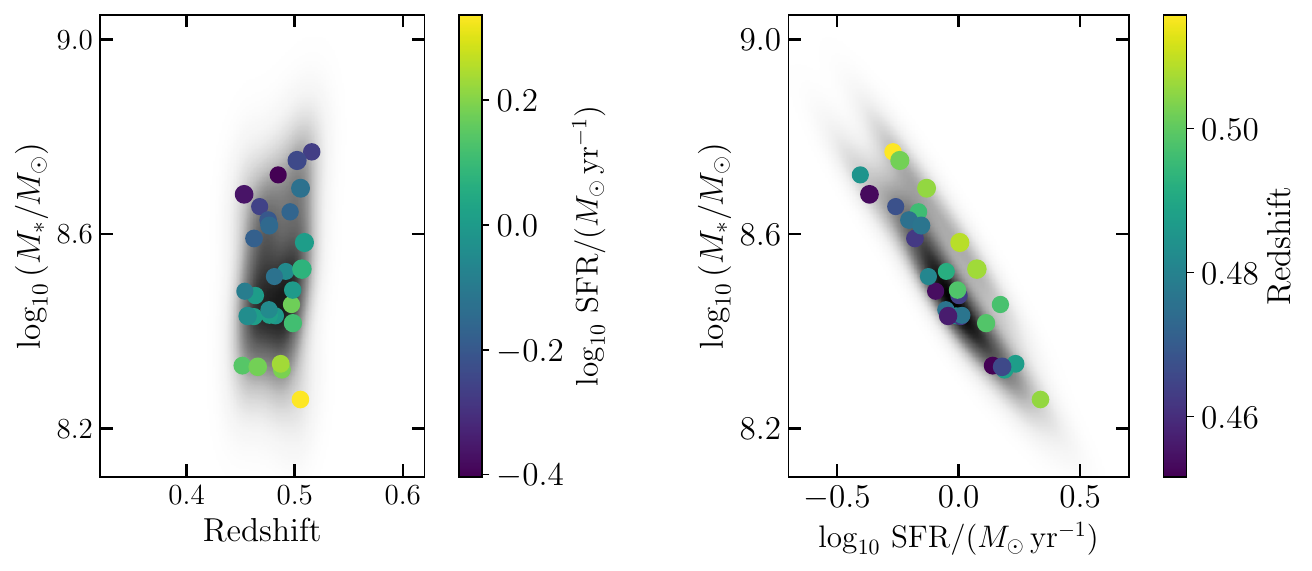}}
  \caption{2D likelihood distribution of $M_*$ versus redshift (left panel) and versus SFR (right panel) for an example galaxy (\texttt{OBJECT\_ID}=$-$549665747293683088). Colour-coded points correspond to the 30 nearest neighbours.}
  \label{f:pp:gal:2dpdf}
\end{figure*}

For each \Euclid Q1 galaxy, \NNPZ identifies the 30 nearest-neighbour reference SED models taking into account Galactic reddening and scaled to best match the object photometry. The IDs of these neighbours, their distances (likelihoods), and amplitude scalings are output as part of our physical property catalogue. The physical properties of the maximum likelihood model are recorded in the catalogue as the mode values, such as \texttt{PHZ\_PP\_MODE\_STELLARMASS}. In addition, the collection of neighbours is used to derive weighted percentile values at $16$, $50$, and $84\%$ for the marginalised posterior of each galaxy property (stored in the \texttt{PHZ\_PP\_MEDIAN\_*} and \texttt{PHZ\_PP\_68\_*} columns). As an example of physical property posteriors obtained by \NNPZ, \cref{f:pp:gal:2dpdf} shows the 2D posterior distribution of stellar mass versus SFR and versus redshift for a particular galaxy.

\subsubsection{Known issues}

Following the full production run of the galaxy physical properties, we identified two areas for improvement for future releases. The first is related to the intrinsic attenuation values $A_V$, which are currently absent from our output catalogue. The correct values are nevertheless present in the reference sample and can still be accessed through the set of neighbour IDs.

A second, more significant aspect concerns a subset of objects displaying very high specific star-formation rates (around $10^{-8}\,$yr$^{-1}$) and very low ages (around $10^7\,$yr). In designing the model parameter ranges, our goal was to avoid imposing hard priors, allowing users to apply their own constraints based on specific scientific needs. However, the choice to include ages as young as $10^7\,$yr led to a problematic outcome: many objects were predominantly fit by models near this lower limit, resulting in very low stellar mass estimates.

Despite these challenges, the majority of Q1 galaxies have well-determined physical properties and remain suitable for scientific analyses, particularly in contexts where completeness is not the primary concern. We are actively investigating the best steps to refine our approach and further improve the pipeline, ensuring greater accuracy and reliability in future analyses. For further discussion of these issues, and strategies to overcome them, we refer the reader to \citet{Q1-SP031}.

\begin{figure}
  \resizebox{\hsize}{!}{\includegraphics{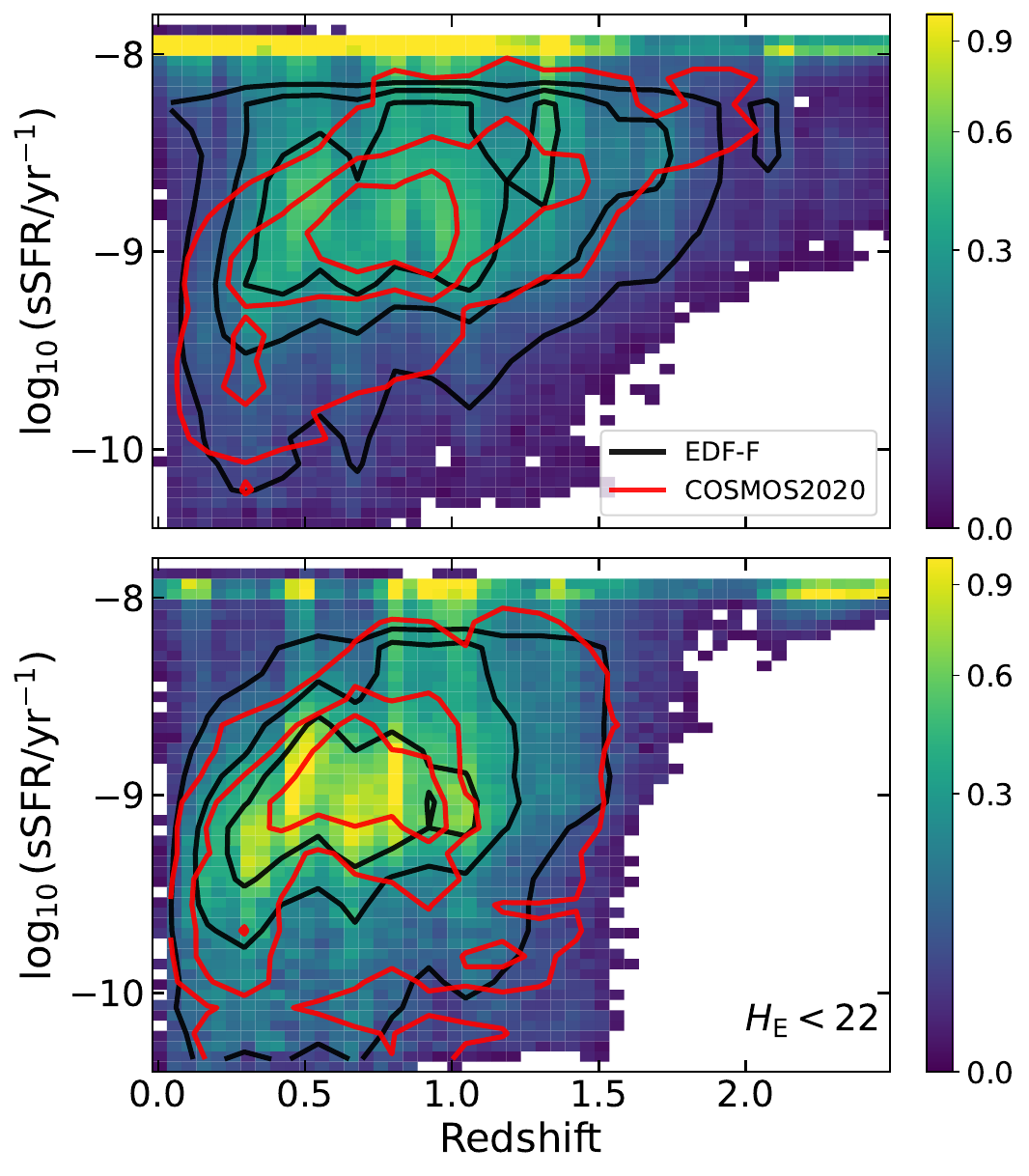}}
  \caption{Density map of sSFR as a function of redshift for galaxies in the EDF-F at $\HE<24$ (top panel) and at $\HE<22$ (bottom panel). Density contours from the COSMOS2020 catalogue (red lines) and from the \Euclid galaxies with $\logten{({\rm sSFR/yr}^{-1})}<-8.2$ (black lines) are superimposed.}
  \label{f:pp:gal:ssfr_z}
\end{figure}

\subsubsection{\label{ssc:pp:gal:v} Validation of galaxy physical properties}

\begin{figure}
  \resizebox{\hsize}{!}{\includegraphics{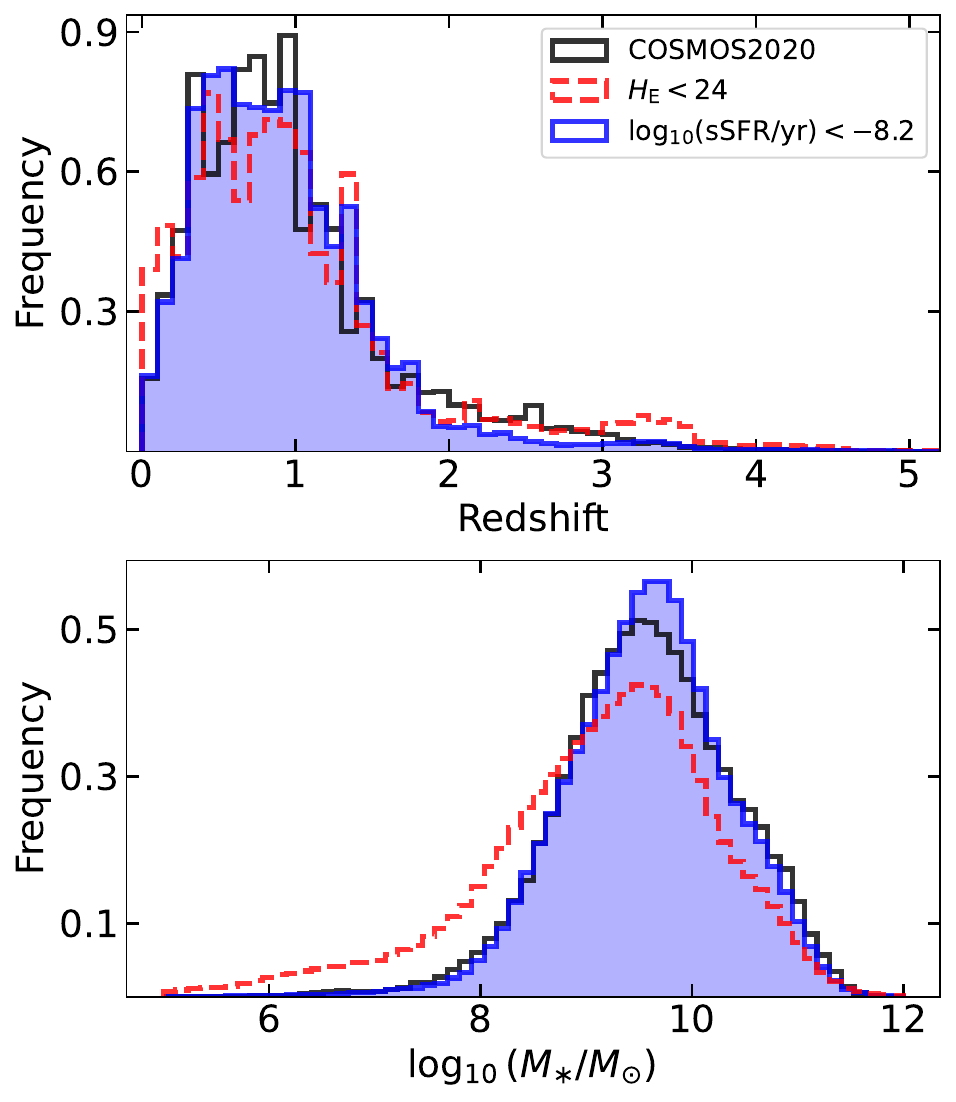}}
  \caption{Redshift (top panel) and stellar mass (bottom panel) distribution for galaxy-classified objects in the EDF-F with $\logten{({\rm sSFR/yr}^{-1})<-8.2}$, compared with the distributions from the COSMOS2020 catalogue (black lines). The red dashed histograms are for all EDF-F sources with $\HE<24$.}
  \label{f:pp:gal:histo}
\end{figure}

\begin{figure}
  \resizebox{\hsize}{!}{\includegraphics{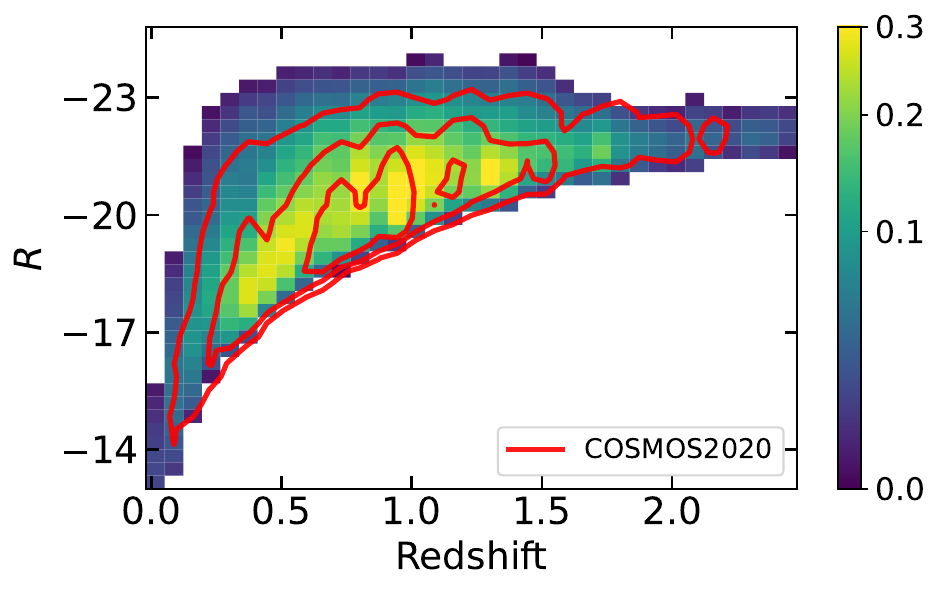}}
  \caption{Density plot of galaxy $R$ absolute magnitude as a function of redshift (excluding the problematic galaxy population). Density contours from the COSMOS2020 catalogue are superimposed.}
  \label{f:pp:gal:r_z}
\end{figure}

Galaxy physical properties derived with \NNPZ are compared with those obtained from the COSMOS2020 catalogue using \texttt{LePhare} (W22). We apply similar magnitude cuts to both data sets, limiting the \Euclid sample to $\HE<24$ and $\IE<24$, and the COSMOS2020 sample to $H_\sfont{VISTA}<24$ and HST/ACS ${\rm F814W}<24.5$. In addition, in order to minimise the number of spurious detections, stars, or QSOs present in the Q1 galaxy-classified sample, we exclude compact sources (\texttt{MUMAX\_MINUS\_MAG}>$-$2.6) and we apply the following MER and PHZ flags: \texttt{MER\_SPURIOUS\_FLAG}=0; \texttt{MER\_DET\_QUALITY\_FLAG}<4; and \texttt{PHZ\_FLAGS}=0.\footnote{\texttt{PHZ\_FLAGS} is a flag present in the photo-$z$ catalogue that says if a source is good for the core science (0=good) 
or not (10=NIR-only, 11=missing bands, 12=too faint).} In the following, we show the results for the EDF-F; no relevant differences are found for the other fields.

\Cref{f:pp:gal:ssfr_z} shows the specific star formation rate as a function of redshift for Q1 sources compared to the COSMOS2020 catalogue. As discussed in the previous subsection, we identify a population of galaxies with unrealistically young ages (${\rm age}<10^8\,$yr) and very high sSFR (${\rm sSFR}\ga10^{-8.2}\,$yr$^{-1}$). They constitute about 40\% of the full sample, but the fraction decreases to 20\% at $\IE<23$. This population is particularly evident in the upper parts of \cref{f:pp:gal:ssfr_z}, which show a stripe around $\log_{10}({\rm sSFR/yr}^{-1})=-8$, both for the samples limited to faint and bright magnitudes.

\begin{figure}
  \resizebox{\hsize}{!}{\includegraphics{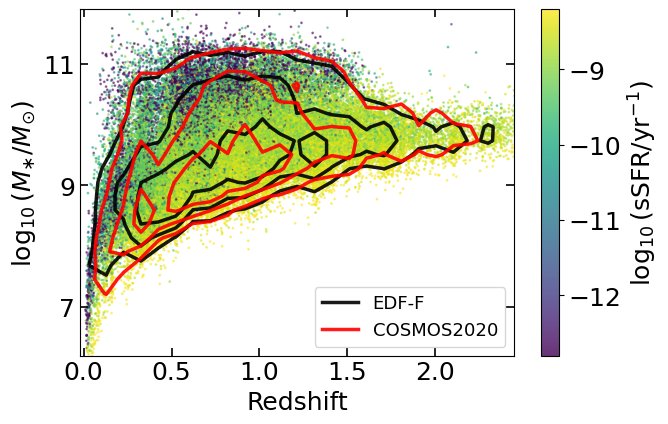}}
  \resizebox{\hsize}{!}{\includegraphics{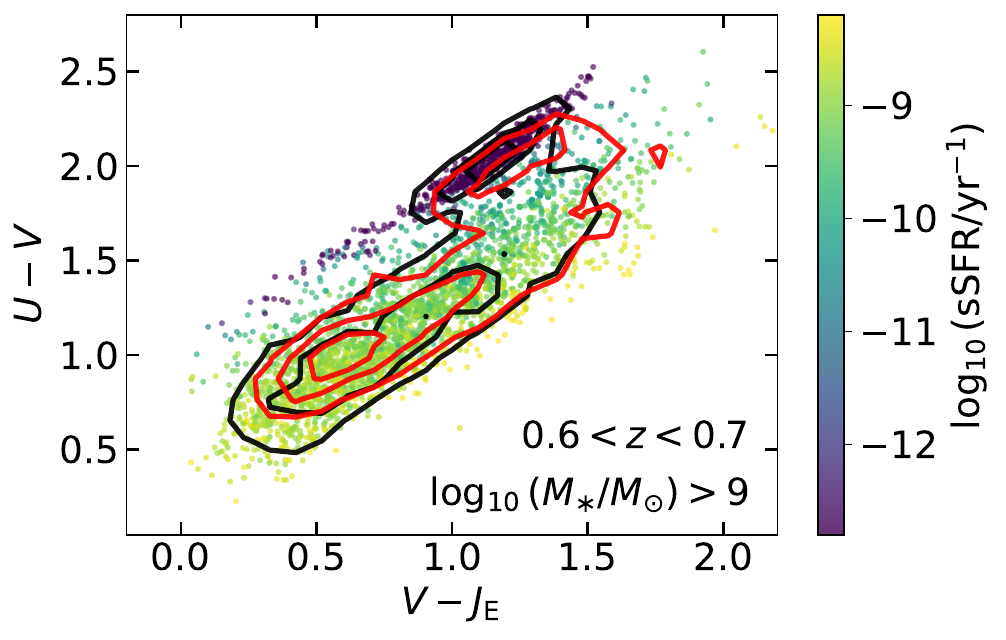}}
  \resizebox{\hsize}{!}{\includegraphics{PP_nuvrj_cut_Q1.pdf}}
  \caption{Scatter plots of the EDF-F galaxies without the problematic population, colour-coded by $\logten{({\rm sSFR}/{\rm yr}^{-1})}$ (for visual reason, only 10\% of the sources are plotted). Contour plots are for the EDF-F galaxies (black lines) and for the COSMOS2020 sample (red lines). {\it (Top Panel)} Stellar mass as a function of redshift; {\it (Central Panel)} $U-V$ versus $V-\JE$ in a limited range of $z$ and $M_*$, as reported in the panel; {\it (Bottom Panel)} NUV$-R$ versus $R-\JE$.}
  \label{f:pp:gal:uvj}
\end{figure}

In \crefrange{f:pp:gal:histo}{f:pp:gal:uvj}, we compare the distributions of physical properties in the two samples after removing the problematic group of galaxies from the \Euclid data set with a cut in sSFR (i.e., ${\rm sSFR}<10^{-8.2}\,$yr$^{-1}$): \cref{f:pp:gal:histo} shows the redshift and the stellar mass distributions; \cref{f:pp:gal:r_z} shows absolute magnitudes in the $R$ band versus redshift; and \cref{f:pp:gal:uvj} (top panel) shows the stellar mass as a function of redshift, colour-coded by sSFR. The two surveys are in good agreement up to high redshift ($z\simeq2.5$) and across the entire range of $R$ magnitudes, with consistent distribution of $z$, stellar mass, and sSFR. \Cref{f:pp:gal:uvj} also compares two rest-frame colour-colour diagrams ($U-V$ versus $V-\JE$ and NUV$-R$ versus $R-\JE$) from the two data sets, which does not show any systematic difference in galaxy colours. The only (small) discrepancy observed in both plots is for quiescent galaxies that in the \Euclid sample tend to appear slightly bluer in $R-\JE$ and $V-\JE$ (and maybe redder in NUV$-R$) compared to those in COSMOS. This is likely due to the difference in stellar population models used in the analysis of the two data sets, which affects the inferred colours. 

To summarise, after removing the problematic objects, the remaining galaxies in both data sets align very closely, suggesting that, under proper conditions, the reconstructed galaxy population in the EWS provide consistent results for the properties of the main galaxy populations that they sample, demonstrating the reliability of the \Euclid data set for studying galaxy properties. 

\begin{figure*}
  \resizebox{\hsize}{!}{\includegraphics{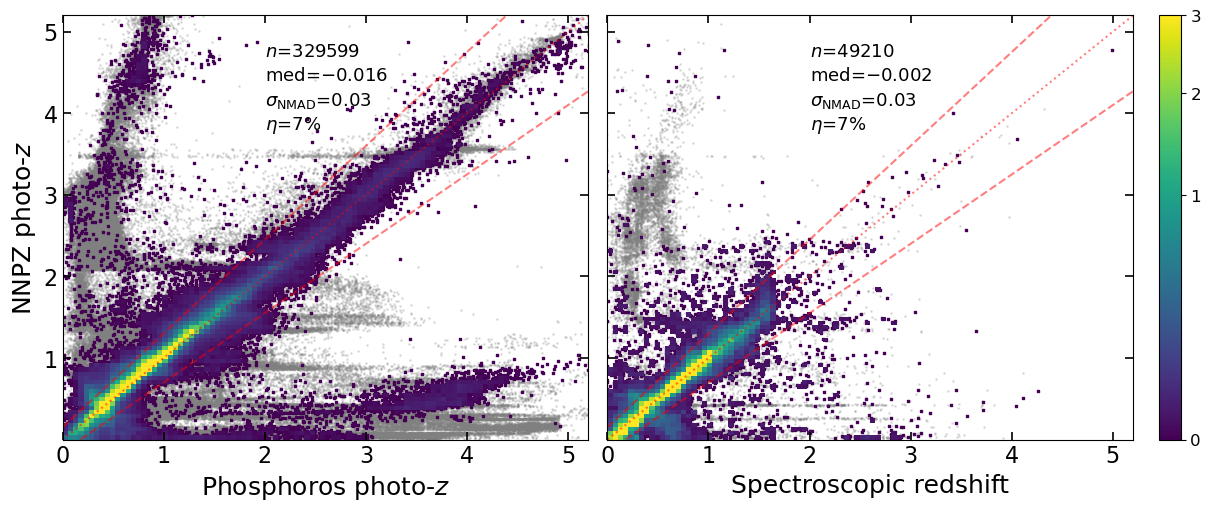}}
  \caption{Density plots comparing photometric redshifts from \NNPZ{} with \Phosphoros{} estimates for galaxies in the EDF-F (left panel) and with spectroscopic redshifts available in public surveys, as in \cref{ssc:phz:valid} (right panel). Grey points correspond to the problematic population.}
  \label{f:pp:gal:phos_nnpz}
\end{figure*}

For the validation of \NNPZ{} products, it is  also important to check the consistency of photometric redshifts obtained by \Phosphoros{} and \NNPZ{}. As shown in \cref{f:pp:gal:phos_nnpz}, if the problematic young galaxy population is excluded, the estimates from the two methods agree typically very well, with some discrepancy only at low redshift. The fraction of `outliers' is around 7\%, which is acceptable, considering the uncertainties present in the \Phosphoros{} estimates. We also compare the \NNPZ{} photometric redshifts with the spectroscopic redshifts of the external surveys: again, after removing the problematic population, we find quality metrics that are similar to those obtained with \Phosphoros, confirming the good performance of the method. 

\subsection{\label{ssc:pp:qso} Determination of QSO parameters}

Objects classified as QSO are treated with the \Phosphoros{} tool, which provides the best-fit parameters for redshift, SED template, luminosity, reddening curve, and intrinsic attenuation. In addition, the $z$PDF is derived by marginalising the posterior distribution in the redshift range [0,\,6] with 0.01 step.

The \Phosphoros{} configuration used for QSOs is the following: the redshift range is from 0 to 6 with 0.05 step; and internal attenuation $E(B-V)$ is considered in the range [0,\,0.4] with 0.1 step, using only the \citet{pre84} reddening curve. The library of templates for the sources classified as QSO is from \citet{sal09} and they have been previously used for the determination of the redshifts of the AGN detected by XMM-{\it Newton} in the COSMOS field \citep{bru10}. The library consists of 30 templates, most of which are hybrids constructed by combining with different ratios templates of QSO and normal galaxies. 

As in \citet{sal09}, a top-hat prior on the absolute magnitude is applied, allowing a range between $-20$ and $-30$. Despite the specific treatment, the quality of the photometric redshifts for the QSO sample is in general not high: only 20\% of QSOs have $\chi^2<10$, to be compared with 46\% and 35\% obtained in the star and galaxy samples, respectively. This is in part related to the problems in the QSO identification discussed in \cref{ssc:class:valid}. Most of the objects in the QSO sample are in fact quite faint, with a classification that is often uncertain. But it is also due to the difficulty in measuring photometric redshifts in these objects, especially with the few filters available in the EWS \citep{sal22}.

\subsection{\label{ssc:pp:star} SED templates for stars}

As for QSOs, stars are handled by the \Phosphoros{} tool using the stellar SED templates and their normalisation as only free parameters, with both redshift and attenuation set to zero. The fluxes are corrected for Galactic extinction assuming that they all lie beyond the Milky Way dust sheet. 

The stellar library is the one that was generated for the \Euclid SGS simulated catalogue \citep{EP-Serrano}. It consists of around 8300 stellar SED templates created from the Basel\,2.2 \citep{las02} template set. These synthetic spectra cover a large range of fundamental stellar parameters: effective temperature ($2000\le T_{\rm eff}\le50\,000$\,K); surface gravity ($-1.02\le\log_{10}(g/{\rm cm\,s^{-2}})\le5.5$); and metallicity ($-5\le [{\rm Fe/H}]\le1$). The library also contains 365 L- and T-type star templates for brown dwarf stars \citep[see ][]{Barnett-EP5} not present in the Basel\,2.2 set.

\begin{figure}
\resizebox{\hsize}{!}{\includegraphics{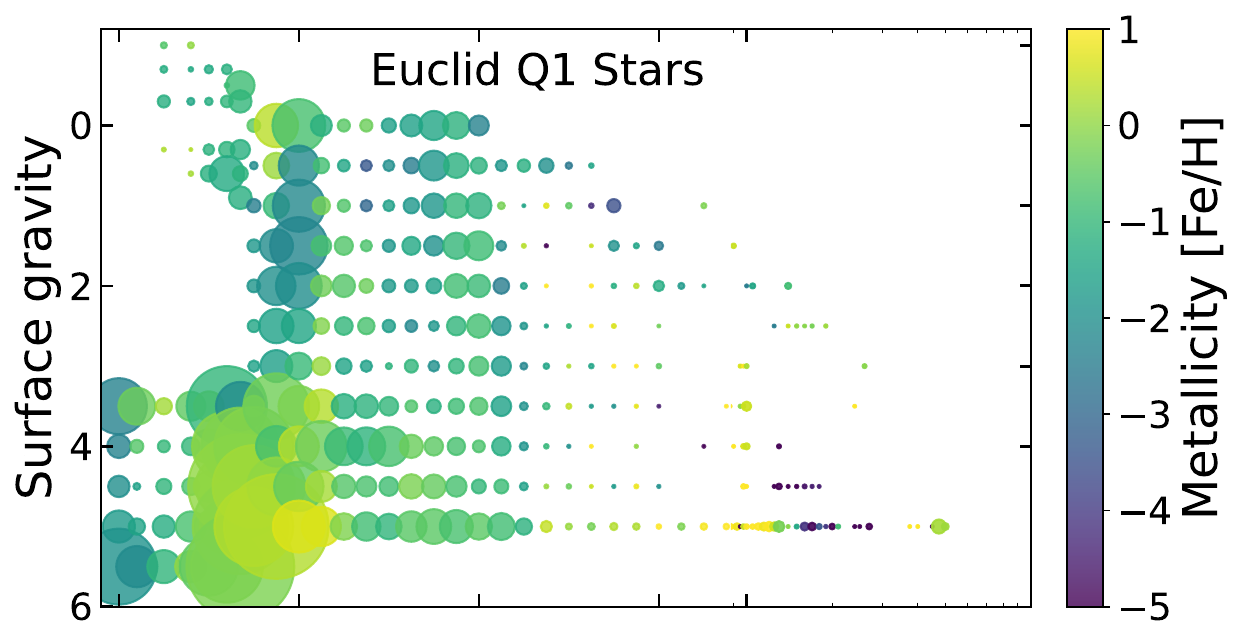}}
\resizebox{\hsize}{!}{\includegraphics{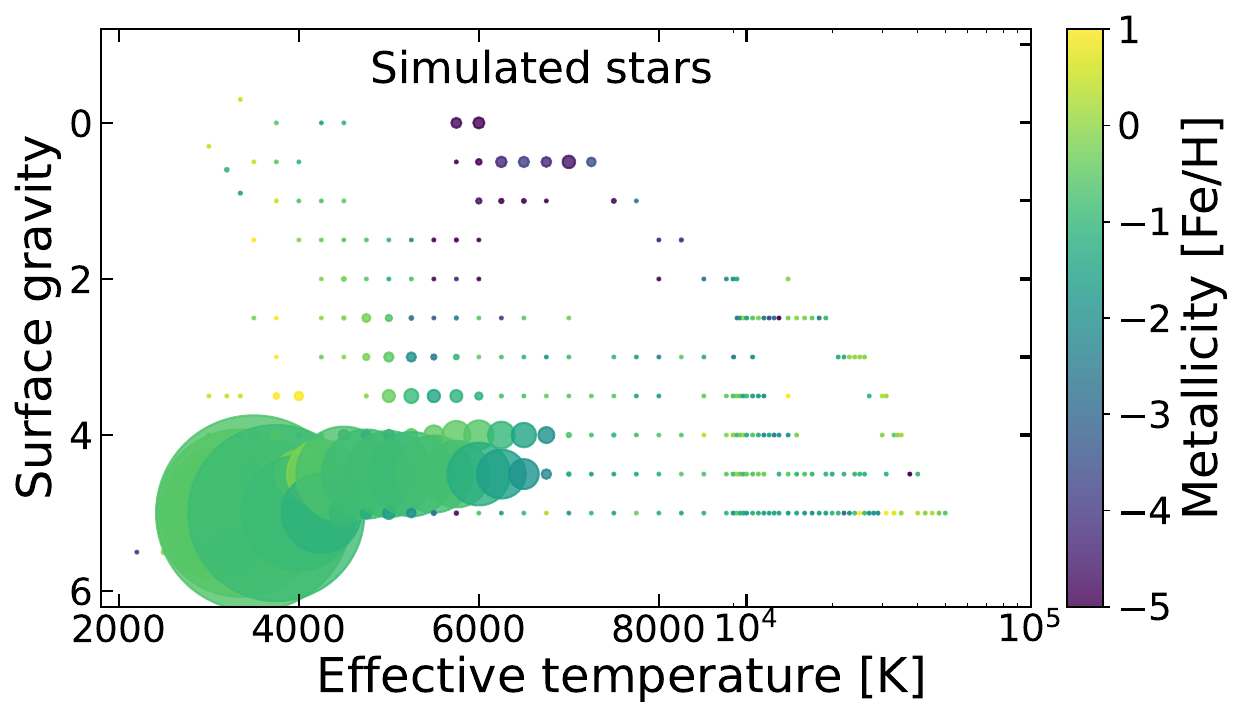}}
\caption{Surface gravity as a function of effective temperature for star-classified objects, as obtained from the \Phosphoros{} best-fit SED (top panel), and for stars from the \Euclid simulated catalogue in the EDFs (bottom panel). Colours of points are related to the mean stellar metallicity, while the size of the points is proportional to the number of objects.}
\label{f:pp_star}
\end{figure}

\Phosphoros{} identifies the star templates that best fit the photometry of star-classified sources, and consequently the associated best-fit physical parameters of stars. In \cref{f:pp_star} we show the distribution of stars in the surface gravity -- effective temperature plane, colour coded by the average metalliticy of stars with the same values of $g$ and $T_{\rm eff}$. This plot can be compared with the one obtained using the \Euclid simulated star catalogue for the EDFs, with the same cut in $\IE$. As expected, most of the stars correspond to main-sequence M and K stars. With respect to simulations, however, we observe much more dispersion in the surface gravity parameter, with many objects at $\log(g)<4$ and $T_{\rm eff}\la6000$\,K not found in the simulated catalogue. Our results therefore overestimate the population of giant and supergiant stars, and a prior should be applied to the $g$ parameter. A small population of white dwarfs and blue stars is also found, in good agreement with simulations.

\subsection{\label{ssc:pp:nir} Physical properties of NIR-only objects}

More than 20\,per\,cent of the sources present in the MER catalogues are detected in the \Euclid NIR stack images but do not pass the detection threshold in VIS. These objects could be either spurious detections or extremely luminous galaxies and quasars at high redshift and brown dwarfs. The physical properties of NIR-only objects are computed by \Phosphoros with the aim to provide useful information for the identification of high-$z$ sources. Along with the usual \Phosphoros parameters, we include in the output products the probability of a source to be at $z>6$ ($p_{z>6}$), computed as the integral of the $z$PDF at redshifts larger than 6.

For NIR-only objects, \Phosphoros is run in a special configuration that combines the above configurations for stars, galaxies, and QSOs, with the only difference that the redshift range is extended up to $z=10$ (from 6 to 10 the step is 0.05). About 6\% of NIR-only sources have $z_{\rm phos}>6$ and high-$z$ probability larger than 0.8. They could be good candidates for high-$z$ sources. Many objects with $z_{\rm phos}>6$ have instead low or intermediate $p_{z>6}$, probably associated with a broad or multiple-peak $z$PDF. We also observe many objects with $z_{\rm phos}=0$ (i.e., best fitted by a star template) and large $p_{z>6}$; they are cases of degeneracy between models of high-$z$ galaxies and brown dwarfs. Indications on the reliability of high-$z$ candidates can also come from the $\chi^2$ associated with the best-fit model. Objects with high $p_{z>6}$ but poorly fitted by the best model (e.g., $\chi^2\gg10$) should be disfavoured as high-$z$ source candidates. The high $\chi^2$ could be in fact an indication of problems in the photometry.

\section{\label{sc:} Conclusions}

This paper describes the PHZ pipeline used for the Q1 release and the related data products. The pipeline is responsible for the determination of photometric redshifts based on the photometry provided in the EWS source catalogues. Other relevant tasks of the PF are the classification of the detected sources and the estimate of their physical properties. The Q1 release allows us to evaluate the performance of the PHZ pipeline directly on \Euclid data. In the paper we have discussed the validation of the pipeline products and their reliability for scientific analyses. In the following we summarize the main results.

\paragraph{--} The classification is performed with the PRF algorithm, on the basis of the source colours. We show that the method is able to provide reliable class probabilities for stars and galaxies, and to extract a pure sample of stars, in agreement with the \Euclid requirements. We also define a galaxy sample with a high level of completeness. The identification of QSOs is instead quite challenging, and the results are not fully reliable yet. For the DR1 release, we aim to improve the training sample, and to review the probability thresholds, especially for the QSO class. In addition, we plan to include morphological information in order to increase the purity of the output samples.

\paragraph{--} Photometric redshifts are computed using the \Phosphoros{} tool and validated with the spectroscopic redshifts from a large set of external surveys. These results show the general goodness of the \Phosphoros{} estimates, with a precision $\sigma_{\rm NMAD}\la0.05$ and an accuracy lower than $0.02(1+z)$, in agreement with the results obtained from the COSMOS2020 catalogue and with previous applications of the tool \citep[e.g.,][]{Desprez-EP10,des23}. Some improvements in the \Phosphoros configuration and in the prior selection will be preformed for DR1 to reduce the fraction of outliers.

\paragraph{--} Physical properties of galaxies, such as luminosity, stellar mass, and star-formation rate, are computed using the \NNPZ{} tool, and validated by comparison with the results from the COSMOS2020 catalogue. The match is good for all the parameters considered if a subset of galaxies spuriously identified as very young with high specific star-formation rate is excluded. We have verified that the application of appropriate priors (i.e., volume and galaxy age priors) is sufficient to obtain reliable physical properties for those objects \citep[see, for more details,][]{Q1-SP031}. These results show the effectiveness of the method in determining galaxy properties. For DR1, the \NNPZ reference sample will be refined, with a careful selection of the galaxy populations, in order to reinforces the reliability of the PHZ products.

\paragraph{}

\begin{acknowledgements}
\AckQone \,
\AckEC  \,
Based on data from UNIONS, a scientific collaboration using three Hawaii-based telescopes: CFHT, Pan-STARRS, and Subaru \url{www.skysurvey.cc}\,.
Based on data from the Dark Energy Camera (DECam) on the Blanco 4-m Telescope at CTIO in Chile \url{https://www.darkenergysurvey.org}\,.
This work uses results from the ESA mission {\it Gaia}, whose data are being processed by the Gaia Data Processing and Analysis Consortium \url{https://www.cosmos.esa.int/gaia}\,.
\end{acknowledgements}

%
%

\bibliography{mybib, Euclid, Q1}

\begin{thebibliography}{109}
\expandafter\ifx\csname natexlab\endcsname\relax\def\natexlab#1{#1}\fi

\bibitem[{{Abbott} {et~al.}(2021){Abbott}, {Adam{\'o}w}, {Aguena}, {Allam},
  {Amon}, {Annis}, {Avila}, {Bacon}, {Banerji}, {Bechtol}, {Becker},
  {Bernstein}, {Bertin}, {Bhargava}, {Bridle}, {Brooks}, {Burke}, {Carnero
  Rosell}, {Carrasco Kind}, {Carretero}, {Castander}, {Cawthon}, {Chang},
  {Choi}, {Conselice}, {Costanzi}, {Crocce}, {da Costa}, {Davis}, {De Vicente},
  {DeRose}, {Desai}, {Diehl}, {Dietrich}, {Drlica-Wagner}, {Eckert},
  {Elvin-Poole}, {Everett}, {Evrard}, {Ferrero}, {Fert{\'e}}, {Flaugher},
  {Fosalba}, {Friedel}, {Frieman}, {Garc{\'\i}a-Bellido}, {Gaztanaga},
  {Gelman}, {Gerdes}, {Giannantonio}, {Gill}, {Gruen}, {Gruendl}, {Gschwend},
  {Gutierrez}, {Hartley}, {Hinton}, {Hollowood}, {Honscheid}, {Huterer},
  {James}, {Jeltema}, {Johnson}, {Kent}, {Kron}, {Kuehn}, {Kuropatkin},
  {Lahav}, {Li}, {Lidman}, {Lin}, {MacCrann}, {Maia}, {Manning}, {Maloney},
  {March}, {Marshall}, {Martini}, {Melchior}, {Menanteau}, {Miquel}, {Morgan},
  {Myles}, {Neilsen}, {Ogando}, {Palmese}, {Paz-Chinch{\'o}n}, {Petravick},
  {Pieres}, {Plazas}, {Pond}, {Rodriguez-Monroy}, {Romer}, {Roodman}, {Rykoff},
  {Sako}, {Sanchez}, {Santiago}, {Scarpine}, {Serrano}, {Sevilla-Noarbe},
  {Smith}, {Smith}, {Soares-Santos}, {Suchyta}, {Swanson}, {Tarle}, {Thomas},
  {To}, {Tremblay}, {Troxel}, {Tucker}, {Turner}, {Varga}, {Walker},
  {Wechsler}, {Weller}, {Wester}, {Wilkinson}, {Yanny}, {Zhang}, {Nikutta},
  {Fitzpatrick}, {Jacques}, {Scott}, {Olsen}, {Huang}, {Herrera}, {Juneau},
  {Nidever}, {Weaver}, {Adean}, {Correia}, {de Freitas}, {Freitas},
  {Singulani}, {Vila-Verde}, \& {Linea Science Server}}]{abb21}
{Abbott}, T.~M.~C., {Adam{\'o}w}, M., {Aguena}, M., {et~al.} 2021, \apjs, 255,
  20

\bibitem[{Ade {et~al.}(2016)Ade, Aghanim, Arnaud, Ashdown, Aumont, Baccigalupi,
  Banday, Barreiro, Bartlett, Bartolo, Battaner, Battye, Benabed, Benoît,
  Benoit-Lévy, Bernard, Bersanelli, Bielewicz, Bock, Bonaldi, Bonavera, Bond,
  Borrill, Bouchet, Boulanger, Bucher, Burigana, Butler, Calabrese, Cardoso,
  Catalano, Challinor, Chamballu, Chary, Chiang, Chluba, Christensen, Church,
  Clements, Colombi, Colombo, Combet, Coulais, Crill, Curto, Cuttaia, Danese,
  Davies, Davis, de~Bernardis, de~Rosa, de~Zotti, Delabrouille, Désert,
  Di~Valentino, Dickinson, Diego, Dolag, Dole, Donzelli, Doré, Douspis,
  Ducout, Dunkley, Dupac, Efstathiou, Elsner, Enßlin, Eriksen, Farhang,
  Fergusson, Finelli, Forni, Frailis, Fraisse, Franceschi, Frejsel, Galeotta,
  Galli, Ganga, Gauthier, Gerbino, Ghosh, Giard, Giraud-Héraud, Giusarma,
  Gjerløw, González-Nuevo, Górski, Gratton, Gregorio, Gruppuso, Gudmundsson,
  Hamann, Hansen, Hanson, Harrison, Helou, Henrot-Versillé,
  Hernández-Monteagudo, Herranz, Hildebrandt, Hivon, Hobson, Holmes,
  Hornstrup, Hovest, Huang, Huffenberger, Hurier, Jaffe, Jaffe, Jones, Juvela,
  Keihänen, Keskitalo, Kisner, Kneissl, Knoche, Knox, Kunz, Kurki-Suonio,
  Lagache, Lähteenmäki, Lamarre, Lasenby, Lattanzi, Lawrence, Leahy,
  Leonardi, Lesgourgues, Levrier, Lewis, Liguori, Lilje, Linden-Vørnle,
  López-Caniego, Lubin, Macías-Pérez, Maggio, Maino, Mandolesi, Mangilli,
  Marchini, Maris, Martin, Martinelli, Martínez-González, Masi, Matarrese,
  McGehee, Meinhold, Melchiorri, Melin, Mendes, Mennella, Migliaccio, Millea,
  Mitra, Miville-Deschênes, Moneti, Montier, Morgante, Mortlock, Moss, Munshi,
  Murphy, Naselsky, Nati, Natoli, Netterfield, Nørgaard-Nielsen, Noviello,
  Novikov, Novikov, Oxborrow, Paci, Pagano, Pajot, Paladini, Paoletti,
  Partridge, Pasian, Patanchon, Pearson, Perdereau, Perotto, Perrotta,
  Pettorino, Piacentini, Piat, Pierpaoli, Pietrobon, Plaszczynski,
  Pointecouteau, Polenta, Popa, Pratt, Prézeau, Prunet, Puget, Rachen, Reach,
  Rebolo, Reinecke, Remazeilles, Renault, Renzi, Ristorcelli, Rocha, Rosset,
  Rossetti, Roudier, Rouillé~d’Orfeuil, Rowan-Robinson, Rubiño-Martín,
  Rusholme, Said, Salvatelli, Salvati, Sandri, Santos, Savelainen, Savini,
  Scott, Seiffert, Serra, Shellard, Spencer, Spinelli, Stolyarov, Stompor,
  Sudiwala, Sunyaev, Sutton, Suur-Uski, Sygnet, Tauber, Terenzi, Toffolatti,
  Tomasi, Tristram, Trombetti, Tucci, Tuovinen, Türler, Umana, Valenziano,
  Valiviita, Van~Tent, Vielva, Villa, Wade, Wandelt, Wehus, White, White,
  Wilkinson, Yvon, Zacchei, \& Zonca}]{planck15}
Ade, P. A.~R., Aghanim, N., Arnaud, M., {et~al.} 2016, \aap, 594, A13

\bibitem[{{Ahumada} {et~al.}(2020){Ahumada}, {Allende Prieto}, {Almeida},
  {Anders}, {Anderson}, {Andrews}, {Anguiano}, {Arcodia}, {Armengaud},
  {Aubert}, {Avila}, {Avila-Reese}, {Badenes}, {Balland}, {Barger},
  {Barrera-Ballesteros}, {Basu}, {Bautista}, {Beaton}, {Beers}, {Benavides},
  {Bender}, {Bernardi}, {Bershady}, {Beutler}, {Bidin}, {Bird}, {Bizyaev},
  {Blanc}, {Blanton}, {Boquien}, {Borissova}, {Bovy}, {Brandt}, {Brinkmann},
  {Brownstein}, {Bundy}, {Bureau}, {Burgasser}, {Burtin}, {Cano-D{\'\i}az},
  {Capasso}, {Cappellari}, {Carrera}, {Chabanier}, {Chaplin}, {Chapman},
  {Cherinka}, {Chiappini}, {Doohyun Choi}, {Chojnowski}, {Chung}, {Clerc},
  {Coffey}, {Comerford}, {Comparat}, {da Costa}, {Cousinou}, {Covey}, {Crane},
  {Cunha}, {Ilha}, {Dai}, {Damsted}, {Darling}, {Davidson}, {Davies}, {Dawson},
  {De}, {de la Macorra}, {De Lee}, {Queiroz}, {Deconto Machado}, {de la Torre},
  {Dell'Agli}, {du Mas des Bourboux}, {Diamond-Stanic}, {Dillon}, {Donor},
  {Drory}, {Duckworth}, {Dwelly}, {Ebelke}, {Eftekharzadeh}, {Davis Eigenbrot},
  {Elsworth}, {Eracleous}, {Erfanianfar}, {Escoffier}, {Fan}, {Farr},
  {Fern{\'a}ndez-Trincado}, {Feuillet}, {Finoguenov}, {Fofie},
  {Fraser-McKelvie}, {Frinchaboy}, {Fromenteau}, {Fu}, {Galbany}, {Garcia},
  {Garc{\'\i}a-Hern{\'a}ndez}, {Garma Oehmichen}, {Ge}, {Geimba Maia},
  {Geisler}, {Gelfand}, {Goddy}, {Gonzalez-Perez}, {Grabowski}, {Green},
  {Grier}, {Guo}, {Guy}, {Harding}, {Hasselquist}, {Hawken}, {Hayes}, {Hearty},
  {Hekker}, {Hogg}, {Holtzman}, {Horta}, {Hou}, {Hsieh}, {Huber}, {Hunt}, {Ider
  Chitham}, {Imig}, {Jaber}, {Jimenez Angel}, {Johnson}, {Jones},
  {J{\"o}nsson}, {Jullo}, {Kim}, {Kinemuchi}, {Kirkpatrick}, {Kite}, {Klaene},
  {Kneib}, {Kollmeier}, {Kong}, {Kounkel}, {Krishnarao}, {Lacerna}, {Lan},
  {Lane}, {Law}, {Le Goff}, {Leung}, {Lewis}, {Li}, {Lian}, {Lin}, {Long},
  {Longa-Pe{\~n}a}, {Lundgren}, {Lyke}, {Mackereth}, {MacLeod}, {Majewski},
  {Manchado}, {Maraston}, {Martini}, {Masseron}, {Masters}, {Mathur},
  {McDermid}, {Merloni}, {Merrifield}, {M{\'e}sz{\'a}ros}, {Miglio}, {Minniti},
  {Minsley}, {Miyaji}, {Mohammad}, {Mosser}, {Mueller}, {Muna},
  {Mu{\~n}oz-Guti{\'e}rrez}, {Myers}, {Nadathur}, {Nair}, {Nandra}, {Correa do
  Nascimento}, {Nevin}, {Newman}, {Nidever}, {Nitschelm}, {Noterdaeme},
  {O'Connell}, {Olmstead}, {Oravetz}, {Oravetz}, {Osorio}, {Pace}, {Padilla},
  {Palanque-Delabrouille}, \& {Palicio}}]{ahu20}
{Ahumada}, R., {Allende Prieto}, C., {Almeida}, A., {et~al.} 2020, \apjs, 249,
  3

\bibitem[{{Amara} \& {R{\'e}fr{\'e}gier}(2007)}]{ama07}
{Amara}, A. \& {R{\'e}fr{\'e}gier}, A. 2007, \mnras, 381, 1018

\bibitem[{{Arnouts} {et~al.}(2002){Arnouts}, {Moscardini}, {Vanzella},
  {Colombi}, {Cristiani}, {Fontana}, {Giallongo}, {Matarrese}, \&
  {Saracco}}]{arn02}
{Arnouts}, S., {Moscardini}, L., {Vanzella}, E., {et~al.} 2002, \mnras, 329,
  355

\bibitem[{{Ben{\'\i}tez}(2000)}]{ben00}
{Ben{\'\i}tez}, N. 2000, \apj, 536, 571

\bibitem[{{Blake} {et~al.}(2016){Blake}, {Amon}, {Childress}, {Erben},
  {Glazebrook}, {Harnois-Deraps}, {Heymans}, {Hildebrandt}, {Hinton},
  {Janssens}, {Johnson}, {Joudaki}, {Klaes}, {Kuijken}, {Lidman}, {Marin},
  {Parkinson}, {Poole}, \& {Wolf}}]{bla2016}
{Blake}, C., {Amon}, A., {Childress}, M., {et~al.} 2016, \mnras, 462, 4240

\bibitem[{{Bolzonella} {et~al.}(2000){Bolzonella}, {Miralles}, \&
  {Pell{\'o}}}]{bol00}
{Bolzonella}, M., {Miralles}, J.~M., \& {Pell{\'o}}, R. 2000, \aap, 363, 476

\bibitem[{{Brammer} {et~al.}(2008){Brammer}, {van Dokkum}, \& {Coppi}}]{bra08}
{Brammer}, G.~B., {van Dokkum}, P.~G., \& {Coppi}, P. 2008, \apj, 686, 1503

\bibitem[{{Breiman}(2001)}]{bre01}
{Breiman}, L. 2001, Machine Learning, 45, 5

\bibitem[{{Brusa} {et~al.}(2010){Brusa}, {Civano}, {Comastri}, {Miyaji},
  {Salvato}, {Zamorani}, {Cappelluti}, {Fiore}, {Hasinger}, {Mainieri},
  {Merloni}, {Bongiorno}, {Capak}, {Elvis}, {Gilli}, {Hao}, {Jahnke},
  {Koekemoer}, {Ilbert}, {Le Floc'h}, {Lusso}, {Mignoli}, {Schinnerer},
  {Silverman}, {Treister}, {Trump}, {Vignali}, {Zamojski}, {Aldcroft},
  {Aussel}, {Bardelli}, {Bolzonella}, {Cappi}, {Caputi}, {Contini},
  {Finoguenov}, {Fruscione}, {Garilli}, {Impey}, {Iovino}, {Iwasawa},
  {Kampczyk}, {Kartaltepe}, {Kneib}, {Knobel}, {Kovac}, {Lamareille},
  {Leborgne}, {Le Brun}, {Le Fevre}, {Lilly}, {Maier}, {McCracken}, {Pello},
  {Peng}, {Perez-Montero}, {de Ravel}, {Sanders}, {Scodeggio}, {Scoville},
  {Tanaka}, {Taniguchi}, {Tasca}, {de la Torre}, {Tresse}, {Vergani}, \&
  {Zucca}}]{bru10}
{Brusa}, M., {Civano}, F., {Comastri}, A., {et~al.} 2010, \apj, 716, 348

\bibitem[{{Bruzual} \& {Charlot}(2003)}]{bru03}
{Bruzual}, G. \& {Charlot}, S. 2003, \mnras, 344, 1000

\bibitem[{{Calzetti} {et~al.}(2000){Calzetti}, {Armus}, {Bohlin}, {Kinney},
  {Koornneef}, \& {Storchi-Bergmann}}]{cal00}
{Calzetti}, D., {Armus}, L., {Bohlin}, R.~C., {et~al.} 2000, \apj, 533, 682

\bibitem[{{Carnall} {et~al.}(2019){Carnall}, {McLure}, {Dunlop}, {Cullen},
  {McLeod}, {Wild}, {Johnson}, {Appleby}, {Dav{\'e}}, {Amorin}, {Bolzonella},
  {Castellano}, {Cimatti}, {Cucciati}, {Gargiulo}, {Garilli}, {Marchi},
  {Pentericci}, {Pozzetti}, {Schreiber}, {Talia}, \& {Zamorani}}]{car19}
{Carnall}, A.~C., {McLure}, R.~J., {Dunlop}, J.~S., {et~al.} 2019, \mnras, 490,
  417

\bibitem[{{Carnall} {et~al.}(2018){Carnall}, {McLure}, {Dunlop}, \&
  {Dav{\'e}}}]{car18}
{Carnall}, A.~C., {McLure}, R.~J., {Dunlop}, J.~S., \& {Dav{\'e}}, R. 2018,
  \mnras, 480, 4379

\bibitem[{{Chambers} {et~al.}(2016){Chambers}, {Magnier}, {Metcalfe},
  {Flewelling}, {Huber}, {Waters}, {Denneau}, {Draper}, {Farrow}, {Finkbeiner},
  {Holmberg}, {Koppenhoefer}, {Price}, {Rest}, {Saglia}, {Schlafly}, {Smartt},
  {Sweeney}, {Wainscoat}, {Burgett}, {Chastel}, {Grav}, {Heasley}, {Hodapp},
  {Jedicke}, {Kaiser}, {Kudritzki}, {Luppino}, {Lupton}, {Monet}, {Morgan},
  {Onaka}, {Shiao}, {Stubbs}, {Tonry}, {White}, {Ba{\~n}ados}, {Bell},
  {Bender}, {Bernard}, {Boegner}, {Boffi}, {Botticella}, {Calamida},
  {Casertano}, {Chen}, {Chen}, {Cole}, {Deacon}, {Frenk}, {Fitzsimmons},
  {Gezari}, {Gibbs}, {Goessl}, {Goggia}, {Gourgue}, {Goldman}, {Grant},
  {Grebel}, {Hambly}, {Hasinger}, {Heavens}, {Heckman}, {Henderson}, {Henning},
  {Holman}, {Hopp}, {Ip}, {Isani}, {Jackson}, {Keyes}, {Koekemoer}, {Kotak},
  {Le}, {Liska}, {Long}, {Lucey}, {Liu}, {Martin}, {Masci}, {McLean}, {Mindel},
  {Misra}, {Morganson}, {Murphy}, {Obaika}, {Narayan}, {Nieto-Santisteban},
  {Norberg}, {Peacock}, {Pier}, {Postman}, {Primak}, {Rae}, {Rai}, {Riess},
  {Riffeser}, {Rix}, {R{\"o}ser}, {Russel}, {Rutz}, {Schilbach}, {Schultz},
  {Scolnic}, {Strolger}, {Szalay}, {Seitz}, {Small}, {Smith}, {Soderblom},
  {Taylor}, {Thomson}, {Taylor}, {Thakar}, {Thiel}, {Thilker}, {Unger},
  {Urata}, {Valenti}, {Wagner}, {Walder}, {Walter}, {Watters}, {Werner},
  {Wood-Vasey}, \& {Wyse}}]{cha16}
{Chambers}, K.~C., {Magnier}, E.~A., {Metcalfe}, N., {et~al.} 2016,
  arXiv:1612.05560

\bibitem[{{Chen} {et~al.}(2018){Chen}, {Brandt}, {Luo}, {Ranalli}, {Yang},
  {Alexander}, {Bauer}, {Kelson}, {Lacy}, {Nyland}, {Tozzi}, {Vito},
  {Cirasuolo}, {Gilli}, {Jarvis}, {Lehmer}, {Paolillo}, {Schneider}, {Shemmer},
  {Smail}, {Sun}, {Tanaka}, {Vaccari}, {Vignali}, {Xue}, {Banerji}, {Chow},
  {H{\"a}u{\ss}ler}, {Norris}, {Silverman}, \& {Trump}}]{che18}
{Chen}, C. T.~J., {Brandt}, W.~N., {Luo}, B., {et~al.} 2018, \mnras, 478, 2132

\bibitem[{{Coil} {et~al.}(2011){Coil}, {Blanton}, {Burles}, {Cool},
  {Eisenstein}, {Moustakas}, {Wong}, {Zhu}, {Aird}, {Bernstein}, {Bolton}, \&
  {Hogg}}]{coi2011}
{Coil}, A.~L., {Blanton}, M.~R., {Burles}, S.~M., {et~al.} 2011, \apj, 741, 8

\bibitem[{{Colless} {et~al.}(2003){Colless}, {Peterson}, {Jackson}, {Peacock},
  {Cole}, {Norberg}, {Baldry}, {Baugh}, {Bland-Hawthorn}, {Bridges}, {Cannon},
  {Collins}, {Couch}, {Cross}, {Dalton}, {De Propris}, {Driver}, {Efstathiou},
  {Ellis}, {Frenk}, {Glazebrook}, {Lahav}, {Lewis}, {Lumsden}, {Maddox},
  {Madgwick}, {Sutherland}, \& {Taylor}}]{col03}
{Colless}, M., {Peterson}, B.~A., {Jackson}, C., {et~al.} 2003,
  arXiv:astro-ph/0306581

\bibitem[{{Conroy} \& {Gunn}(2010)}]{con10}
{Conroy}, C. \& {Gunn}, J.~E. 2010, \apj, 712, 833

\bibitem[{{Cool} {et~al.}(2013){Cool}, {Moustakas}, {Blanton}, {Burles},
  {Coil}, {Eisenstein}, {Wong}, {Zhu}, {Aird}, {Bernstein}, {Bolton}, {Hogg},
  \& {Mendez}}]{coi2013}
{Cool}, R.~J., {Moustakas}, J., {Blanton}, M.~R., {et~al.} 2013, \apj, 767, 118

\bibitem[{Cover \& Hart(1967)}]{cov67}
Cover, T. \& Hart, P. 1967, IEEE Transactions on Information Theory, 13, 21

\bibitem[{{Cropper} {et~al.}(2013){Cropper}, {Hoekstra}, {Kitching}, {Massey},
  {Amiaux}, {Miller}, {Mellier}, {Rhodes}, {Rowe}, {Pires}, {Saxton}, \&
  {Scaramella}}]{cro13}
{Cropper}, M., {Hoekstra}, H., {Kitching}, T., {et~al.} 2013, \mnras, 431, 3103

\bibitem[{Dawid(1984)}]{daw84}
Dawid, A.~P. 1984, J. R. Stat. Soc., A (General), 147, 278

\bibitem[{{DESI Collaboration} {et~al.}(2024){DESI Collaboration}, {Adame},
  {Aguilar}, {Ahlen}, {Alam}, {Aldering}, {Alexander}, {Alfarsy}, {Allende
  Prieto}, {Alvarez}, {Alves}, {Anand}, {Andrade-Oliveira}, {Armengaud},
  {Asorey}, {Avila}, {Aviles}, {Bailey}, {Balaguera-Antol{\'\i}nez},
  {Ballester}, {Baltay}, {Bault}, {Bautista}, {Behera}, {Beltran}, {BenZvi},
  {Beraldo e Silva}, {Bermejo-Climent}, {Berti}, {Besuner}, {Beutler},
  {Bianchi}, {Blake}, {Blum}, {Bolton}, {Brieden}, {Brodzeller}, {Brooks},
  {Brown}, {Buckley-Geer}, {Burtin}, {Cabayol-Garcia}, {Cai}, {Canning},
  {Cardiel-Sas}, {Carnero Rosell}, {Castander}, {Cervantes-Cota}, {Chabanier},
  {Chaussidon}, {Chaves-Montero}, {Chen}, {Chen}, {Chuang}, {Claybaugh},
  {Cole}, {Cooper}, {Cuceu}, {Davis}, {Dawson}, {de Belsunce}, {de la Cruz},
  {de la Macorra}, {Della Costa}, {de Mattia}, {Demina}, {Demirbozan},
  {DeRose}, {Dey}, {Dey}, {Dhungana}, {Ding}, {Ding}, {Doel}, {Doshi},
  {Douglass}, {Edge}, {Eftekharzadeh}, {Eisenstein}, {Elliott}, {Ereza},
  {Escoffier}, {Fagrelius}, {Fan}, {Fanning}, {Fawcett}, {Ferraro}, {Flaugher},
  {Font-Ribera}, {Forero-Romero}, {Forero-S{\'a}nchez}, {Frenk},
  {G{\"a}nsicke}, {Garc{\'\i}a}, {Garc{\'\i}a-Bellido}, {Garcia-Quintero},
  {Garrison}, {Gil-Mar{\'\i}n}, {Golden-Marx}, {Gontcho A Gontcho},
  {Gonzalez-Morales}, {Gonzalez-Perez}, {Gordon}, {Graur}, {Green}, {Gruen},
  {Guy}, {Hadzhiyska}, {Hahn}, {Han}, {Hanif}, {Herrera-Alcantar}, {Honscheid},
  {Hou}, {Howlett}, {Huterer}, {Ir{\v{s}}i{\v{c}}}, {Ishak}, {Jacques}, {Jana},
  {Jiang}, {Jimenez}, {Jing}, {Joudaki}, {Joyce}, {Jullo}, {Juneau},
  {Kara{\c{c}}ayl{\i}}, {Karim}, {Kehoe}, {Kent}, {Khederlarian}, {Kim},
  {Kirkby}, {Kisner}, {Kitaura}, {Kizhuprakkat}, {Kneib}, {Koposov},
  {Kov{\'a}cs}, {Kremin}, {Krolewski}, {L'Huillier}, {Lahav}, {Lambert},
  {Lamman}, {Lan}, {Landriau}, {Lang}, {Lange}, {Lasker}, {Leauthaud}, {Le
  Guillou}, {Levi}, {Li}, {Linder}, {Lyons}, {Magneville}, {Manera}, {Manser},
  {Margala}, {Martini}, {McDonald}, {Medina}, {Medina-Varela}, {Meisner},
  {Mena-Fern{\'a}ndez}, {Meneses-Rizo}, {Mezcua}, {Miquel}, {Montero-Camacho},
  {Moon}, {Moore}, {Moustakas}, {Mueller}, {Mundet}, {Mu{\~n}oz-Guti{\'e}rrez},
  {Myers}, {Nadathur}, {Napolitano}, {Neveux}, {Newman}, {Nie}, {Nikutta},
  {Niz}, {Norberg}, {Noriega}, {Paillas}, {Palanque-Delabrouille}, {Palmese},
  {Pan}, {Parkinson}, {Penmetsa}, {Percival}, {P{\'e}rez-Fern{\'a}ndez},
  {P{\'e}rez-R{\`a}fols}, {Pieri}, {Poppett}, {Porredon}, {Pothier}, {Prada},
  {Pucha}, {Raichoor}, {Ram{\'\i}rez-P{\'e}rez}, {Ramirez-Solano},
  {Rashkovetskyi}, {Ravoux}, {Rocher}, {Rockosi}, {Ross}, {Rossi}, {Ruggeri},
  {Ruhlmann-Kleider}, {Sabiu}, {Said}, {Saintonge}, {Samushia}, {Sanchez},
  {Saulder}, {Schaan}, {Schlafly}, {Schlegel}, {Scholte}, {Schubnell}, {Seo},
  {Shafieloo}, {Sharples}, {Sheu}, {Silber}, {Sinigaglia}, {Siudek}, {Slepian},
  {Smith}, {Soumagnac}, {Sprayberry}, {Stephey}, {Su{\'a}rez-P{\'e}rez}, {Sun},
  {Tan}, {Tarl{\'e}}, {Tojeiro}, {Ure{\~n}a-L{\'o}pez}, {Vaisakh}, {Valcin},
  {Valdes}, {Valluri}, {Vargas-Maga{\~n}a}, {Variu}, {Verde}, {Walther},
  {Wang}, {Wang}, {Weaver}, {Weaverdyck}, {Wechsler}, {White}, {Xie}, {Yang},
  {Y{\`e}che}, {Yu}, {Yuan}, {Zhang}, {Zhang}, {Zhao}, {Zheng}, {Zhou}, {Zhou},
  {Zou}, {Zou}, \& {Zu}}]{desi24}
{DESI Collaboration}, {Adame}, A.~G., {Aguilar}, J., {et~al.} 2024, \aj, 168,
  58

\bibitem[{{Desprez} {et~al.}(2023){Desprez}, {Picouet}, {Moutard}, {Arnouts},
  {Sawicki}, {Coupon}, {Gwyn}, {Chen}, {Huang}, {Golob}, {Furusawa}, {Ikeda},
  {Paltani}, {Cheng}, {Hartley}, {Hsieh}, {Ilbert}, {Kauffmann}, {McCracken},
  {Shuntov}, {Tanaka}, {Toft}, {Tresse}, \& {Weaver}}]{des23}
{Desprez}, G., {Picouet}, V., {Moutard}, T., {et~al.} 2023, \aap, 670, A82

\bibitem[{{D'Eugenio} {et~al.}(2024){D'Eugenio}, {Cameron}, {Scholtz},
  {Carniani}, {Willott}, {Curtis-Lake}, {Bunker}, {Parlanti}, {Maiolino},
  {Willmer}, {Jakobsen}, {Robertson}, {Johnson}, {Tacchella}, {Cargile},
  {Rawle}, {Arribas}, {Chevallard}, {Curti}, {Egami}, {Eisenstein}, {Kumari},
  {Looser}, {Rieke}, {Rodr{\'\i}guez Del Pino}, {Saxena}, {{\"U}bler},
  {Venturi}, {Witstok}, {Baker}, {Bhatawdekar}, {Bonaventura}, {Boyett},
  {Charlot}, {Danhaive}, {Hainline}, {Hausen}, {Helton}, {Ji}, {Ji}, {Jones},
  {Joud{\v{z}}balis}, {Maseda}, {P{\'e}rez-Gonz{\'a}lez}, {Perna},
  {Pusk{\'a}s}, {Shivaei}, {Silcock}, {Simmonds}, {Smit}, {Sun}, {Villanueva},
  {Williams}, \& {Zhu}}]{deu2024}
{D'Eugenio}, F., {Cameron}, A.~J., {Scholtz}, J., {et~al.} 2024,
  arXiv:2404.06531

\bibitem[{{D'Isanto} \& {Polsterer}(2018)}]{dis18}
{D'Isanto}, A. \& {Polsterer}, K.~L. 2018, \aap, 609, A111

\bibitem[{{Eisenstein} {et~al.}(2023){Eisenstein}, {Willott}, {Alberts},
  {Arribas}, {Bonaventura}, {Bunker}, {Cameron}, {Carniani}, {Charlot},
  {Curtis-Lake}, {D'Eugenio}, {Endsley}, {Ferruit}, {Giardino}, {Hainline},
  {Hausen}, {Jakobsen}, {Johnson}, {Maiolino}, {Rieke}, {Rieke}, {Rix},
  {Robertson}, {Stark}, {Tacchella}, {Williams}, {Willmer}, {Baker}, {Baum},
  {Bhatawdekar}, {Boyett}, {Chen}, {Chevallard}, {Circosta}, {Curti},
  {Danhaive}, {DeCoursey}, {de Graaff}, {Dressler}, {Egami}, {Helton},
  {Hviding}, {Ji}, {Jones}, {Kumari}, {L{\"u}tzgendorf}, {Laseter}, {Looser},
  {Lyu}, {Maseda}, {Nelson}, {Parlanti}, {Perna}, {Pusk{\'a}s}, {Rawle},
  {Rodr{\'\i}guez Del Pino}, {Sandles}, {Saxena}, {Scholtz}, {Sharpe},
  {Shivaei}, {Silcock}, {Simmonds}, {Skarbinski}, {Smit}, {Stone}, {Suess},
  {Sun}, {Tang}, {Topping}, {{\"U}bler}, {Villanueva}, {Wallace}, {Whitler},
  {Witstok}, \& {Woodrum}}]{eis2023}
{Eisenstein}, D.~J., {Willott}, C., {Alberts}, S., {et~al.} 2023,
  arXiv:2306.02465

\bibitem[{{Euclid Collaboration: Aussel} {et~al.}(2025){Euclid Collaboration:
  Aussel}, {Tereno}, {Schirmer}, {et~al.}}]{Q1-TP001}
{Euclid Collaboration: Aussel}, H., {Tereno}, I., {Schirmer}, M., {et~al.}
  2025, \aap, submitted

\bibitem[{{Euclid Collaboration: Barnett} {et~al.}(2019){Euclid Collaboration:
  Barnett}, {Warren}, {Mortlock}, {et~al.}}]{Barnett-EP5}
{Euclid Collaboration: Barnett}, R., {Warren}, S.~J., {Mortlock}, D.~J.,
  {et~al.} 2019, \aap, 631, A85

\bibitem[{{Euclid Collaboration: Bisigello} {et~al.}(2024){Euclid
  Collaboration: Bisigello}, {Massimo}, {Tortora}, {et~al.}}]{EP-Bisigello}
{Euclid Collaboration: Bisigello}, L., {Massimo}, M., {Tortora}, C., {et~al.}
  2024, \aap, 691, A1

\bibitem[{{Euclid Collaboration: Bisigello} {et~al.}(2025){Euclid
  Collaboration: Bisigello}, {Rodighiero}, {Fotopoulou}, {et~al.}}]{Q1-SP011}
{Euclid Collaboration: Bisigello}, L., {Rodighiero}, G., {Fotopoulou}, S.,
  {et~al.} 2025, \aap, submitted

\bibitem[{{Euclid Collaboration: Cleland} {et~al.}(2025){Euclid Collaboration:
  Cleland}, {Mei}, {De Lucia}, {et~al.}}]{Q1-SP017}
{Euclid Collaboration: Cleland}, C., {Mei}, S., {De Lucia}, G., {et~al.} 2025,
  \aap, submitted

\bibitem[{{Euclid Collaboration: Cropper} {et~al.}(2024){Euclid Collaboration:
  Cropper}, {Al Bahlawan}, {Amiaux}, {et~al.}}]{EuclidSkyVIS}
{Euclid Collaboration: Cropper}, M., {Al Bahlawan}, A., {Amiaux}, J., {et~al.}
  2024, \aap, accepted, arXiv:2405.13492

\bibitem[{{Euclid Collaboration: Desprez} {et~al.}(2020){Euclid Collaboration:
  Desprez}, {Paltani}, {Coupon}, {et~al.}}]{Desprez-EP10}
{Euclid Collaboration: Desprez}, G., {Paltani}, S., {Coupon}, J., {et~al.}
  2020, \aap, 644, A31

\bibitem[{{Euclid Collaboration: Enia} {et~al.}(2024){Euclid Collaboration:
  Enia}, {Bolzonella}, {Pozzetti}, {et~al.}}]{EP-Enia}
{Euclid Collaboration: Enia}, A., {Bolzonella}, M., {Pozzetti}, L., {et~al.}
  2024, \aap, 691, A175

\bibitem[{{Euclid Collaboration: Enia} {et~al.}(2025){Euclid Collaboration:
  Enia}, {Pozzetti}, {Bolzonella}, {et~al.}}]{Q1-SP031}
{Euclid Collaboration: Enia}, A., {Pozzetti}, L., {Bolzonella}, M., {et~al.}
  2025, \aap, submitted

\bibitem[{{Euclid Collaboration: Girardi} {et~al.}(2025){Euclid Collaboration:
  Girardi}, {Rodighiero}, {Bisigello}, {et~al.}}]{Q1-SP016}
{Euclid Collaboration: Girardi}, G., {Rodighiero}, G., {Bisigello}, L.,
  {et~al.} 2025, \aap, submitted

\bibitem[{{Euclid Collaboration: Jahnke} {et~al.}(2024){Euclid Collaboration:
  Jahnke}, {Gillard}, {Schirmer}, {et~al.}}]{EuclidSkyNISP}
{Euclid Collaboration: Jahnke}, K., {Gillard}, W., {Schirmer}, M., {et~al.}
  2024, \aap, accepted, arXiv:2405.13493

\bibitem[{{Euclid Collaboration: La Marca} {et~al.}(2025){Euclid Collaboration:
  La Marca}, {Wang}, {Margalef-Bentabol}, {et~al.}}]{Q1-SP013}
{Euclid Collaboration: La Marca}, A., {Wang}, L., {Margalef-Bentabol}, B.,
  {et~al.} 2025, \aap, submitted

\bibitem[{{Euclid Collaboration: Le Brun} {et~al.}(2025){Euclid Collaboration:
  Le Brun}, {Bethermin}, {et~al.}}]{Q1-TP007}
{Euclid Collaboration: Le Brun}, V., {Bethermin}, B., {et~al.} 2025, \aap,
  submitted

\bibitem[{{Euclid Collaboration: Margalef-Bentabol} {et~al.}(2025){Euclid
  Collaboration: Margalef-Bentabol}, {Wang}, {La Marca}, {et~al.}}]{Q1-SP015}
{Euclid Collaboration: Margalef-Bentabol}, B., {Wang}, L., {La Marca}, A.,
  {et~al.} 2025, \aap, submitted

\bibitem[{{Euclid Collaboration: Matamoro Zatarain} {et~al.}(2025){Euclid
  Collaboration: Matamoro Zatarain}, {Fotopoulou}, {Ricci},
  {et~al.}}]{Q1-SP027}
{Euclid Collaboration: Matamoro Zatarain}, T., {Fotopoulou}, S., {Ricci}, F.,
  {et~al.} 2025, \aap, submitted

\bibitem[{{Euclid Collaboration: McCracken} {et~al.}(2025){Euclid
  Collaboration: McCracken}, {Benson}, {et~al.}}]{Q1-TP002}
{Euclid Collaboration: McCracken}, H., {Benson}, K., {et~al.} 2025, \aap,
  submitted

\bibitem[{{Euclid Collaboration: Mellier} {et~al.}(2024){Euclid Collaboration:
  Mellier}, {Abdurro'uf}, {Acevedo~Barroso}, {et~al.}}]{EuclidSkyOverview}
{Euclid Collaboration: Mellier}, Y., {Abdurro'uf}, {Acevedo~Barroso}, J.,
  {et~al.} 2024, \aap, accepted, arXiv:2405.13491

\bibitem[{{Euclid Collaboration: Paltani} {et~al.}(2024){Euclid Collaboration:
  Paltani}, {Coupon}, {Hartley}, {et~al.}}]{EP-Paltani}
{Euclid Collaboration: Paltani}, S., {Coupon}, J., {Hartley}, W.~G., {et~al.}
  2024, \aap, 681, A66

\bibitem[{{Euclid Collaboration: Polenta} {et~al.}(2025){Euclid Collaboration:
  Polenta}, {Frailis}, {Alavi}, {et~al.}}]{Q1-TP003}
{Euclid Collaboration: Polenta}, G., {Frailis}, M., {Alavi}, A., {et~al.} 2025,
  \aap, submitted

\bibitem[{{Euclid Collaboration: Romelli} {et~al.}(2025){Euclid Collaboration:
  Romelli}, {K\"ummel}, {Dole}, {et~al.}}]{Q1-TP004}
{Euclid Collaboration: Romelli}, E., {K\"ummel}, M., {Dole}, H., {et~al.} 2025,
  \aap, submitted

\bibitem[{{Euclid Collaboration: Roster} {et~al.}(2025){Euclid Collaboration:
  Roster}, {Salvato}, {Buchner}, {et~al.}}]{Q1-SP003}
{Euclid Collaboration: Roster}, W., {Salvato}, M., {Buchner}, J., {et~al.}
  2025, \aap, submitted

\bibitem[{{Euclid Collaboration: Scaramella} {et~al.}(2022){Euclid
  Collaboration: Scaramella}, {Amiaux}, {Mellier}, {et~al.}}]{Scaramella-EP1}
{Euclid Collaboration: Scaramella}, R., {Amiaux}, J., {Mellier}, Y., {et~al.}
  2022, \aap, 662, A112

\bibitem[{{Euclid Collaboration: Serrano} {et~al.}(2024){Euclid Collaboration:
  Serrano}, {Hudelot}, {Seidel}, {et~al.}}]{EP-Serrano}
{Euclid Collaboration: Serrano}, S., {Hudelot}, P., {Seidel}, G., {et~al.}
  2024, \aap, 690, A103

\bibitem[{{Euclid Collaboration: Siudek} {et~al.}(2025){Euclid Collaboration:
  Siudek}, {Huertas-Company}, {Smith}, {et~al.}}]{Q1-SP049}
{Euclid Collaboration: Siudek}, M., {Huertas-Company}, M., {Smith}, M.,
  {et~al.} 2025, \aap, submitted

\bibitem[{{Euclid Collaboration: Stevens} {et~al.}(2025){Euclid Collaboration:
  Stevens}, {Fotopoulou}, {Bremer}, {et~al.}}]{Q1-SP009}
{Euclid Collaboration: Stevens}, G., {Fotopoulou}, S., {Bremer}, M., {et~al.}
  2025, \aap, submitted

\bibitem[{{Euclid Collaboration: Tarsitano} {et~al.}(2025){Euclid
  Collaboration: Tarsitano}, {Fotopoulou}, {Banerji}, {et~al.}}]{Q1-SP023}
{Euclid Collaboration: Tarsitano}, F., {Fotopoulou}, S., {Banerji}, M.,
  {et~al.} 2025, \aap, submitted

\bibitem[{{Euclid Quick Release Q1}(2025)}]{Q1cite}
{Euclid Quick Release Q1}. 2025, \url{https://doi.org/10.57780/esa-2853f3b}

\bibitem[{{Gaia Collaboration} {et~al.}(2023){Gaia Collaboration}, {Vallenari},
  {Brown}, {Prusti}, {de Bruijne}, {Arenou}, {Babusiaux}, {Biermann},
  {Creevey}, {Ducourant}, {Evans}, {Eyer}, {Guerra}, {Hutton}, {Jordi},
  {Klioner}, {Lammers}, {Lindegren}, {Luri}, {Mignard}, {Panem}, {Pourbaix},
  {Randich}, {Sartoretti}, {Soubiran}, {Tanga}, {Walton}, {Bailer-Jones},
  {Bastian}, {Drimmel}, {Jansen}, {Katz}, {Lattanzi}, {van Leeuwen}, {Bakker},
  {Cacciari}, {Casta{\~n}eda}, {De Angeli}, {Fabricius}, {Fouesneau},
  {Fr{\'e}mat}, {Galluccio}, {Guerrier}, {Heiter}, {Masana}, {Messineo},
  {Mowlavi}, {Nicolas}, {Nienartowicz}, {Pailler}, {Panuzzo}, {Riclet}, {Roux},
  {Seabroke}, {Sordo}, {Th{\'e}venin}, {Gracia-Abril}, {Portell}, {Teyssier},
  {Altmann}, {Andrae}, {Audard}, {Bellas-Velidis}, {Benson}, {Berthier},
  {Blomme}, {Burgess}, {Busonero}, {Busso}, {C{\'a}novas}, {Carry}, {Cellino},
  {Cheek}, {Clementini}, {Damerdji}, {Davidson}, {de Teodoro}, {Nu{\~n}ez
  Campos}, {Delchambre}, {Dell'Oro}, {Esquej}, {Fern{\'a}ndez-Hern{\'a}ndez},
  {Fraile}, {Garabato}, {Garc{\'\i}a-Lario}, {Gosset}, {Haigron}, {Halbwachs},
  {Hambly}, {Harrison}, {Hern{\'a}ndez}, {Hestroffer}, {Hodgkin}, {Holl},
  {Jan{\ss}en}, {Jevardat de Fombelle}, {Jordan}, {Krone-Martins}, {Lanzafame},
  {L{\"o}ffler}, {Marchal}, {Marrese}, {Moitinho}, {Muinonen}, {Osborne},
  {Pancino}, {Pauwels}, {Recio-Blanco}, {Reyl{\'e}}, {Riello}, {Rimoldini},
  {Roegiers}, {Rybizki}, {Sarro}, {Siopis}, {Smith}, {Sozzetti}, {Utrilla},
  {van Leeuwen}, {Abbas}, {{\'A}brah{\'a}m}, {Abreu Aramburu}, {Aerts},
  {Aguado}, {Ajaj}, {Aldea-Montero}, {Altavilla}, {{\'A}lvarez}, {Alves},
  {Anders}, {Anderson}, {Anglada Varela}, {Antoja}, {Baines}, {Baker},
  {Balaguer-N{\'u}{\~n}ez}, {Balbinot}, {Balog}, {Barache}, {Barbato},
  {Barros}, {Barstow}, {Bartolom{\'e}}, {Bassilana}, {Bauchet}, {Becciani},
  {Bellazzini}, {Berihuete}, {Bernet}, {Bertone}, {Bianchi}, {Binnenfeld},
  {Blanco-Cuaresma}, {Blazere}, {Boch}, {Bombrun}, {Bossini}, {Bouquillon},
  {Bragaglia}, {Bramante}, {Breedt}, {Bressan}, {Brouillet}, {Brugaletta},
  {Bucciarelli}, {Burlacu}, {Butkevich}, {Buzzi}, {Caffau}, {Cancelliere},
  {Cantat-Gaudin}, {Carballo}, {Carlucci}, {Carnerero}, {Carrasco},
  {Casamiquela}, {Castellani}, {Castro-Ginard}, {Chaoul}, {Charlot}, {Chemin},
  {Chiaramida}, {Chiavassa}, {Chornay}, {Comoretto}, {Contursi}, {Cooper},
  {Cornez}, {Cowell}, {Crifo}, {Cropper}, {Crosta}, {Crowley}, {Dafonte},
  {Dapergolas}, {David}, {David}, {de Laverny}, {De Luise}, \& {De
  March}}]{gaiadr3}
{Gaia Collaboration}, {Vallenari}, A., {Brown}, A.~G.~A., {et~al.} 2023, \aap,
  674, A1

\bibitem[{{Galametz} {et~al.}(2017){Galametz}, {Saglia}, {Paltani},
  {Apostolakos}, \& {Dubath}}]{gal17}
{Galametz}, A., {Saglia}, R., {Paltani}, S., {Apostolakos}, N., \& {Dubath}, P.
  2017, \aap, 598, A20

\bibitem[{{Garilli} {et~al.}(2021){Garilli}, {McLure}, {Pentericci},
  {Franzetti}, {Gargiulo}, {Carnall}, {Cucciati}, {Iovino}, {Amorin},
  {Bolzonella}, {Bongiorno}, {Castellano}, {Cimatti}, {Cirasuolo}, {Cullen},
  {Dunlop}, {Elbaz}, {Finkelstein}, {Fontana}, {Fontanot}, {Fumana}, {Guaita},
  {Hartley}, {Jarvis}, {Juneau}, {Maccagni}, {McLeod}, {Nandra}, {Pompei},
  {Pozzetti}, {Scodeggio}, {Talia}, {Calabr{\`o}}, {Cresci}, {Fynbo}, {Hathi},
  {Hibon}, {Koekemoer}, {Magliocchetti}, {Salvato}, {Vietri}, {Zamorani},
  {Almaini}, {Balestra}, {Bardelli}, {Begley}, {Brammer}, {Bell}, {Bowler},
  {Brusa}, {Buitrago}, {Caputi}, {Cassata}, {Charlot}, {Citro}, {Cristiani},
  {Curtis-Lake}, {Dickinson}, {Fazio}, {Ferguson}, {Fiore}, {Franco},
  {Georgakakis}, {Giavalisco}, {Grazian}, {Hamadouche}, {Jung}, {Kim},
  {Khusanova}, {Le F{\`e}vre}, {Longhetti}, {Lotz}, {Mannucci}, {Maltby},
  {Matsuoka}, {Mendez-Hernandez}, {Mendez-Abreu}, {Mignoli}, {Moresco},
  {Nonino}, {Pannella}, {Papovich}, {Popesso}, {Roberts-Borsani}, {Rosario},
  {Saldana-Lopez}, {Santini}, {Saxena}, {Schaerer}, {Schreiber}, {Stark},
  {Tasca}, {Thomas}, {Vanzella}, {Wild}, {Williams}, \& {Zucca}}]{gar21}
{Garilli}, B., {McLure}, R., {Pentericci}, L., {et~al.} 2021, \aap, 647, A150

\bibitem[{{Gordon} {et~al.}(2023){Gordon}, {Clayton}, {Decleir}, {Fitzpatrick},
  {Massa}, {Misselt}, \& {Tollerud}}]{gor23}
{Gordon}, K.~D., {Clayton}, G.~C., {Decleir}, M., {et~al.} 2023, \apj, 950, 86

\bibitem[{{Guarneri} {et~al.}(2022){Guarneri}, {Calderone}, {Cristiani},
  {Porru}, {Fontanot}, {Boutsia}, {Cupani}, {Grazian}, {D'Odorico}, {Murphy},
  {Bongiorno}, {Saccheo}, \& {Nicastro}}]{gua22}
{Guarneri}, F., {Calderone}, G., {Cristiani}, S., {et~al.} 2022, \mnras, 517,
  2436

\bibitem[{{Halton}(1960)}]{hal60}
{Halton}, J.~H. 1960, Numerische Mathematik, 2(1), 84

\bibitem[{{Hartley} {et~al.}(2022){Hartley}, {Choi}, {Amon}, {Gruendl},
  {Sheldon}, {Harrison}, {Bernstein}, {Sevilla-Noarbe}, {Yanny}, {Eckert},
  {Diehl}, {Alarcon}, {Banerji}, {Bechtol}, {Buchs}, {Cantu}, {Conselice},
  {Cordero}, {Davis}, {Davis}, {Dodelson}, {Drlica-Wagner}, {Everett},
  {Fert{\'e}}, {Gruen}, {Honscheid}, {Jarvis}, {Johnson}, {Kokron}, {MacCrann},
  {Myles}, {Pace}, {Palmese}, {Paz-Chinch{\'o}n}, {Pereira}, {Plazas}, {Prat},
  {Rodriguez-Monroy}, {Rykoff}, {Samuroff}, {S{\'a}nchez}, {Secco},
  {Tarsitano}, {Tong}, {Troxel}, {Vasquez}, {Wang}, {Zhou}, {Abbott}, {Aguena},
  {Allam}, {Annis}, {Bacon}, {Bertin}, {Bhargava}, {Brooks}, {Burke}, {Carnero
  Rosell}, {Carrasco Kind}, {Carretero}, {Castander}, {Costanzi}, {Crocce}, {da
  Costa}, {De Vicente}, {DeRose}, {Desai}, {Dietrich}, {Eifler}, {Elvin-Poole},
  {Ferrero}, {Flaugher}, {Fosalba}, {Garc{\'\i}a-Bellido}, {Gaztanaga},
  {Gerdes}, {Gschwend}, {Gutierrez}, {Hinton}, {Hollowood}, {Huterer}, {James},
  {Kent}, {Krause}, {Kuehn}, {Kuropatkin}, {Lahav}, {Lin}, {Maia}, {March},
  {Marshall}, {Martini}, {Melchior}, {Menanteau}, {Miquel}, {Mohr}, {Morgan},
  {Neilsen}, {Ogando}, {Pandey}, {Romer}, {Roodman}, {Sako}, {Sanchez},
  {Scarpine}, {Serrano}, {Smith}, {Soares-Santos}, {Suchyta}, {Swanson},
  {Tarle}, {Thomas}, {To}, {Varga}, {Walker}, {Wester}, {Wilkinson}, {Zuntz},
  {Zuntz}, \& {DES Collaboration}}]{har22}
{Hartley}, W.~G., {Choi}, A., {Amon}, A., {et~al.} 2022, \mnras, 509, 3547

\bibitem[{{Hoaglin} {et~al.}(1983){Hoaglin}, {Mosteller}, \& {Tukey}}]{hoa83}
{Hoaglin}, D.~C., {Mosteller}, F., \& {Tukey}, J.~W. 1983, Wiley, edited by
  Hoaglin, David C.; Mosteller, Frederick; Tukey, John W.

\bibitem[{{Huchra} {et~al.}(2012){Huchra}, {Macri}, {Masters}, {Jarrett},
  {Berlind}, {Calkins}, {Crook}, {Cutri}, {Erdo{\v{g}}du}, {Falco}, {George},
  {Hutcheson}, {Lahav}, {Mader}, {Mink}, {Martimbeau}, {Schneider},
  {Skrutskie}, {Tokarz}, \& {Westover}}]{huc2012}
{Huchra}, J.~P., {Macri}, L.~M., {Masters}, K.~L., {et~al.} 2012, \apjs, 199,
  26

\bibitem[{{Huterer} {et~al.}(2006){Huterer}, {Takada}, {Bernstein}, \&
  {Jain}}]{hut06}
{Huterer}, D., {Takada}, M., {Bernstein}, G., \& {Jain}, B. 2006, \mnras, 366,
  101

\bibitem[{{Ibata} {et~al.}(2017){Ibata}, {McConnachie}, {Cuillandre}, {Fantin},
  {Haywood}, {Martin}, {Bergeron}, {Beckmann}, {Bernard}, {Bonifacio},
  {Caffau}, {Carlberg}, {C{\^o}t{\'e}}, {Cabanac}, {Chapman}, {Duc}, {Durret},
  {Famaey}, {Fabbro}, {Gwyn}, {Hammer}, {Hill}, {Hudson}, {Lan{\c{c}}on},
  {Lewis}, {Malhan}, {di Matteo}, {McCracken}, {Mei}, {Mellier}, {Navarro},
  {Pires}, {Pritchet}, {Reyl{\'e}}, {Richer}, {Robin}, {S{\'a}nchez-Janssen},
  {Sawicki}, {Scott}, {Scottez}, {Spekkens}, {Starkenburg}, {Thomas}, \&
  {Venn}}]{iba17}
{Ibata}, R.~A., {McConnachie}, A., {Cuillandre}, J.-C., {et~al.} 2017, \apj,
  848, 128

\bibitem[{{Ilbert} {et~al.}(2006){Ilbert}, {Arnouts}, {McCracken},
  {Bolzonella}, {Bertin}, {Le F{\`e}vre}, {Mellier}, {Zamorani}, {Pell{\`o}},
  {Iovino}, {Tresse}, {Le Brun}, {Bottini}, {Garilli}, {Maccagni}, {Picat},
  {Scaramella}, {Scodeggio}, {Vettolani}, {Zanichelli}, {Adami}, {Bardelli},
  {Cappi}, {Charlot}, {Ciliegi}, {Contini}, {Cucciati}, {Foucaud}, {Franzetti},
  {Gavignaud}, {Guzzo}, {Marano}, {Marinoni}, {Mazure}, {Meneux}, {Merighi},
  {Paltani}, {Pollo}, {Pozzetti}, {Radovich}, {Zucca}, {Bondi}, {Bongiorno},
  {Busarello}, {de La Torre}, {Gregorini}, {Lamareille}, {Mathez}, {Merluzzi},
  {Ripepi}, {Rizzo}, \& {Vergani}}]{ilb06}
{Ilbert}, O., {Arnouts}, S., {McCracken}, H.~J., {et~al.} 2006, \aap, 457, 841

\bibitem[{{Ilbert} {et~al.}(2009){Ilbert}, {Capak}, {Salvato}, {Aussel},
  {McCracken}, {Sanders}, {Scoville}, {Kartaltepe}, {Arnouts}, {Le Floc'h},
  {Mobasher}, {Taniguchi}, {Lamareille}, {Leauthaud}, {Sasaki}, {Thompson},
  {Zamojski}, {Zamorani}, {Bardelli}, {Bolzonella}, {Bongiorno}, {Brusa},
  {Caputi}, {Carollo}, {Contini}, {Cook}, {Coppa}, {Cucciati}, {de la Torre},
  {de Ravel}, {Franzetti}, {Garilli}, {Hasinger}, {Iovino}, {Kampczyk},
  {Kneib}, {Knobel}, {Kovac}, {Le Borgne}, {Le Brun}, {Le F{\`e}vre}, {Lilly},
  {Looper}, {Maier}, {Mainieri}, {Mellier}, {Mignoli}, {Murayama}, {Pell{\`o}},
  {Peng}, {P{\'e}rez-Montero}, {Renzini}, {Ricciardelli}, {Schiminovich},
  {Scodeggio}, {Shioya}, {Silverman}, {Surace}, {Tanaka}, {Tasca}, {Tresse},
  {Vergani}, \& {Zucca}}]{ilb09}
{Ilbert}, O., {Capak}, P., {Salvato}, M., {et~al.} 2009, \apj, 690, 1236

\bibitem[{{Ilbert} {et~al.}(2005){Ilbert}, {Tresse}, {Zucca}, {Bardelli},
  {Arnouts}, {Zamorani}, {Pozzetti}, {Bottini}, {Garilli}, {Le Brun}, {Le
  F{\`e}vre}, {Maccagni}, {Picat}, {Scaramella}, {Scodeggio}, {Vettolani},
  {Zanichelli}, {Adami}, {Arnaboldi}, {Bolzonella}, {Cappi}, {Charlot},
  {Contini}, {Foucaud}, {Franzetti}, {Gavignaud}, {Guzzo}, {Iovino},
  {McCracken}, {Marano}, {Marinoni}, {Mathez}, {Mazure}, {Meneux}, {Merighi},
  {Paltani}, {Pello}, {Pollo}, {Radovich}, {Bondi}, {Bongiorno}, {Busarello},
  {Ciliegi}, {Lamareille}, {Mellier}, {Merluzzi}, {Ripepi}, \& {Rizzo}}]{ilb05}
{Ilbert}, O., {Tresse}, L., {Zucca}, E., {et~al.} 2005, \aap, 439, 863

\bibitem[{{Inoue} {et~al.}(2014){Inoue}, {Shimizu}, {Iwata}, \&
  {Tanaka}}]{ino14}
{Inoue}, A.~K., {Shimizu}, I., {Iwata}, I., \& {Tanaka}, M. 2014, \mnras, 442,
  1805

\bibitem[{{Ivezi{\'c}} {et~al.}(2019){Ivezi{\'c}}, {Kahn}, {Tyson}, {Abel},
  {Acosta}, {Allsman}, {Alonso}, {AlSayyad}, {Anderson}, {Andrew}, {Angel},
  {Angeli}, {Ansari}, {Antilogus}, {Araujo}, {Armstrong}, {Arndt}, {Astier},
  {Aubourg}, {Auza}, {Axelrod}, {Bard}, {Barr}, {Barrau}, {Bartlett}, {Bauer},
  {Bauman}, {Baumont}, {Bechtol}, {Bechtol}, {Becker}, {Becla}, {Beldica},
  {Bellavia}, {Bianco}, {Biswas}, {Blanc}, {Blazek}, {Blandford}, {Bloom},
  {Bogart}, {Bond}, {Booth}, {Borgland}, {Borne}, {Bosch}, {Boutigny},
  {Brackett}, {Bradshaw}, {Brandt}, {Brown}, {Bullock}, {Burchat}, {Burke},
  {Cagnoli}, {Calabrese}, {Callahan}, {Callen}, {Carlin}, {Carlson},
  {Chandrasekharan}, {Charles-Emerson}, {Chesley}, {Cheu}, {Chiang}, {Chiang},
  {Chirino}, {Chow}, {Ciardi}, {Claver}, {Cohen-Tanugi}, {Cockrum}, {Coles},
  {Connolly}, {Cook}, {Cooray}, {Covey}, {Cribbs}, {Cui}, {Cutri}, {Daly},
  {Daniel}, {Daruich}, {Daubard}, {Daues}, {Dawson}, {Delgado}, {Dellapenna},
  {de Peyster}, {de Val-Borro}, {Digel}, {Doherty}, {Dubois},
  {Dubois-Felsmann}, {Durech}, {Economou}, {Eifler}, {Eracleous}, {Emmons},
  {Fausti Neto}, {Ferguson}, {Figueroa}, {Fisher-Levine}, {Focke}, {Foss},
  {Frank}, {Freemon}, {Gangler}, {Gawiser}, {Geary}, {Gee}, {Geha}, {Gessner},
  {Gibson}, {Gilmore}, {Glanzman}, {Glick}, {Goldina}, {Goldstein}, {Goodenow},
  {Graham}, {Gressler}, {Gris}, {Guy}, {Guyonnet}, {Haller}, {Harris},
  {Hascall}, {Haupt}, {Hernandez}, {Herrmann}, {Hileman}, {Hoblitt}, {Hodgson},
  {Hogan}, {Howard}, {Huang}, {Huffer}, {Ingraham}, {Innes}, {Jacoby}, {Jain},
  {Jammes}, {Jee}, {Jenness}, {Jernigan}, {Jevremovi{\'c}}, {Johns}, {Johnson},
  {Johnson}, {Jones}, {Juramy-Gilles}, {Juri{\'c}}, {Kalirai}, {Kallivayalil},
  {Kalmbach}, {Kantor}, {Karst}, {Kasliwal}, {Kelly}, {Kessler}, {Kinnison},
  {Kirkby}, {Knox}, {Kotov}, {Krabbendam}, {Krughoff}, {Kub{\'a}nek},
  {Kuczewski}, {Kulkarni}, {Ku}, {Kurita}, {Lage}, {Lambert}, {Lange},
  {Langton}, {Le Guillou}, {Levine}, {Liang}, {Lim}, {Lintott}, {Long},
  {Lopez}, {Lotz}, {Lupton}, {Lust}, {MacArthur}, {Mahabal}, {Mandelbaum},
  {Markiewicz}, {Marsh}, {Marshall}, {Marshall}, {May}, {McKercher}, {McQueen},
  {Meyers}, {Migliore}, {Miller}, {Mills}, {Miraval}, {Moeyens}, {Moolekamp},
  {Monet}, {Moniez}, {Monkewitz}, {Montgomery}, {Morrison}, {Mueller},
  {Muller}, {Mu{\~n}oz Arancibia}, {Neill}, {Newbry}, {Nief}, {Nomerotski},
  {Nordby}, {O'Connor}, {Oliver}, {Olivier}, {Olsen}, {O'Mullane}, {Ortiz},
  {Osier}, {Owen}, {Pain}, {Palecek}, {Parejko}, {Parsons}, {Pease},
  {Peterson}, {Peterson}, {Petravick}, {Libby Petrick}, {Petry},
  {Pierfederici}, {Pietrowicz}, {Pike}, {Pinto}, {Plante}, {Plate}, {Plutchak},
  {Price}, {Prouza}, {Radeka}, {Rajagopal}, {Rasmussen}, {Regnault}, {Reil},
  {Reiss}, {Reuter}, {Ridgway}, {Riot}, {Ritz}, {Robinson}, {Roby}, {Roodman},
  {Rosing}, {Roucelle}, {Rumore}, {Russo}, {Saha}, {Sassolas}, {Schalk},
  {Schellart}, {Schindler}, {Schmidt}, {Schneider}, {Schneider}, {Schoening},
  {Schumacher}, {Schwamb}, {Sebag}, {Selvy}, {Sembroski}, {Seppala}, {Serio},
  {Serrano}, {Shaw}, {Shipsey}, {Sick}, {Silvestri}, {Slater}, {Smith},
  {Smith}, {Sobhani}, {Soldahl}, {Storrie-Lombardi}, {Stover}, {Strauss},
  {Street}, {Stubbs}, {Sullivan}, {Sweeney}, {Swinbank}, {Szalay}, {Takacs},
  {Tether}, {Thaler}, {Thayer}, {Thomas}, {Thornton}, {Thukral}, {Tice},
  {Trilling}, {Turri}, {Van Berg}, {Vanden Berk}, {Vetter}, {Virieux},
  {Vucina}, {Wahl}, {Walkowicz}, {Walsh}, {Walter}, {Wang}, {Wang}, {Warner},
  {Wiecha}, {Willman}, {Winters}, {Wittman}, {Wolff}, {Wood-Vasey}, {Wu},
  {Xin}, {Yoachim}, \& {Zhan}}]{ive19}
{Ivezi{\'c}}, {\v{Z}}., {Kahn}, S.~M., {Tyson}, J.~A., {et~al.} 2019, \apj,
  873, 111

\bibitem[{{Jarvis} {et~al.}(2013){Jarvis}, {Bonfield}, {Bruce}, {Geach},
  {McAlpine}, {McLure}, {Gonz{\'a}lez-Solares}, {Irwin}, {Lewis}, {Yoldas},
  {Andreon}, {Cross}, {Emerson}, {Dalton}, {Dunlop}, {Hodgkin}, {Le},
  {Karouzos}, {Meisenheimer}, {Oliver}, {Rawlings}, {Simpson}, {Smail},
  {Smith}, {Sullivan}, {Sutherland}, {White}, \& {Zwart}}]{jar13}
{Jarvis}, M.~J., {Bonfield}, D.~G., {Bruce}, V.~A., {et~al.} 2013, \mnras, 428,
  1281

\bibitem[{{Jones} {et~al.}(2009){Jones}, {Read}, {Saunders}, {Colless},
  {Jarrett}, {Parker}, {Fairall}, {Mauch}, {Sadler}, {Watson}, {Burton},
  {Campbell}, {Cass}, {Croom}, {Dawe}, {Fiegert}, {Frankcombe}, {Hartley},
  {Huchra}, {James}, {Kirby}, {Lahav}, {Lucey}, {Mamon}, {Moore}, {Peterson},
  {Prior}, {Proust}, {Russell}, {Safouris}, {Wakamatsu}, {Westra}, \&
  {Williams}}]{jon2009}
{Jones}, D.~H., {Read}, M.~A., {Saunders}, W., {et~al.} 2009, \mnras, 399, 683

\bibitem[{{Katz} {et~al.}(2023){Katz}, {Sartoretti}, {Guerrier}, {Panuzzo},
  {Seabroke}, {Th{\'e}venin}, {Cropper}, {Benson}, {Blomme}, {Haigron},
  {Marchal}, {Smith}, {Baker}, {Chemin}, {Damerdji}, {David}, {Dolding},
  {Fr{\'e}mat}, {Gosset}, {Jan{\ss}en}, {Jasniewicz}, {Lobel}, {Plum},
  {Samaras}, {Snaith}, {Soubiran}, {Vanel}, {Zwitter}, {Antoja}, {Arenou},
  {Babusiaux}, {Brouillet}, {Caffau}, {Di Matteo}, {Fabre}, {Fabricius},
  {Fragkoudi}, {Haywood}, {Huckle}, {Hottier}, {Lasne}, {Leclerc},
  {Mastrobuono-Battisti}, {Royer}, {Teyssier}, {Zorec}, {Crifo}, {Jean-Antoine
  Piccolo}, {Turon}, \& {Viala}}]{kat23}
{Katz}, D., {Sartoretti}, P., {Guerrier}, A., {et~al.} 2023, \aap, 674, A5

\bibitem[{{Kennicutt}(1998)}]{ken98}
{Kennicutt}, Robert~C., J. 1998, \araa, 36, 189

\bibitem[{{Kinson} {et~al.}(2021){Kinson}, {Oliveira}, \& {van Loon}}]{kin21}
{Kinson}, D.~A., {Oliveira}, J.~M., \& {van Loon}, J.~T. 2021, \mnras, 507,
  5106

\bibitem[{{Kitching} {et~al.}(2008){Kitching}, {Miller}, {Heymans}, {van
  Waerbeke}, \& {Heavens}}]{kit08}
{Kitching}, T.~D., {Miller}, L., {Heymans}, C.~E., {van Waerbeke}, L., \&
  {Heavens}, A.~F. 2008, \mnras, 390, 149

\bibitem[{{Kriek} {et~al.}(2015){Kriek}, {Shapley}, {Reddy}, {Siana}, {Coil},
  {Mobasher}, {Freeman}, {de Groot}, {Price}, {Sanders}, {Shivaei}, {Brammer},
  {Momcheva}, {Skelton}, {van Dokkum}, {Whitaker}, {Aird}, {Azadi}, {Kassis},
  {Bullock}, {Conroy}, {Dav{\'e}}, {Kere{\v{s}}}, \& {Krumholz}}]{kri2015}
{Kriek}, M., {Shapley}, A.~E., {Reddy}, N.~A., {et~al.} 2015, \apjs, 218, 15

\bibitem[{{Lastennet} {et~al.}(2002){Lastennet}, {Lejeune}, {Oblak}, {Westera},
  \& {Buser}}]{las02}
{Lastennet}, E., {Lejeune}, T., {Oblak}, E., {Westera}, P., \& {Buser}, R.
  2002, \apss, 280, 83

\bibitem[{{Le F{\`e}vre} {et~al.}(2013){Le F{\`e}vre}, {Cassata}, {Cucciati},
  {Garilli}, {Ilbert}, {Le Brun}, {Maccagni}, {Moreau}, {Scodeggio}, {Tresse},
  {Zamorani}, {Adami}, {Arnouts}, {Bardelli}, {Bolzonella}, {Bondi},
  {Bongiorno}, {Bottini}, {Cappi}, {Charlot}, {Ciliegi}, {Contini}, {de la
  Torre}, {Foucaud}, {Franzetti}, {Gavignaud}, {Guzzo}, {Iovino}, {Lemaux},
  {L{\'o}pez-Sanjuan}, {McCracken}, {Marano}, {Marinoni}, {Mazure}, {Mellier},
  {Merighi}, {Merluzzi}, {Paltani}, {Pell{\`o}}, {Pollo}, {Pozzetti},
  {Scaramella}, {Tasca}, {Vergani}, {Vettolani}, {Zanichelli}, \&
  {Zucca}}]{lef13}
{Le F{\`e}vre}, O., {Cassata}, P., {Cucciati}, O., {et~al.} 2013, \aap, 559,
  A14

\bibitem[{{Lidman} {et~al.}(2020){Lidman}, {Tucker}, {Davis}, {Uddin},
  {Asorey}, {Bolejko}, {Brout}, {Calcino}, {Carollo}, {Carr}, {Childress},
  {Hoormann}, {Foley}, {Galbany}, {Glazebrook}, {Hinton}, {Kessler}, {Kim},
  {King}, {Kremin}, {Kuehn}, {Lagattuta}, {Lewis}, {Macaulay}, {Malik},
  {March}, {Martini}, {M{\"o}ller}, {Mudd}, {Nichol}, {Panther}, {Parkinson},
  {Pursiainen}, {Sako}, {Swann}, {Scalzo}, {Scolnic}, {Sharp}, {Smith},
  {Sommer}, {Sullivan}, {Webb}, {Wiseman}, {Yu}, {Yuan}, {Zhang}, {Abbott},
  {Aguena}, {Allam}, {Annis}, {Avila}, {Bertin}, {Bhargava}, {Brooks}, {Carnero
  Rosell}, {Carrasco Kind}, {Carretero}, {Castander}, {Costanzi}, {da Costa},
  {De Vicente}, {Doel}, {Eifler}, {Everett}, {Fosalba}, {Frieman},
  {Garc{\'\i}a-Bellido}, {Gaztanaga}, {Gruen}, {Gruendl}, {Gschwend},
  {Gutierrez}, {Hartley}, {Hollowood}, {Honscheid}, {James}, {Kuropatkin},
  {Li}, {Lima}, {Lin}, {Maia}, {Marshall}, {Melchior}, {Menanteau}, {Miquel},
  {Palmese}, {Paz-Chinch{\'o}n}, {Plazas}, {Roodman}, {Rykoff}, {Sanchez},
  {Santiago}, {Scarpine}, {Schubnell}, {Serrano}, {Sevilla-Noarbe}, {Suchyta},
  {Swanson}, {Tarle}, {Tucker}, {Varga}, {Walker}, {Wester}, {Wilkinson}, \&
  {DES Collaboration}}]{lid20}
{Lidman}, C., {Tucker}, B.~E., {Davis}, T.~M., {et~al.} 2020, \mnras, 496, 19

\bibitem[{{Lilly} {et~al.}(2009){Lilly}, {Le Brun}, {Maier}, {Mainieri},
  {Mignoli}, {Scodeggio}, {Zamorani}, {Carollo}, {Contini}, {Kneib}, {Le
  F{\`e}vre}, {Renzini}, {Bardelli}, {Bolzonella}, {Bongiorno}, {Caputi},
  {Coppa}, {Cucciati}, {de la Torre}, {de Ravel}, {Franzetti}, {Garilli},
  {Iovino}, {Kampczyk}, {Kovac}, {Knobel}, {Lamareille}, {Le Borgne}, {Pello},
  {Peng}, {P{\'e}rez-Montero}, {Ricciardelli}, {Silverman}, {Tanaka}, {Tasca},
  {Tresse}, {Vergani}, {Zucca}, {Ilbert}, {Salvato}, {Oesch}, {Abbas},
  {Bottini}, {Capak}, {Cappi}, {Cassata}, {Cimatti}, {Elvis}, {Fumana},
  {Guzzo}, {Hasinger}, {Koekemoer}, {Leauthaud}, {Maccagni}, {Marinoni},
  {McCracken}, {Memeo}, {Meneux}, {Porciani}, {Pozzetti}, {Sanders},
  {Scaramella}, {Scarlata}, {Scoville}, {Shopbell}, \& {Taniguchi}}]{lil09}
{Lilly}, S.~J., {Le Brun}, V., {Maier}, C., {et~al.} 2009, \apjs, 184, 218

\bibitem[{{Lupton} {et~al.}(1999){Lupton}, {Gunn}, \& {Szalay}}]{lup99}
{Lupton}, R.~H., {Gunn}, J.~E., \& {Szalay}, A.~S. 1999, \aj, 118, 1406

\bibitem[{{Ma} {et~al.}(2006){Ma}, {Hu}, \& {Huterer}}]{ma06}
{Ma}, Z., {Hu}, W., \& {Huterer}, D. 2006, \apj, 636, 21

\bibitem[{{Maciaszek} {et~al.}(2022){Maciaszek}, {Ealet}, {Gillard}, {Jahnke},
  {Barbier}, {Prieto}, {Bon}, {Bonnefoi}, {Caillat}, {Carle}, {Costille},
  {Ducret}, {Fabron}, {Foulon}, {Gimenez}, {Grassi}, {Jaquet}, {Le Mignant},
  {Martin}, {Pamplona}, {Sanchez}, {Cl{\'e}mens}, {Caillat}, {Niclas},
  {Secroun}, {Kubik}, {Ferriol}, {Berthe}, {Barri{\`e}re}, {Fontignie},
  {Valenziano}, {Auricchio}, {Battaglia}, {De Rosa}, {Farinelli}, {Franceschi},
  {Medinaceli}, {Morgante}, {Sortino}, {Trifoglio}, {Corcione}, {Capobianco},
  {Ligori}, {Dusini}, {Borsato}, {Dal Corso}, {Laudisio}, {Sirignano},
  {Stanco}, {Ventura}, {Patrizii}, {Chiarusi}, {Fornari}, {Giacomini},
  {Margiotta}, {Mauri}, {Pasqualini}, {Sirri}, {Spurio}, {Tenti}, {Travaglini},
  {Bonoli}, {Bortoletto}, {Balestra}, {Dalessandro}, {Grupp}, {Penka},
  {Steinwagner}, {Hormuth}, {Schirmer}, {Seidel}, {Padilla}, {Casas}, {Lloro},
  {Toledo-Moreo}, {Gomez}, {Colodro-Conde}, {Liz{\'a}n}, {Diaz}, {Lilje},
  {Andersen}, {Andersen}, {S{\o}rensen}, {Hornstrup}, {Jessen}, {Thizy},
  {Holmes}, {Pniel}, {Jhabvala}, {Pravdo}, {Seiffert}, {Waczynski}, {Laureij},
  {Racca}, {Salvignol}, {Boenke}, {Strada}, \& {Mellier}}]{Maciaszek22}
{Maciaszek}, T., {Ealet}, A., {Gillard}, W., {et~al.} 2022, in Society of
  Photo-Optical Instrumentation Engineers (SPIE) Conference Series, Vol. 12180,
  Space Telescopes and Instrumentation 2022: Optical, Infrared, and Millimeter
  Wave, ed. L.~E. {Coyle}, S.~{Matsuura}, \& M.~D. {Perrin}, arXiv:2210.10112

\bibitem[{{Maraston}(2005)}]{mar05}
{Maraston}, C. 2005, \mnras, 362, 799

\bibitem[{{Marchesi} {et~al.}(2016){Marchesi}, {Civano}, {Elvis}, {Salvato},
  {Brusa}, {Comastri}, {Gilli}, {Hasinger}, {Lanzuisi}, {Miyaji}, {Treister},
  {Urry}, {Vignali}, {Zamorani}, {Allevato}, {Cappelluti}, {Cardamone},
  {Finoguenov}, {Griffiths}, {Karim}, {Laigle}, {LaMassa}, {Jahnke}, {Ranalli},
  {Schawinski}, {Schinnerer}, {Silverman}, {Smolcic}, {Suh}, \&
  {Trakhtenbrot}}]{mar16}
{Marchesi}, S., {Civano}, F., {Elvis}, M., {et~al.} 2016, \apj, 817, 34

\bibitem[{{Masters} {et~al.}(2015){Masters}, {Capak}, {Stern}, {Ilbert},
  {Salvato}, {Schmidt}, {Longo}, {Rhodes}, {Paltani}, {Mobasher}, {Hoekstra},
  {Hildebrandt}, {Coupon}, {Steinhardt}, {Speagle}, {Faisst}, {Kalinich},
  {Brodwin}, {Brescia}, \& {Cavuoti}}]{mas15}
{Masters}, D., {Capak}, P., {Stern}, D., {et~al.} 2015, \apj, 813, 53

\bibitem[{{McCracken} {et~al.}(2012){McCracken}, {Milvang-Jensen}, {Dunlop},
  {Franx}, {Fynbo}, {Le F{\`e}vre}, {Holt}, {Caputi}, {Goranova}, {Buitrago},
  {Emerson}, {Freudling}, {Hudelot}, {L{\'o}pez-Sanjuan}, {Magnard}, {Mellier},
  {M{\o}ller}, {Nilsson}, {Sutherland}, {Tasca}, \& {Zabl}}]{mcc12}
{McCracken}, H.~J., {Milvang-Jensen}, B., {Dunlop}, J., {et~al.} 2012, \aap,
  544, A156

\bibitem[{{Merlin} {et~al.}(2016){Merlin}, {Bourne}, {Castellano}, {Ferguson},
  {Wang}, {Derriere}, {Dunlop}, {Elbaz}, \& {Fontana}}]{mer16}
{Merlin}, E., {Bourne}, N., {Castellano}, M., {et~al.} 2016, \aap, 595, A97

\bibitem[{{Merlin} {et~al.}(2015){Merlin}, {Fontana}, {Ferguson}, {Dunlop},
  {Elbaz}, {Bourne}, {Bruce}, {Buitrago}, {Castellano}, {Schreiber}, {Grazian},
  {McLure}, {Okumura}, {Shu}, {Wang}, {Amor{\'\i}n}, {Boutsia}, {Cappelluti},
  {Comastri}, {Derriere}, {Faber}, \& {Santini}}]{mer15}
{Merlin}, E., {Fontana}, A., {Ferguson}, H.~C., {et~al.} 2015, \aap, 582, A15

\bibitem[{{Miyazaki} {et~al.}(2018){Miyazaki}, {Komiyama}, {Kawanomoto}, {Doi},
  {Furusawa}, {Hamana}, {Hayashi}, {Ikeda}, {Kamata}, {Karoji}, {Koike},
  {Kurakami}, {Miyama}, {Morokuma}, {Nakata}, {Namikawa}, {Nakaya}, {Nariai},
  {Obuchi}, {Oishi}, {Okada}, {Okura}, {Tait}, {Takata}, {Tanaka}, {Tanaka},
  {Terai}, {Tomono}, {Uraguchi}, {Usuda}, {Utsumi}, {Yamada}, {Yamanoi},
  {Aihara}, {Fujimori}, {Mineo}, {Miyatake}, {Oguri}, {Uchida}, {Tanaka},
  {Yasuda}, {Takada}, {Murayama}, {Nishizawa}, {Sugiyama}, {Chiba}, {Futamase},
  {Wang}, {Chen}, {Ho}, {Liaw}, {Chiu}, {Ho}, {Lai}, {Lee}, {Jeng}, {Iwamura},
  {Armstrong}, {Bickerton}, {Bosch}, {Gunn}, {Lupton}, {Loomis}, {Price},
  {Smith}, {Strauss}, {Turner}, {Suzuki}, {Miyazaki}, {Muramatsu}, {Yamamoto},
  {Endo}, {Ezaki}, {Ito}, {Kawaguchi}, {Sofuku}, {Taniike}, {Akutsu}, {Dojo},
  {Kasumi}, {Matsuda}, {Imoto}, {Miwa}, {Suzuki}, {Takeshi}, \&
  {Yokota}}]{miy18}
{Miyazaki}, S., {Komiyama}, Y., {Kawanomoto}, S., {et~al.} 2018, \pasj, 70, S1

\bibitem[{{Momcheva} {et~al.}(2016){Momcheva}, {Brammer}, {van Dokkum},
  {Skelton}, {Whitaker}, {Nelson}, {Fumagalli}, {Maseda}, {Leja}, {Franx},
  {Rix}, {Bezanson}, {Da Cunha}, {Dickey}, {F{\"o}rster Schreiber},
  {Illingworth}, {Kriek}, {Labb{\'e}}, {Ulf Lange}, {Lundgren}, {Magee},
  {Marchesini}, {Oesch}, {Pacifici}, {Patel}, {Price}, {Tal}, {Wake}, {van der
  Wel}, \& {Wuyts}}]{mom2016}
{Momcheva}, I.~G., {Brammer}, G.~B., {van Dokkum}, P.~G., {et~al.} 2016, \apjs,
  225, 27

\bibitem[{{Mu{\v{z}}i{\'c}} {et~al.}(2022){Mu{\v{z}}i{\'c}}, {Almendros-Abad},
  {Bouy}, {Kubiak}, {Pe{\~n}a Ram{\'\i}rez}, {Krone-Martins}, {Moitinho}, \&
  {Concei{\c{c}}{\~a}o}}]{muz22}
{Mu{\v{z}}i{\'c}}, K., {Almendros-Abad}, V., {Bouy}, H., {et~al.} 2022, \aap,
  668, A19

\bibitem[{{Ni} {et~al.}(2021){Ni}, {Brandt}, {Chen}, {Luo}, {Nyland}, {Yang},
  {Zou}, {Aird}, {Alexander}, {Bauer}, {Lacy}, {Lehmer}, {Mallick}, {Salvato},
  {Schneider}, {Tozzi}, {Traulsen}, {Vaccari}, {Vignali}, {Vito}, {Xue},
  {Banerji}, {Chow}, {Comastri}, {Del Moro}, {Gilli}, {Mullaney}, {Paolillo},
  {Schwope}, {Shemmer}, {Sun}, {Timlin}, \& {Trump}}]{ni21}
{Ni}, Q., {Brandt}, W.~N., {Chen}, C.-T., {et~al.} 2021, \apjs, 256, 21

\bibitem[{{Planck Collaboration} {et~al.}(2014){Planck Collaboration},
  {Abergel}, {Ade}, {Aghanim}, \& et~al.}]{planckdust}
{Planck Collaboration}, {Abergel}, A., {Ade}, P.~A.~R., {Aghanim}, N., \&
  et~al. 2014, \aap, 571, A11

\bibitem[{{Polletta} {et~al.}(2007){Polletta}, {Tajer}, {Maraschi},
  {Trinchieri}, {Lonsdale}, {Chiappetti}, {Andreon}, {Pierre}, {Le F{\`e}vre},
  {Zamorani}, {Maccagni}, {Garcet}, {Surdej}, {Franceschini}, {Alloin},
  {Shupe}, {Surace}, {Fang}, {Rowan-Robinson}, {Smith}, \& {Tresse}}]{pol07}
{Polletta}, M., {Tajer}, M., {Maraschi}, L., {et~al.} 2007, \apj, 663, 81

\bibitem[{{Prevot} {et~al.}(1984){Prevot}, {Lequeux}, {Maurice}, {Prevot}, \&
  {Rocca-Volmerange}}]{pre84}
{Prevot}, M.~L., {Lequeux}, J., {Maurice}, E., {Prevot}, L., \&
  {Rocca-Volmerange}, B. 1984, \aap, 132, 389

\bibitem[{{Reis} {et~al.}(2019){Reis}, {Baron}, \& {Shahaf}}]{rei19}
{Reis}, I., {Baron}, D., \& {Shahaf}, S. 2019, \aj, 157, 16

\bibitem[{{Robin} {et~al.}(2003){Robin}, {Reyl{\'e}}, {Derri{\`e}re}, \&
  {Picaud}}]{rob03}
{Robin}, A.~C., {Reyl{\'e}}, C., {Derri{\`e}re}, S., \& {Picaud}, S. 2003,
  \aap, 409, 523

\bibitem[{{Rodrigues} {et~al.}(2023){Rodrigues}, {Raul Abramo}, {Queiroz},
  {Mart{\'\i}nez-Solaeche}, {P{\'e}rez-R{\`a}fols}, {Bonoli}, {Chaves-Montero},
  {Pieri}, {Gonz{\'a}lez Delgado}, {Morrison}, {Marra}, {M{\'a}rquez},
  {Hern{\'a}n-Caballero}, {D{\'\i}az-Garc{\'\i}a}, {Ben{\'\i}tez}, {Cenarro},
  {Dupke}, {Ederoclite}, {L{\'o}pez-Sanjuan}, {Mar{\'\i}n-Franch}, {Mendes de
  Oliveira}, {Moles}, {Sodr{\'e}}, {Varela}, {V{\'a}zquez Rami{\'o}}, \&
  {Taylor}}]{rod23}
{Rodrigues}, N. V.~N., {Raul Abramo}, L., {Queiroz}, C., {et~al.} 2023, \mnras,
  520, 3494

\bibitem[{{Salvato} {et~al.}(2009){Salvato}, {Hasinger}, {Ilbert}, {Zamorani},
  {Brusa}, {Scoville}, {Rau}, {Capak}, {Arnouts}, {Aussel}, {Bolzonella},
  {Buongiorno}, {Cappelluti}, {Caputi}, {Civano}, {Cook}, {Elvis}, {Gilli},
  {Jahnke}, {Kartaltepe}, {Impey}, {Lamareille}, {Le Floc'h}, {Lilly},
  {Mainieri}, {McCarthy}, {McCracken}, {Mignoli}, {Mobasher}, {Murayama},
  {Sasaki}, {Sanders}, {Schiminovich}, {Shioya}, {Shopbell}, {Silverman},
  {Smol{\v{c}}i{\'c}}, {Surace}, {Taniguchi}, {Thompson}, {Trump}, {Urry}, \&
  {Zamojski}}]{sal09}
{Salvato}, M., {Hasinger}, G., {Ilbert}, O., {et~al.} 2009, \apj, 690, 1250

\bibitem[{{Salvato} {et~al.}(2019){Salvato}, {Ilbert}, \& {Hoyle}}]{sal19}
{Salvato}, M., {Ilbert}, O., \& {Hoyle}, B. 2019, Nature Astronomy, 3, 212

\bibitem[{{Salvato} {et~al.}(2022){Salvato}, {Wolf}, {Dwelly}, {Georgakakis},
  {Brusa}, {Merloni}, {Liu}, {Toba}, {Nandra}, {Lamer}, {Buchner}, {Schneider},
  {Freund}, {Rau}, {Schwope}, {Nishizawa}, {Klein}, {Arcodia}, {Comparat},
  {Musiimenta}, {Nagao}, {Brunner}, {Malyali}, {Finoguenov}, {Anderson},
  {Shen}, {Ibarra-Medel}, {Trump}, {Brandt}, {Urry}, {Rivera}, {Krumpe},
  {Urrutia}, {Miyaji}, {Ichikawa}, {Schneider}, {Fresco}, {Boller}, {Haase},
  {Brownstein}, {Lane}, {Bizyaev}, \& {Nitschelm}}]{sal22}
{Salvato}, M., {Wolf}, J., {Dwelly}, T., {et~al.} 2022, \aap, 661, A3

\bibitem[{{Scodeggio} {et~al.}(2018){Scodeggio}, {Guzzo}, {Garilli}, {Granett},
  {Bolzonella}, {de la Torre}, {Abbas}, {Adami}, {Arnouts}, {Bottini}, {Cappi},
  {Coupon}, {Cucciati}, {Davidzon}, {Franzetti}, {Fritz}, {Iovino}, {Krywult},
  {Le Brun}, {Le F{\`e}vre}, {Maccagni}, {Ma{\l}ek}, {Marchetti}, {Marulli},
  {Polletta}, {Pollo}, {Tasca}, {Tojeiro}, {Vergani}, {Zanichelli}, {Bel},
  {Branchini}, {De Lucia}, {Ilbert}, {McCracken}, {Moutard}, {Peacock},
  {Zamorani}, {Burden}, {Fumana}, {Jullo}, {Marinoni}, {Mellier}, {Moscardini},
  \& {Percival}}]{sco18}
{Scodeggio}, M., {Guzzo}, L., {Garilli}, B., {et~al.} 2018, \aap, 609, A84

\bibitem[{{Tanaka} {et~al.}(2018){Tanaka}, {Coupon}, {Hsieh}, {Mineo},
  {Nishizawa}, {Speagle}, {Furusawa}, {Miyazaki}, \& {Murayama}}]{tan18}
{Tanaka}, M., {Coupon}, J., {Hsieh}, B.-C., {et~al.} 2018, \pasj, 70, S9

\bibitem[{{Tasca} {et~al.}(2017){Tasca}, {Le F{\`e}vre}, {Ribeiro}, {Thomas},
  {Moreau}, {Cassata}, {Garilli}, {Le Brun}, {Lemaux}, {Maccagni},
  {Pentericci}, {Schaerer}, {Vanzella}, {Zamorani}, {Zucca}, {Amorin},
  {Bardelli}, {Cassar{\`a}}, {Castellano}, {Cimatti}, {Cucciati}, {Durkalec},
  {Fontana}, {Giavalisco}, {Grazian}, {Hathi}, {Ilbert}, {Paltani}, {Pforr},
  {Scodeggio}, {Sommariva}, {Talia}, {Tresse}, {Vergani}, {Capak}, {Charlot},
  {Contini}, {de la Torre}, {Dunlop}, {Fotopoulou}, {Guaita}, {Koekemoer},
  {L{\'o}pez-Sanjuan}, {Mellier}, {Salvato}, {Scoville}, {Taniguchi}, \&
  {Wang}}]{tas17}
{Tasca}, L.~A.~M., {Le F{\`e}vre}, O., {Ribeiro}, B., {et~al.} 2017, \aap, 600,
  A110

\bibitem[{{Weaver} {et~al.}(2022){Weaver}, {Kauffmann}, {Ilbert}, {McCracken},
  {Moneti}, {Toft}, {Brammer}, {Shuntov}, {Davidzon}, {Hsieh}, {Laigle},
  {Anastasiou}, {Jespersen}, {Vinther}, {Capak}, {Casey}, {McPartland},
  {Milvang-Jensen}, {Mobasher}, {Sanders}, {Zalesky}, {Arnouts}, {Aussel},
  {Dunlop}, {Faisst}, {Franx}, {Furtak}, {Fynbo}, {Gould}, {Greve}, {Gwyn},
  {Kartaltepe}, {Kashino}, {Koekemoer}, {Kokorev}, {Le F{\`e}vre}, {Lilly},
  {Masters}, {Magdis}, {Mehta}, {Peng}, {Riechers}, {Salvato}, {Sawicki},
  {Scarlata}, {Scoville}, {Shirley}, {Silverman}, {Sneppen}, {Smolc̆i{\'c}},
  {Steinhardt}, {Stern}, {Tanaka}, {Taniguchi}, {Teplitz}, {Vaccari}, {Wang},
  \& {Zamorani}}]{wea22}
{Weaver}, J.~R., {Kauffmann}, O.~B., {Ilbert}, O., {et~al.} 2022, \apjs, 258,
  11

\end{thebibliography}

%

\begin{appendix}

\section{\label{app:preprocess} Data pre-processing for classification}

Before applying the classification to \Euclid detected objects, a few
pre-processing steps are required.

\subsection{\label{app:ssc:gred} Galactic reddening correction}

For the determination of photometric redshifts and physical properties, both \Phosphoros{} and \NNPZ{} perform an SED-dependent Galactic reddening correction following \citet{gal17}. Such a correction is needed for classification as well. This is done in the following way both in the training sample and in the target catalogues.

We employ an empirical SED-dependent approach in which the SED of a source is directly derived from observed photometry by a cubic spline interpolation of fluxes in the $(u),\,g,\,r,\,i,\,z,\,\YE,\,\JE,$ and $\HE$ bands. The knots of the interpolation correspond to the reference wavelengths\footnote{The reference wavelength for a filter transmission $T$ is commonly defined as $\lambda_{\rm ref}=\int\lambda\, T(\lambda)\,\diff\lambda\,/\int T(\lambda)\,\diff\lambda$.} for these bands. SEDs are also linearly extrapolated outside the boundaries of the wavelength coverage. For a given source, the correction for the Galactic extinction in the band $k$ is then estimated as the ratio between the integrated fluxes without and with Galactic extinction:
\be
C^k_{\rm red}=\frac{\int\lambda\,\bar{f}_{\lambda}\,T_k(\lambda)\,\diff\lambda}{\int
  \lambda\,\bar{f}_{\lambda}\,10^{-0.4A_{\lambda,{\rm MW}}}\,T_k(\lambda)\,\diff\lambda}\,,
\ee
where $T_k$ is the filter transmission for the band $k$, $A_{\lambda,{\rm MW}}$ is the Galactic extinction associated with the source, and $\bar{f}_{\lambda}$ is the estimated SED. Based on the \Euclid simulations, we verified that such a correction differs by less than 1\% from the actual correction, and it guarantees the same classification performance similar to those obtained in the absence of Galactic extinction.

\subsection{\label{app:ssc:impute} Data imputation}

Unlike most of machine-learning algorithms, the \PRF{} can handle missing data in a quite natural way (an object with a non-measured feature will just propagate both to the left and the right of a tree node with 50\% probability). However, tests on the \Euclid simulations highlighted that the imputation of missing data based on object SEDs improves the performance of the \PRF{} classifiers, adding useful information with respect to no data.

Our method for data imputation is based on the $k$-nearest neighbours
(kNN) algorithm \citep{cov67}. Given a source catalogue, objects without missing photometry are taken as training sample for the kNN algorithm. The imputed fluxes in a given band are the average values from the $k=15$ nearest neighbours of the training set. Uniform weights are used in the average. The uncertainties of imputed values are computed as a squared sum of the standard deviation of the neighbour fluxes and the median of the neighbour flux errors,
$$
\sigma^2(F_{kNN})=\sigma^2(F_{\rm neigh})+{\rm median}(\sigma_{\rm neigh})^2\,.
$$
The second term has been introduced in order to avoid that errors coming from objects with similar fluxes but low S/N, would be underestimated.

Tests on simulations have shown that the method is able to estimate one missing photometric flux per object with a good accuracy, without significant bias even for noisy objects. The typical error on the imputed magnitudes is $\sigma_{\rm mag}\la0.1$ for objects with ${\rm AB\,mag}<23$, and increases to 0.2--0.4 at magnitudes 24.5, depending on the band. The largest errors are observed in $u$ and $g$ bands. Increasing the number of missing bands the method becomes less and less precise, and objects with more than three missing fluxes are not classified.

If all or the majority of objects in a catalogue have missing data, imputation is performed as a simple linear interpolation or extrapolation of measured fluxes. In this case, uncertainties are again the squared sum of two terms: the error due to the linear interpolation; and an error proportional to the imputed flux. The latter guarantees a maximum S/N, which is chosen to be 5 in the case of interpolation and 1 for extrapolation.

\subsection{\label{app:asinhmag} Asinh magnitudes and colours}

In order to handle negative fluxes, which can often be found for faint objects, we convert the observed fluxes, $F_{\rm obs}$, into `asinh' magnitudes \citep{lup99}, which can be defined even for negative fluxes and converge to standard magnitudes at high flux. They are defined as:
\be
\mu(F_{\rm obs}) = -a[\sinh^{-1}(F_{\rm obs}/2b)+\ln b-m_0]\,,
\ee
where $a=2.5\logten {\rm e}$ and $m_0$ is the magnitude zero point, which is $\ln(3.631\times 10^9)$ for fluxes measured in $\mu$Jy. The term $b$ is an arbitrary softening parameter that defines at which scale `asinh' magnitudes transition from logarithmic to linear behaviour. The optimal value of $b$ should be close to the standard deviation of the observed fluxes. For each photometric band, we set $b$ equal to the median of the flux error distribution. Errors on $\mu(F)$ are defined consequently: $\sigma(\mu)=a\,\sigma(F)/\sqrt{F^2+4b^2}$. 

Colours are hence defined as the difference between the `asinh' magnitudes in two bands. In the classification, we consider all possible colour combinations, so, for the eight(nine) bands we have a total of 28(36) colours in the EDF-F and EDF-S (EDF-N).

\subsection{\label{app:ssc:bandphot} Photometry reconstruction}

For the training sample of classification we need to compute the fluxes in the UNIONS and \Euclid bands, based on DES and VIDEO/UltraVISTA photometry. To do this, we take advantage of the \Phosphoros{} algorithm. For each source, we determine its best-fitting SED, using the proper configuration for the object class, and we compute the corrective factors to convert DES and VIDEO/UltraVISTA photometry into UNIONS and \Euclid photometry. In particular, UNIONS photometry is obtained from
\be
F_{X_U}=F_{X_{\rm DES}}\,\frac{\tilde{F}_{X_U}}{\tilde{F}_{X_{\rm DES}}}\,,
\ee
where $X_U$ is one of the UNIONS $g,\,r,\,i,\,z\,$ bands and $\tilde{F}_X$ are the fluxes computed from the best-fit SED in the UNIONS and DES band. The same correction is applied to the uncertainties. Because DES $u$-band photometry was not available in all fields, we compute the CFIS $u$ fluxes directly from the best-fit SED, that is $F_u=\tilde{F}_u$. Due to the possible large uncertainties in their estimates, we set the $u$ flux errors equal to $\tilde{F}_u$ too.

The estimate of fluxes in the \Euclid bands is a bit more complex: the VIS passband covers a large wavelength range, between $0.53<\lambda/\mu{\rm m}<0.92$, while the NISP passbands are broader than the VISTA ones. We proceed therefore as above, but combining fluxes from different filters:
\begin{flalign}
  F_{\IE}&=\frac{0.6F_{r_{\rm DES}}+0.4F_{i_{\rm DES}}+0.05F_{z_{\rm DES}}}{0.6\tilde{F}_{r_{\rm DES}}+0.4\tilde{F}_{i_{\rm DES}}+0.05\tilde{F}_{z_{\rm DES}}}\,\tilde{F}_{\IE}\,;
  \nonumber \\
  F_{\YE}&=\frac{0.75F_{Y_{\rm VISTA}}+0.25F_{J_{\rm VISTA}}}{0.75\tilde{F}_{Y_{\rm VISTA}}+0.25\tilde{F}_{J_{\rm VISTA}}}\,\tilde{F}_{\YE}\,;
  \nonumber \\
  F_{\JE}&=\frac{0.75F_{J_{\rm VISTA}}+0.25F_{H_{\rm VISTA}}}{0.75\tilde{F}_{J_{\rm VISTA}}+0.25\tilde{F}_{H_{\rm VISTA}}}\,\tilde{F}_{\JE}\,;
  \nonumber \\
  F_{\HE}&=F_{H_{\rm VISTA}}\,\frac{\tilde{F}_{\HE}}{\tilde{F}_{H_{\rm VISTA}}}\,.
\end{flalign}
The above filter combinations are chosen based on simulated data in order to minimise the scatter between the true and estimated fluxes. The flux uncertainties are computed consequently. 

\section{\label{app:sc:other} Other PHZ data products}

\subsection{\label{app:sc:sed} SED reconstruction}

Due to the chromatic dependence of the VIS PSF, galaxy SEDs in the wavelength range covered by the VIS band must be taken into account in order to obtain unbiased measurements of their shape \citep{cro13}. Similarly, a successful determination of the PSF modelling requires knowledge of the star SEDs used for the PSF calibration. Star and galaxy SEDs are computed inside the Production pipeline (see the core science branch of \cref{f:prod}) using the \NNPZ{} algorithm. They are output in specific catalogues (\texttt{catalogue.phz\_star\_sed} and \texttt{catalogue.phz\_galaxy\_sed} in the EAS) as a vector of 55 idealised narrow-band filters of 10\,nm width, spanning the wavelength range $450<\lambda/{\rm nm}<1000$. Similarly to the physical properties, the narrow-band fluxes are recovered from \NNPZ{} as weighted means of the 30 closest neighbours (where the weight is the $\chi^2$ distance metric in the broad-band flux space).

\NNPZ{} reference samples for the SED modelling are built by OU-PHZ, using a different strategy for stars and galaxies. For stars, they are created using the {\it Gaia} BP/RP spectrophotometry of bright stars with \Euclid counterpart, under the hypothesis that {\it Gaia} stars across the \Euclid footprint cover all the relevant stellar types and metallicities, without any significant difference in terms of Galactic reddening between {\it Gaia} and \Euclid stars. Narrow-band fluxes are computed by integrating the {\it Gaia} spectra in each of the 55 filters.

The reference sample for galaxies cannot rely on spectrophotometric data, as for stars. Galaxy spectroscopic samples are in fact typically incomplete and much shallower than \Euclid, and their use could introduce biases in the SED reconstruction. Currently, the only resources for measuring galaxy SEDs are from broad- and intermediate-band photometry present in the source catalogues of special fields like COSMOS and CDFS. To reconstruct galaxy SEDs, we follow the approach developed in Euclid Collaboration: Tarsitano et al. (in prep.), which combines template fitting and Gaussian processes. The method is validated through a specific metric that evaluates the accuracy of the reconstructed SEDs as a function of redshift bins, and in comparison with simulations. This analysis suggests that the method is robust and passes the requirements at almost all redshifts.

\subsection{\label{app:sc:tom} Tomographic binning}
 
The PHZ pipeline attributes to each object a tomographic bin based on the median point estimate of photometric redshifts computed by Phosphoros. Two different definitions of tomographic bins are considered assuming a configuration of 13 redshift bins between 0.2 and 2.5: equidistant bins that are uniformly distributed in the redshift range; equipopulated bins that should contain a similar number of sources. The latter are defined using the \Euclid simulations. This information is provided in the photo-$z$ catalogue, but it is not used in the Q1 release.

\section{\label{app:sc:cat} Description of the output catalogues}

The data products of the PHZ pipeline consist in eight catalogues with more than 500 columns. A description of the contents of each catalogue can be found in \href{http://st-dm.pages.euclid-sgs.uk/data-product-doc/dmq1/phzdpd/dpcards/phz_phzpfoutputcatalog.html}{Products for core science} and in \href{http://st-dm.pages.euclid-sgs.uk/data-product-doc/dmq1/phzdpd/dpcards/phz_phzpfoutputforl3.html}{Products for non-cosmological science}.

\end{appendix}

\end{document}